  \documentclass[journal,11pt,draftcls,onecolumn]{IEEEtran}
   \newcommand{\DRAFT}[1]{#1}
  \newcommand{\FINAL}[1]{}

\usepackage{here}
\usepackage{color}

\usepackage[noadjust]{cite}      

\usepackage{graphicx}  

\usepackage{subfigure} 

\usepackage{stfloats}  
\usepackage{amsmath}   

\usepackage{xspace,bbm,epic,eepic,amssymb}
\usepackage[latin1]{inputenc}
\usepackage{theorem}
\usepackage{citesort}

\def\XS{\xspace}
\DeclareMathAlphabet{\mathb}{OML}{cmm}{b}{it}
\def\sbm#1{\ensuremath{\mathb{#1}}}       
\def\sbmm#1{\ensuremath{\boldsymbol{#1}}} 
\def\sbv#1{\ensuremath{\mathbf{#1}}}      
\def\scu#1{\ensuremath{\mathcal{#1\XS}}}  
\def\sbl#1{\ensuremath{\mathbbm{#1}}}     

\def\Ab{{\sbm{A}}\XS}  
\def\Bb{{\sbm{B}}\XS}
\def\Cb{{\sbm{C}}\XS}
\def\Ib{{\sbm{I}}\XS}  
\def\Pb{{\sbm{P}}\XS}  
\def\Xb{{\sbm{X}}\XS}  

\def\ab{{\sbm{a}}\XS}
\def\bb{{\sbm{b}}\XS} 
\def\cb{{\sbm{c}}\XS} 
\def\gb{{\sbm{g}}\XS}
\def\hb{{\sbm{h}}\XS}
\def\rb{{\sbm{r}}\XS}
\def\tb{{\sbm{t}}\XS}
\def\vb{{\sbm{v}}\XS}
\def\xb{{\sbm{x}}\XS}
\def\yb{{\sbm{y}}\XS}
\def\zb{{\sbm{z}}\XS}
\def\unb     {{\sbv{1}}\XS}

\def\betab {{\sbmm{\beta}}\XS}
\def\Deltab {{\sbmm{\Delta}}\XS}

\def\varepsilonb {{\sbmm{\varepsilon}}\XS}

\def\zerob   {{\sbv{0}}\XS}  

\def\Cc{{\scu{C}}\XS}   
\def\Nc{{\scu{N}}\XS}   
\def\Qc{{\scu{Q}}\XS}

\def\Rbb{{\sbl{R}}\XS}

\RequirePackage{ifthen}
\RequirePackage{calc}

\newcommand{\taille}[1][\scad]{%
\ifthenelse{#1 = -5}{}{}%
\ifthenelse{#1 = -4}{\tiny}{}%
\ifthenelse{#1 = -3}{\scriptsize}{}%
\ifthenelse{#1 = -2}{\footnotesize}{}%
\ifthenelse{#1 = -1}{\small}{}%
\ifthenelse{#1 = 0}{\normalsize}{}%
\ifthenelse{#1 = 1}{\large}{}%
\ifthenelse{#1 = 2}{\Large}{}%
\ifthenelse{#1 = 3}{\LARGE}{}%
\ifthenelse{#1 = 4}{\huge}{}%
\ifthenelse{#1 = 5}{\Huge}{}}

\newboolean{@serif}
\setboolean{@serif}{true}
\def\scad{-5} 

\newcounter{taille}
\newcommand{\sca}[2][\scad]{\setcounter{taille}{#1}%
  \ifthenelse{\boolean{@serif}}
  {{\taille[\thetaille]\textsc{#2}}}
  {\setcounter{taille}{\value{taille}-1}{\uppercase{\taille[\thetaille]#2}}}}

\def\numero{n°}

\def\eg{\textit{e.g.,}\XS}

\def\ie{\textit{i.e.,}\XS}
\def\IFF{\textit{if and only if}\XS}

\def\vs{\textit{vs}\XS}

\def\CRAN{Centre de Recherche en Automatique de Nancy\XS}
\newcommand{\cran}[1][\scad]{\sca[#1]{cran, umr 7039},
   Universit\'e de Lorraine, \sca[#1]{cnrs}\XS}

\newcommand{\adresseCRAN}[1][\scad]{Campus Sciences, B.P. 70239, 
F-5   4506 Vand{\oe}uvre-l\`es-Nancy, France\XS}

                \def\stdpth#1{(#1)}
              \def\stdacc#1{\{#1\}}
\def\cro#1{\left[#1\right]}                \def\stdcro#1{[#1]}
\def\bars#1{\left|#1\right|}               \def\stdbars#1{|#1|}
             \def\stdnorm#1{\|#1\|}
   \def\stdscal#1{\langle#1\rangle}
 
\def\bigpth#1{\bigl(#1\bigr)}              
            
\def\bigcro#1{\bigl[#1\bigr]}              \def\biggcro#1{\biggl[#1\biggr]}
\def\bigbars#1{\bigl|#1\bigr|}             
\def\bignorm#1{\bigl\|#1\bigr\|}

\def\Bigpth#1{\Bigl(#1\Bigr)}              
            \def\Biggacc#1{\Biggl\{#1\Biggr\}}
\def\Bigcro#1{\Bigl[#1\Bigr]}              
             \def\Biggbars#1{\Biggl|#1\Biggr|}

\def\sgn{{\mathrm{sgn}}}

\def\spansub#1{{\mathrm{span}}\stdpth{#1}}

\def\Card#1{{\mathrm{Card}}\cro{#1}}

\def\spark#1{{\mathrm{spark}}\stdpth{#1}}

\def\argmax{\mathop{\mathrm{arg\,max}}} 
\def\argmin{\mathop{\mathrm{arg\,min}}} 

\def\froc#1#2{{#1/#2}}                  

\newtheorem{definition}{Definition}
\newtheorem{example}{Example}
\newtheorem{remark}{Remark}
\newtheorem{proposition}{Proposition}
\newtheorem{lemma}{Lemma}
\newtheorem{theorem}{Theorem}
\newtheorem{corollary}{Corollary}

\setboolean{@serif}{false}

\begin{document}
\title{Joint $k$-step analysis of Orthogonal Matching Pursuit and
  Orthogonal Least Squares}
\author{Charles~Soussen$^\star$,~\IEEEmembership{}%
~R\'emi Gribonval,~\IEEEmembership{}
~J\'er\^ome~Idier,~\IEEEmembership{}
~and~C\'edric Herzet~\IEEEmembership{}
  \thanks{C.~Soussen is with the \CRAN (\cran). Campus Sciences, B.P.
    70239, F-54506 Vand{\oe}uvre-l\`es-Nancy, France\DRAFT{. Tel:
      (+33)-3 83 68 44 71, Fax: (+33)-3 83 68 44 62} (e-mail:
    Charles.Soussen@cran.uhp-nancy.fr.)
    This work was carried out in part while C. Soussen was visiting 
    IRCCyN during the academic year 2010-2011 with the financial support 
    of CNRS.}
  \thanks{R.~Gribonval and C.~Herzet are with INRIA Rennes - Bretagne
    Atlantique, Campus de Beaulieu, F-35042 Rennes Cedex, France
    \DRAFT{Tel: (+33)-2 99 84 25 06/73 50, Fax: (+33)-2 99 84 71 71}
    (e-mail: Remi.Gribonval@inria.fr; Cedric.Herzet@inria.fr).
    R.~Gribonval acknowledges the partial support of the European
    Union's FP7-FET program, SMALL project, under grant agreement
    \numero~225913. }
  \thanks{J.~Idier is with the Institut de Recherche en Communications
    et Cybernétique de Nantes (IRCCyN, UMR CNRS 6597), BP 92101, 1 rue
    de la No\"e, 44321 Nantes Cedex~3, France\DRAFT{. Tel: (+33)-2 40
      37 69 09, Fax: (+33)-2 40 37 69 30} (e-mail:
    Jerome.Idier@irccyn.ec-nantes.fr).}%
}
\markboth{Soussen, Gribonval, Idier, Herzet: Technical Report%
\FINAL{, \today}}{Soussen, Gribonval, Idier, Herzet: Using the Document Class IEEEtran.cls}

\maketitle

\begin{abstract}
  Tropp's analysis of Orthogonal Matching Pursuit (OMP) using the
  Exact Recovery Condition (ERC)~\cite{Tropp04} is extended to a first
  exact recovery analysis of Orthogonal Least Squares (OLS). We show
  that when the ERC is met, OLS is guaranteed to exactly recover the
  unknown support in at most $k$ iterations. Moreover, we provide a
  closer look at the analysis of both OMP and OLS when the ERC is not
  fulfilled. The existence of dictionaries for which some subsets are
  never recovered by OMP is proved. This phenomenon also appears with
  basis pursuit where support recovery depends on the sign patterns,
  but it does not occur for OLS. Finally, numerical experiments show
  that none of the considered algorithms is uniformly better than the
  other but for correlated dictionaries, guaranteed exact recovery may
  be obtained after fewer iterations for OLS than for OMP.
\end{abstract}
\begin{IEEEkeywords}
  ERC exact recovery condition; Orthogonal Matching Pursuit;
  Orthogonal Least Squares; Order Recursive Matching Pursuit;
  Optimized Orthogonal Matching Pursuit; forward selection.
\end{IEEEkeywords}

\DRAFT{\newpage\tableofcontents}

\section{Introduction}
\label{sec:intro}
\IEEEPARstart{C}{lassical} greedy subset selection algorithms include,
by increasing order of complexity: Matching Pursuit
(MP)~\cite{Mallat93}, Orthogonal Matching Pursuit (OMP)~\cite{Pati93}
and Orthogonal Least Squares (OLS)~\cite{Chen89,Natarajan95}. OLS is
indeed relatively expensive in comparison with OMP since OMP performs
one linear inversion per iteration whereas OLS performs as many linear
inversions as there are non-active atoms. We refer the reader to the
technical report~\cite{Blumensath07} for a comprehensive review on the
difference between OMP and OLS.

OLS is referred to using many other names in the literature. It is
known as forward selection in statistical regression~\cite{Miller02}
and as the greedy algorithm~\cite{Natarajan95}, Order Recursive
Matching Pursuit (ORMP)~\cite{Cotter99} and Optimized Orthogonal
Matching Pursuit (OOMP)~\cite{RebolloNeira02} in the signal processing
literature, all these algorithms being actually the same. It is worth
noticing that the above-mentioned algorithms were introduced by
following either an optimization~\cite{Chen89,Miller02} or an
orthogonal projection methodology~\cite{Natarajan95}, or
both~\cite{Cotter99,RebolloNeira02}. In the optimization viewpoint,
the atom yielding the largest decrease of the approximation error is
selected. This leads to a greedy sub-optimal algorithm dedicated to
the minimization of the approximation error. In the orthogonal
projection viewpoint, the atom selection rule is defined as an
extension of the OMP rule: the data vector and the dictionary atoms
are being projected onto the subspace that is orthogonal to the span
of the active atoms, and the \emph{normalized} projected atom having
the largest inner product with the data residual is selected. As the
number of active atoms increases by one at any iteration, the
projections are done on a subspace whose dimension is decreasing.

\subsection{Main objective of the paper}
Our primary goal is to address the OLS exact recovery analysis from
noise-free data and to investigate the connection between the OMP and
OLS exact recovery conditions. In the literature, much attention was
paid to the exact recovery analysis of sparse algorithms that are
faster than OLS, \eg thresholding algorithms and simpler greedy
algorithms like OMP~\cite{Tropp10}. But to the best of our knowledge,
no exact recovery result is available for OLS. In their recent
paper~\cite{Davies12}, Davies and Eldar mention this issue and state
that the relation between OMP and OLS remains unclear.

\subsection{Existing results for OMP}
\label{sec:exist_omp}
Our starting point is the existing $k$-step analysis of OMP whose
structure is somewhat close to OLS. The notion of $k$-step solution
property was defined in~\cite{Donoho08}: ``any vector with at most $k$
nonzeros can be recovered from the related noise-free observation in
at most $k$ iterations.'' The $k$-step property will also be referred
to as the ``exact support recovery'' in the following. Exact recovery
studies of OMP rely on alternate methodologies.

Tropp's Exact Recovery Condition (ERC)~\cite{Tropp04} is a necessary
and sufficient condition of exact support recovery in a worst case
analysis. On the one hand, if a subset of $k$ atoms satisfies the ERC,
then it can be recovered from any linear combination of the $k$ atoms
in at most $k$ steps. On the other hand, when the ERC is not
satisfied, one can generate a counterexample (\ie a specific
combination of the $k$ atoms) for which OMP fails, \ie OMP selects a
wrong atom during its first $k$ iterations. Specifically, the atom
selected in the \emph{first} iteration is a wrong one.

Davenport and Wakin~\cite{Davenport10} used another analysis to show
that OMP yields exact support recovery under certain Restricted
Isometry Property (RIP) assumptions, and several improvements of their
condition were proposed more recently~\cite{Liu12,Mo12}. Actually, the
ERC necessarily holds when the latter conditions are fulfilled since
the ERC is a sufficient and worst case necessary condition of exact
recovery.

\subsection{Generalization of Tropp's condition}
We propose to extend Tropp's condition to OLS. We remark that the very
first iteration of OLS is identical to that of OMP: the first selected
atom is the one whose inner product with the input vector is maximal.
Therefore, when the ERC does not hold, the counterexample for which
the first iteration of OMP fails also yields a failure of the first
iteration of OLS. Hence one cannot expect to derive an exact recovery
condition for OLS that would be weaker than the ERC at the first
iteration. We show that the ERC indeed ensures the success of OLS.

We further address the case where the ERC does not hold, \ie the first
iteration of OMP/OLS is not guaranteed to always succeed but
nevertheless succeeds for a given vector. In practice, even for non
random dictionaries, this phenomenon is likely to occur since the ERC
is a worst case necessary condition. The purpose of a large part of
the paper is specifically to analyze what is going on in the remaining
iterations for these vectors. With $\ell_1$ minimization, the
situation is clearer because support recovery depends on the sign
patterns~\cite[Theorem~2]{Plumbley07} and one can predict whether a
specific vector will be recovered independently of the support
amplitudes. For greedy algorithms, things are more tricky and it is
one of the purpose of the paper to analyze this. We introduce weaker
conditions than the ERC which guarantee that an exact support recovery
will occur in the subsequent iterations. These extended recovery
conditions coincide with the ERC at the first iteration but differ
from it afterwards.

Our main results state that:
\begin{itemize}
\item The ERC is a sufficient condition of exact recovery for OLS in
  at most $k$ steps (Theorem~\ref{th:tropp_ols}).
\item When the early iterations of OMP/OLS have all succeeded, we
  derive two sufficient conditions, named ERC-OMP and ERC-OLS, for the
  recovery of the remaining true atoms
  (Theorem~\ref{th:suffic_omp_ols}). This result is a ($k-q$)-step
  property, where $q$ stands for the number of iterations which have
  been already performed.
\item Moreover, we show that our conditions are, in some sense,
  necessary (Theorems~\ref{th:necess_ols} and~\ref{th:necess_omp}).
\end{itemize}
The criteria we provide might not necessarily be directly useful for
practitioners working in the field. In fact, just as many other
theoretical success guarantees, they are rather ``motivational'': by
proving that the considered algorithms are guaranteed to perform well
in a restricted regime, they strengthen our confidence that the
heuristics behind the algorithms are reasonably grounded.
Practitioners know that the algorithms indeed work much beyond the
considered restricted regime, but proving this fact would typically
require probabilistic arguments, based on models of random dictionary
or random input signals~\cite{Tropp07,Fletcher09}. Despite their
potential interest, the theoretical results that can be foreseen in
this spirit would be highly dependent on the adequacy of such models
to the actual distribution of data from the real world.

\subsection{Organization of the paper}
In Section~\ref{sec:prereq}, we recall the principle of OMP and OLS
and their interpretation in terms of orthogonal projections. Then, we
properly define the notions of successful support recovery and support
recovery failure. Section~\ref{sec:overview} is dedicated to the
analysis of OMP and OLS at any iteration where the most technical
developments and proofs are omitted for readability reasons. These
important elements can be found in the appendix section~\ref{sec:cns}.
In Section~\ref{sec:empirical}, we show using Monte Carlo simulations
that there is no systematic implication between the ERC-OMP and
ERC-OLS conditions but we exhibit some elements of discrimination in
favor of OLS.

\section{Notations and prerequisites}
\label{sec:prereq}
The following notations will be used in this paper.
$\stdscal{\,.\,,\,.\,}$ refers to the inner product between vectors,
and $\|\,.\,\|$ and $\|\,.\,\|_1$ stand for the Euclidean norm and the
$\ell_1$ norm, respectively. $.^\dag$ denotes the pseudo-inverse of a
matrix. For a full rank and undercomplete matrix, we have
$\Xb^\dag=(\Xb^t\Xb)^{-1}\Xb^t$ where $.^t$ stands for the matrix
transposition. When \Xb is overcomplete, $\spark{\Xb}$ denotes the
minimum number of columns from \Xb that are linearly
dependent~\cite{Donoho03b}. The letter \Qc denotes some subset of the
column indices, and $\Xb_{\Qc}$ is the submatrix of \Xb gathering the
columns indexed by \Qc. Finally, $\Pb_\Qc=\Xb_\Qc\Xb_\Qc^\dag$ and
$\Pb_\Qc^{\perp}=\Ib-\Pb_\Qc$ denote the orthogonal projection
operators on $\spansub{\Xb_\Qc}$ and $\spansub{\Xb_\Qc}^{\perp}$,
where $\spansub{\Xb}$ stands for the column span of \Xb,
$\spansub{\Xb}^{\perp}$ is the orthogonal complement of
$\spansub{\Xb}$ and \Ib is the identity matrix whose dimension is
equal to the number of rows in \Xb.

\subsection{Subset selection}
Let $\Ab=\stdcro{\ab_1,\ldots,\ab_n}$ denote the dictionary gathering
normalized atoms $\ab_i\in\Rbb^m$. \Ab is a matrix of size $m\times
n$. Assuming that the atoms are normalized is actually not necessary
for OLS as the behavior of OLS is unchanged whether the atoms are
normalized or not~\cite{Blumensath07}. On the contrary, OMP is highly
sensitive to the normalization of atoms since its selection rule
involves the inner products between the current residual and the
non-selected atoms.

We consider a subset $\Qc^\star$ of $\stdacc{1,\ldots,n}$ of
cardinality $k\triangleq\Card{\Qc^\star}<\min(m,n)$ and study the
behavior of OMP and OLS \emph{for all} inputs
$\yb\in\spansub{\Ab_{\Qc^\star}}$, \ie for any combination
$\yb=\Ab_{\Qc^\star}\tb$ where the submatrix $\Ab_{\Qc^\star}$ is of
size $m\times k$ and the weight vector $\tb\in\Rbb^k$. The $k$ atoms
$\stdacc{\ab_i,\,i\in\Qc^\star}$ indexed by $\Qc^\star$ will be
referred to as the ``true'' atoms while for the remaining (``wrong'')
atoms $\stdacc{\ab_j,\,j\notin\Qc^\star}$, we will use the subscript
notation $j$. The forward greedy algorithms considered in this paper
start from the empty support and select a new atom per iteration. At
intermediate iterations $q\in\stdacc{0,\ldots,k-1}$, we denote by \Qc
the current support (with $\Card{\Qc}=q$).

Throughout the paper, we make the general assumption that
$\Ab_{\Qc^\star}$ is full rank. Note that the representation
$\yb=\Ab_{\Qc^\star}\tb$ is not guaranteed to be unique under this
assumption: there may be another $k$-term representation
$\yb=\Ab_{\Qc'}\tb'$ where $\Ab_{\Qc'}$ includes some wrong atoms
$\ab_j$. The stronger assumption $\spark{\Ab}> 2k$ is a necessary and
sufficient condition for uniqueness of any $k$-term
representation~\cite{Donoho03b}. Therefore, when $\spark{\Ab}> 2k$,
the selection of a wrong atom by a greedy algorithm disables a
$k$-term representation of \yb in $k$ steps~\cite{Tropp04}. We make
the weak assumption that $\Ab_{\Qc^\star}$ is full rank because it is
sufficient to elaborate our exact recovery conditions under which no
wrong atom is selected in the first $k$ iterations.

\subsection{OMP and  OLS algorithms}
The common feature between OMP and OLS is that they both perform an
orthogonal projection whenever the support \Qc is updated: the data
approximation reads $\Pb_\Qc\yb$ and the residual error is defined by
\begin{equation*}
  \rb_\Qc\triangleq\yb-\Pb_\Qc\yb=\Pb_{\Qc}^{\perp}\yb.
\end{equation*}
Let us now recall how the selection rule of OLS differs from that of
OMP.

At each iteration of OLS, the atom $\ab_\ell$ yielding the minimum
least-square error $\|\rb_{\Qc\cup\stdacc{\ell}}\|^2$ is selected:
\begin{equation*}
  \ell^{\textrm{OLS}}\in\argmin_{i\notin\Qc}\|\rb_{\Qc\cup\stdacc{i}}\|^2
  \label{eq:ols_rule_ai}
\end{equation*}
and $n-\Card{\Qc}$ least-square problems are being solved to compute
$\|\rb_{\Qc\cup\stdacc{i}}\|^2$ for all $i\notin\Qc$ (\footnote{Our
  purpose is not to focus on the OLS implementation. However, let us
  just mention that in the typical implementation, the least-square
  problems are solved recursively using the Gram Schmidt
  orthonormalization procedure~\cite{Chen89}.})~\cite{Chen89}. On the
contrary, OMP adopts the simpler rule
\begin{equation*}
  \ell^{\textrm{OMP}}\in\argmax_{i\notin\Qc}\stdbars{\stdscal{\rb_\Qc,\ab_i}}
  \label{eq:omp_rule_ai}
\end{equation*}
to select the new atom $\ab_\ell$ and then solves only one
least-square problem to update
$\rb_{\Qc\cup\stdacc{\ell}}$~\cite{Blumensath07}. Depending on the
application, the OMP and OLS stopping rules can involve a maximum
number of atoms and/or a residual threshold. Note that when the data
are noise-free (they read as $\yb=\Ab_{\Qc^\star}\tb$) and no wrong
atom is selected, the squared error $\|\rb_{\Qc}\|^2$ is equal to 0
after at most $k$ iterations. Therefore, we will consider no more than
$k$ iterations in the following.

\subsection{Geometric interpretation}
\label{sec:geom}
A geometric interpretation in terms of orthogonal projections will be
useful for deriving recovery conditions. It is essentially inspired by
the technical report of Blumensath and Davies~\cite{Blumensath07} and
by Davenport and Wakin's analysis of OMP under the RIP
assumption~\cite{Davenport10}.

We introduce the notation $\tilde{\ab}_i=\Pb_{\Qc}^\perp\ab_i$ for the
projected atoms onto $\spansub{\Ab_\Qc}^\perp$ where for simplicity,
the dependence upon \Qc is omitted. When there is a risk of confusion,
we will use $\tilde{\ab}_i^{\Qc}$ instead of $\tilde{\ab}_i$. Notice
that $\tilde{\ab}_i=\zerob$ \IFF $\ab_i\in\spansub{\Ab_\Qc}$. In
particular, $\tilde{\ab}_i=\zerob$ for $i\in\Qc$. Finally, we define
the normalized vectors
\begin{align*}
  \tilde{\bb}_i=\left\{
        \setlength{\arraycolsep}{0.25cm}
        \begin{array}{cl}
          \froc{\tilde{\ab}_i}{\|\tilde{\ab}_i\|}&
          \textrm{if}~\tilde{\ab}_i\ne \zerob,\\
          \zerob&\textrm{otherwise}.
          \end{array}
        \right .
        \label{eq:bi}
\end{align*}
Again, we will use $\tilde{\bb}_i^{\Qc}$ when there is a risk of
confusion.

We now emphasize that the projected atoms $\tilde{\ab}_i$ (or
$\tilde{\bb}_i$) play a central role in the analysis of both OMP and
OLS. Because the residual $\rb_\Qc=\Pb_{\Qc}^{\perp}\yb$ lays in
$\spansub{\Ab_\Qc}^{\perp}$,
$\stdscal{\rb_\Qc,\ab_i}=\stdscal{\rb_\Qc,\tilde{\ab}_i}$ and the OMP
selection rule rereads:
\begin{equation}
  \ell^{\textrm{OMP}}\in\argmax_{i\notin\Qc}\stdbars{\stdscal{\rb_\Qc,\tilde{\ab}_i}}
  \label{eq:omp_rule}
\end{equation}
whereas for OLS, minimizing $\|\rb_{\Qc\cup\stdacc{i}}\|^2$ with
respect to $i\notin\Qc$ is equivalent to maximizing
$\|\rb_{\Qc}\|^2-\|\rb_{\Qc\cup\stdacc{i}}\|^2=
\stdscal{\rb_\Qc,\tilde{\bb}_i}^2$ (see \eg~\cite{RebolloNeira02} for
a complete calculation):
\begin{equation}
  \ell^{\textrm{OLS}}\in\argmax_{i\notin\Qc}
  \stdbars{\stdscal{\rb_\Qc,\tilde{\bb}_i}}.
  \label{eq:ols_rule}
\end{equation}
We notice that~\eqref{eq:omp_rule} and~\eqref{eq:ols_rule} only rely
on the vectors $\rb_\Qc$ and $\tilde{\ab}_i$ belonging to the subspace
$\spansub{\Ab_\Qc}^{\perp}$. OMP maximizes the inner product
$\stdbars{\stdscal{\rb_\Qc,\tilde{\ab}_i}}$ whereas OLS minimizes the
angle between $\rb_\Qc$ and $\tilde{\ab}_i$ (this difference was
already stressed and graphically illustrated in~\cite{Blumensath07}).
When the dictionary is close to orthogonal, \eg for dictionaries
satisfying the RIP assumption, this does not make a strong difference
since $\|\tilde{\ab}_i\|$ is close to 1 for all
atoms~\cite{Davenport10}. But in the general case, $\|\tilde{\ab}_i\|$
may have wider variations between 0 and 1 leading to substantial
differences between the behavior of OMP and OLS.

\subsection{Definition of successful recovery and failure}
\label{sec:def_success}
Throughout the paper, we will use the common acronym Oxx in statements
that apply to both OMP and OLS. Moreover, we define the unifying
notation:
\begin{equation*}
  \tilde{\cb}_i\triangleq
  \left\{
    \begin{array}{ll}
      \tilde{\ab}_i & \textrm{for OMP},\\
      \tilde{\bb}_i & \textrm{for OLS}.
    \end{array}
  \right.
\end{equation*}

We first stress that in special cases where the Oxx selection rule
yields multiple solutions including a wrong atom, \ie when
\begin{align}
  \max_{i\in\Qc^\star\backslash\Qc}\stdbars{\stdscal{\rb_\Qc,\tilde{\cb}_i}}&=
  \max_{j\notin\Qc^\star}\stdbars{\stdscal{\rb_\Qc,\tilde{\cb}_{j}}},
  \label{eq:exaequo}
\end{align}
we consider that Oxx automatically makes the wrong decision. Tropp
used this convention for OMP and showed that when the upper bound on
his ERC condition (see Section~\ref{sec:tropp_omp}) is reached, the
limit situation~\eqref{eq:exaequo} occurs, hence a wrong atom is
selected at the first iteration~\cite{Tropp04}. Let us now properly
define the $k$-step property for successful support recovery.
\begin{definition}
  Oxx with $\yb\in\spansub{\Ab_{\Qc^\star}}$ as input succeeds \IFF no
  wrong atom is selected and the residual $\rb_\Qc$ is equal to \zerob
  after at most $k$ iterations.
  \label{def:wide_success_recov}
\end{definition}
When a successful recovery occurs, the subset \Qc yielded by Oxx
satisfies $\Qc_\yb\subseteq\Qc\subseteq\Qc^\star$ where $\Qc_\yb$ is
the subset indexed by the nonzero weights $t_i$'s in the decomposition
$\yb=\Ab_{\Qc^\star}\tb$. When all $t_i$'s are nonzero, $\Qc_\yb$
identifies with $\Qc^\star$ and a successful recovery cannot occur in
less than $k$ iterations.

The word ``failure'' refers to the exact contrary of successful recovery. 
\begin{definition}
  Oxx with $\yb\in\spansub{\Ab_{\Qc^\star}}$ as input fails when at
  least one wrong atom is selected during the first $k$ iterations. In
  particular, Oxx fails when~\eqref{eq:exaequo} occurs with
  $\rb_\Qc\ne\zerob$.
  \label{def:wide_failure_recov}
\end{definition}
The notion of successful recovery may be defined in a weaker sense:
Plumbley~\cite[Corollary~4]{Plumbley07} pointed out that there exist
problems for which the ERC fails but nevertheless, a ``delayed
recovery'' occurs after more than $k$ steps, in that a larger support
including $\Qc^\star$ is found, but all atoms which do not belong to
$\Qc^\star$ are weighted by 0 in the solution vector. Recently, a
delayed recovery analysis of OMP using RIP assumptions was proposed
in~\cite{Zhang11b}, and then extended to the weak OMP
algorithm~\cite{Foucart11}. In the present paper, no more than $k$
steps are performed, thus delayed recovery is considered as a recovery
failure.

\section{Overview of our recovery analysis of OMP and OLS}
\label{sec:overview}
In this section, we present our main concepts and results regarding
the sparse recovery guarantees with OLS, their connection with the
existing OMP results and the new results regarding OMP. For clarity
reasons, we place the technical analysis including most of the proofs
in the main appendix section~\ref{sec:cns}. Let us first recall
Tropp's ERC condition for OMP which is our starting point.

\subsection{Tropp's ERC condition for OMP}
\label{sec:tropp_omp}
\begin{theorem}{\textbf{[ERC is a sufficient recovery condition for
      OMP and a necessary condition at the first
      iteration~\cite[Theorems~3.1~and~3.10]{Tropp04}]}}
  If $\Ab_{\Qc^\star}$ is full rank and
  \begin{equation}
    \max_{j\notin\Qc^\star}\stdacc{
      F_{\Qc^\star}(\ab_j)\triangleq\bignorm{\Ab_{\Qc^\star}^\dag\ab_j}_1}<1,
    \tag*{ERC($\Ab,\Qc^\star$)}
    \label{eq:erc_tropp}
  \end{equation}
  then OMP succeeds for any input $\yb\in\spansub{\Ab_{\Qc^\star}}$.
  Furthermore, when ERC($\Ab,\Qc^\star$) does not hold, there exists
  $\yb\in\spansub{\Ab_{\Qc^\star}}$ for which some wrong atom is
  selected at the first iteration of OMP. When $\spark{\Ab}>2k$, this
  implies that OMP cannot recover the (unique) $k$-term representation
  of \yb.
  \label{th:tropp_omp} 
\end{theorem}
Note that ERC($\Ab,\Qc^\star$) involves the dictionary atoms but not
their weights as it results from a worst case analysis: if
ERC($\Ab,\Qc^\star$) holds, then a successful recovery occurs with
$\yb=\Ab_{\Qc^\star}\tb$ \emph{whatever} $\tb\in\Rbb^k$.

\subsection{Main theorem}
\label{sec:main_theorem}
A theorem similar to Theorem~\ref{th:tropp_omp} applies to OLS. 
\begin{theorem}{\textbf{[ERC is a sufficient recovery condition for OLS and a
    necessary condition at the first iteration]}}
If $\Ab_{\Qc^\star}$ is full rank and ERC($\Ab,\Qc^\star$) holds, then
OLS succeeds for any input $\yb\in\spansub{\Ab_{\Qc^\star}}$.
Furthermore, when ERC($\Ab,\Qc^\star$) does not hold, there exists
$\yb\in\spansub{\Ab_{\Qc^\star}}$ for which some wrong atom is
selected at the first iteration of OLS. When $\spark{\Ab}> 2k$, this
implies that OLS cannot recover the (unique) $k$-term representation
of \yb.
  \label{th:tropp_ols} 
\end{theorem}
The necessary condition result is obvious since the very first
iteration of OLS coincides with that of OMP and the ERC is a worst
case necessary condition for OMP. The core of our contribution is the
sufficient condition result for OLS. We now introduce the main
concepts on which our analysis relies. They also lead to a more
precise analysis of OMP from the second iteration.

\subsection{Main concepts}
\label{sec:main_concepts}
Let us keep in mind that the ERC is a worst case necessary
condition \emph{at the first iteration}. But what happens when the ERC
is not met but nevertheless, the first $q$ iterations of Oxx select
$q$ true atoms ($q<k$)? Can we characterize the exact recovery
conditions at the $(q+1)$-th iteration? We will answer to these
questions and provide:
\begin{enumerate}
\item an extension of the ERC condition to the $q$-th iteration of
  OMP;
\item a new necessary and sufficient condition dedicated to the $q$-th
  iteration of OLS.
\end{enumerate}
This will allow us to prove Theorem~\ref{th:tropp_ols} as a special
case of the latter condition when $q=0$.

In the following two paragraphs, we introduce useful notations for a
single wrong atom $\ab_j$ and then define our new exact recovery
conditions by considering all wrong atoms together. \Qc plays the role
of the subset found by Oxx after the first $q$ iterations.

\subsubsection{Notations related to a single wrong atom}
For $\Qc\subsetneq\Qc^\star$ and $j\notin\Qc^\star$, we define:
\begin{align}
  F_{\Qc^\star,\Qc}^{\mathrm{OMP}}(\ab_j)&
  \triangleq\sum_{i\in\Qc^\star\backslash\Qc}
  \,\bigbars{\bigpth{\Ab_{\Qc^\star}^\dag\ab_j}(i)}
  \label{eq:Ferc_omp}\\
  F_{\Qc^\star,\Qc}^{\mathrm{OLS}}(\ab_j)& \triangleq
  \sum_{i\in\Qc^\star\backslash\Qc}
  \frac{\|\tilde{\ab}_{i}\|}{\|\tilde{\ab}_{j}\|}
  \,\bigbars{\bigpth{\Ab_{\Qc^\star}^\dag\ab_j}(i)}
  \label{eq:Ferc_ols}
\end{align}
when $\tilde{\ab}_{j}\ne\zerob$ and
$F_{\Qc^\star,\Qc}^{\mathrm{Oxx}}(\ab_j)=0$ when
$\tilde{\ab}_{j}=\zerob$ (we recall that
$\tilde{\ab}_i=\Pb_{\Qc}^\perp\ab_i$ and
$\tilde{\ab}_{j}=\Pb_{\Qc}^\perp\ab_j$ depend on \Qc). Up to some
manipulations on orthogonal projections,~\eqref{eq:Ferc_omp}
and~\eqref{eq:Ferc_ols} can be rewritten as follows.
\begin{lemma}
  Assume that $\Ab_{\Qc^\star}$ is full rank. For
  $\Qc\subsetneq\Qc^\star$ and $j\notin\Qc^\star$,
  $F_{\Qc^\star,\Qc}^{\mathrm{OMP}}(\ab_j)$ and
  $F_{\Qc^\star,\Qc}^{\mathrm{OLS}}(\ab_j)$ also read
  \begin{align}
    F_{\Qc^\star,\Qc}^{\mathrm{OMP}}(\ab_j) &=\stdnorm{
      \tilde{\Ab}_{\Qc^\star\backslash\Qc}^\dag\tilde{\ab}_{j}}_1
    \label{eq:Ferc_omp_bis}\\
    F_{\Qc^\star,\Qc}^{\mathrm{OLS}}(\ab_j) &=\stdnorm{
      \tilde{\Bb}_{\Qc^\star\backslash\Qc}^\dag\tilde{\bb}_{j}}_1
    \label{eq:Ferc_ols_bis}
  \end{align}
  where the matrices $\tilde{\Ab}_{\Qc^\star\backslash\Qc}
  =\stdacc{\tilde{\ab}_{i},\,i\in\Qc^\star\backslash\Qc}$ and
  $\tilde{\Bb}_{\Qc^\star\backslash\Qc}=
  \stdacc{\tilde{\bb}_{i},\,i\in\Qc^\star\backslash\Qc}$ of size
  $m\times(k-q)$ are full rank.
  \label{lem:reexpress_erc}
\end{lemma}
Lemma~\ref{lem:reexpress_erc} is proved in
Appendix~\ref{app:reexpress_erc}.

\subsubsection{ERC-Oxx conditions for the whole dictionary}
We define four binary conditions by considering all the wrong atoms
together:
\begin{align}
  \max_{j\notin\Qc^\star}& \,F_{\Qc^\star,\Qc}^{\mathrm{OMP}}(\ab_j)<1
  \tag*{ERC-OMP($\Ab,\Qc^\star,\Qc$)}
  \label{eq:erc_omp_q}\\
  \max_{j\notin\Qc^\star}&\,F_{\Qc^\star,\Qc}^{\mathrm{OLS}}(\ab_j)<1
  \tag*{ERC-OLS($\Ab,\Qc^\star,\Qc$)}
  \label{eq:erc_omp_j}\\
  \max_{\substack{\Qc\subsetneq\Qc^\star\\\Card{\Qc}=q}}&
  \max_{j\notin\Qc^\star} F_{\Qc^\star,\Qc}^{\mathrm{OMP}}(\ab_j) <1
  \tag*{ERC-OMP($\Ab,\Qc^\star,q$)}
  \label{eq:erc_ols_q}\\
  \max_{\substack{\Qc\subsetneq\Qc^\star\\\Card{\Qc}=q}}&
  \max_{j\notin\Qc^\star} F_{\Qc^\star,\Qc}^{\mathrm{OLS}}(\ab_j) <1
  \tag*{ERC-OLS($\Ab,\Qc^\star,q$)}
  \label{eq:erc_ols_j}
\end{align}
We will use the common notations
$F_{\Qc^\star,\Qc}^{\mathrm{Oxx}}(\ab_j)$,
ERC-Oxx($\Ab,\Qc^\star,\Qc$) and ERC-Oxx($\Ab,\Qc^\star,q$) for
statements that are common to both OMP and OLS.
\begin{remark}
  $F_{\Qc^\star,\emptyset}^{\mathrm{OMP}}(\ab_j)$ and
  $F_{\Qc^\star,\emptyset}^{\mathrm{OLS}}(\ab_j)$ both reread
  $F_{\Qc^\star}(\ab_j)=\bignorm{\Ab_{\Qc^\star}^\dag\ab_j}_1$ since
  $\tilde{\ab}_i^\emptyset$ reduces to $\ab_i$ which is of unit norm.
  Thus, ERC-Oxx($\Ab,\Qc^\star,\emptyset$) and
  ERC-Oxx($\Ab,\Qc^\star,0$) all identify with ERC($\Ab,\Qc^\star$).
\label{rk:erc0}
\end{remark}

\subsection{Sufficient conditions of exact recovery at any iteration}
\label{sec:overview_sufficient}
The sufficient conditions of Theorems~\ref{th:tropp_omp}
and~\ref{th:tropp_ols} reread as special cases of the following
theorem where $\Qc=\emptyset$.
\begin{theorem}
  {\textbf{[Sufficient recovery condition for Oxx after $q$ successful
      iterations]}} Assume that $\Ab_{\Qc^\star}$ is full rank. If Oxx
  with $\yb\in\spansub{\Ab_{\Qc^\star}}$ as input selects
  $\Qc\subsetneq\Qc^\star$ and ERC-Oxx($\Ab,\Qc^\star,\Qc$) holds,
  then Oxx succeeds in at most $k$ steps.
  \label{th:suffic_omp_ols}
\end{theorem}
The following corollary is a straightforward adaptation of
Theorem~\ref{th:suffic_omp_ols} to ERC-Oxx($\Ab,\Qc^\star,q$).
\begin{corollary}
  Assume that $\Ab_{\Qc^\star}$ is full rank. If Oxx with
  $\yb\in\spansub{\Ab_{\Qc^\star}}$ as input selects true atoms during
  the first $q<k$ iterations and ERC-Oxx($\Ab,\Qc^\star,q$) holds,
  then Oxx succeeds in at most $k$ iterations.
  \label{cor:suffic_omp_ols}
\end{corollary}

The key element which enables us to establish
Theorem~\ref{th:suffic_omp_ols} is a recursive relation linking
$F_{\Qc^\star,\Qc}^{\mathrm{Oxx}}(\ab_j)$ with
$F_{\Qc^\star,\Qc'}^{\mathrm{Oxx}}(\ab_j)$ when \Qc is
increased by one element of $\Qc^\star\backslash\Qc$, resulting in
subset $\Qc'$. This leads to the main technical novelty of the paper,
stated in Lemma~\ref{lem:erc_omp_ols_rec} (see
Appendix~\ref{sec:sufficient}). From the thorough analysis of this
recursive relation, we elaborate the following lemma which guarantees
the monotonic decrease of
$F_{\Qc^\star,\Qc}^{\mathrm{Oxx}}(\ab_j)$ when
$\Qc\subsetneq\Qc^\star$ is growing.
\begin{lemma}
  Assume that $\Ab_{\Qc^\star}$ is full rank. Let
  $\Qc\subsetneq\Qc'\subsetneq\Qc^\star$. For any $j\notin\Qc^\star$,
  \begin{align}
    F_{\Qc^\star,\Qc'}^{\mathrm{OMP}}(\ab_j)\leqslant
    F_{\Qc^\star,\Qc}^{\mathrm{OMP}}(\ab_j)
    \label{eq:omp_decrease}\\
    F_{\Qc^\star,\Qc}^{\mathrm{OLS}}(\ab_j)<1
    \Rightarrow
    F_{\Qc^\star,\Qc'}^{\mathrm{OLS}}(\ab_j)\leqslant
    F_{\Qc^\star,\Qc}^{\mathrm{OLS}}(\ab_j)
    \label{eq:ols_decrease}
  \end{align}
  \label{lem:erc_decrease}
\end{lemma}
We refer the reader to Appendix~\ref{sec:sufficient} for the proofs of
Lemmas~\ref{lem:erc_omp_ols_rec} and~\ref{lem:erc_decrease}, and then
Theorem~\ref{th:suffic_omp_ols}.

\subsection{Necessary conditions of exact recovery at any iteration}
\label{sec:overview_necessary}
We recall that the ERC is a worst case necessary condition
guaranteed for the selection of a true atom by OMP and OLS in their
very first iteration. We provide extended results stating that ERC-Oxx
are worst case necessary conditions when the first iterations of Oxx
have succeeded, up to a ``reachability assumption'' defined hereafter,
for OMP.
\begin{definition}{\textbf{[Reachability]}} 
  \Qc is reachable \IFF there exists an input $\yb=\Ab_\Qc\tb$ where
  $t_i\ne 0$ for all $i$, for which Oxx recovers \Qc in $\Card{\Qc}$
  iterations. Specifically, the selection
  rule~\eqref{eq:omp_rule}-\eqref{eq:ols_rule} always yields a unique
  maximum.
  \label{def:reachability}
\end{definition}
We start with the OLS condition which is simpler.

\subsubsection{OLS necessary condition}
\label{sec:overview_necess_results_OLS}
\begin{theorem}
  {\textbf{[Necessary condition for OLS after $q$ iterations]}} Let
  $\Qc\subsetneq\Qc^\star$ be a subset of cardinality $q$. Assume that
  $\Ab_{\Qc^\star}$ is full rank and $\spark{\Ab}\geqslant (q+2)$. If
  ERC-OLS($\Ab,\Qc^\star,\Qc$) does not hold, then there exists
  $\yb\in\spansub{\Ab_{\Qc^\star}}$ for which OLS selects \Qc in the
  first $q$ iterations and then a wrong atom $j\notin\Qc^\star$ at
  iteration $(q+1)$.
  \label{th:necess_ols}
\end{theorem}
Theorem~\ref{th:necess_ols} is proved in Appendix~\ref{sec:necessary}.
An obvious corollary can be obtained by replacing \Qc with $q$ akin to
the derivation of Corollary~\ref{cor:suffic_omp_ols} from
Theorem~\ref{th:suffic_omp_ols}. From now on, such obvious corollaries
will not be explicitly stated.

\subsubsection{Reachability issues}
\label{sec:overview_necess_reach}
The reader may have noticed that Theorem~\ref{th:necess_ols} implies
that \Qc can be reached by OLS at least for some input
$\yb\in\spansub{\Ab_{\Qc^\star}}$. In Appendix~\ref{sec:necessary}, we
establish a stronger result: 
\begin{lemma}[\textbf{Reachability by OLS}]
  Any subset \Qc with $\Card{\Qc}\leqslant\spark{\Ab}-2$ can be
  reached by OLS with some input $\yb\in\spansub{\Ab_\Qc}$.
  \label{lem:ols_reach}
\end{lemma}

Perhaps surprisingly, this result does not remain valid for OMP
although it holds under certain RIP
assumptions~\cite[Theorem~4.1]{Davenport10}. We refer the reader to
subsection~\ref{sec:brc_omp} for a simple counterexample where \Qc
cannot be reached by OMP not only for $\yb\in\spansub{\Ab_\Qc}$ but
also for any $\yb\in\Rbb^m$.

The same somewhat surprising phenomenon of non-reachability may also
occur with $\ell_1$ minimization, associated to certain $k$-faces of
the $\ell_1$ ball in $\Rbb^n$ whose projection through \Ab yields
interior faces~\cite{Donoho04b}. Specifically, for a given \xb
supported by \Qc, Fuchs' necessary and sufficient condition for exact
support recovery from $\yb=\Ab\xb$~\cite{Fuchs04,Plumbley07} involves
the signs of the nonzero amplitudes (denoted by
$\varepsilonb\triangleq\sgn(\xb)\in\stdacc{-1,1}^q$) but not their
values. Either Fuchs' condition is met \emph{for any} vector having
support \Qc and signs \varepsilonb, thus all these vectors will be
correctly recovered, or no vector \xb having support $\Qc$ and signs
$\varepsilonb$ can be recovered. It follows that \Qc is non-reachable
with $\ell_1$ minimization when Fuchs' condition is simultaneously
false for all possible signs \varepsilonb. We refer the reader to
Appendix~\ref{app:brc_l1} for further details.

\subsubsection{OMP necessary condition including reachability assumptions}
\label{sec:overview_necess_results_OMP}
Our necessary condition for OMP is somewhat tricky because we must
assume that \Qc is reachable by OMP using some input in
$\spansub{\Ab_{\Qc}}$.
\begin{theorem}
  {\textbf{[Necessary condition for OMP after $q$ iterations]}} Assume
  that $\Ab_{\Qc^\star}$ is full rank and $\Qc\subsetneq\Qc^\star$ is
  reachable. If ERC-OMP($\Ab,\Qc^\star,\Qc$) does not hold, then there
  exists $\yb\in\spansub{\Ab_{\Qc^\star}}$ for which OMP selects \Qc
  in the first $q$ iterations and then a wrong atom $j\notin\Qc^\star$
  at iteration $(q+1)$.
  \label{th:necess_omp}
\end{theorem}
Theorem~\ref{th:necess_omp} is proved together with
Theorem~\ref{th:necess_ols} in Appendix~\ref{sec:necessary}. Setting
aside the reachability issues, the principle of the proof is common to
OMP and OLS. We proceed the proof of the sufficient condition
(Theorem~\ref{th:suffic_omp_ols}) backwards, as was done
in~\cite[Theorem~3.10]{Tropp04} in the case $\Qc=\emptyset$.

In the special case where $q=1$, Theorem~\ref{th:necess_omp}
simplifies to a corollary similar to the OLS necessary condition
(Theorem~\ref{th:necess_ols}) because any subset \Qc of cardinality 1
is obviously reachable using the atom indexed by \Qc as input vector.
\begin{corollary}
  {\textbf{[Necessary condition for OMP in the second iteration]}}
  Assume that $\Ab_{\Qc^\star}$ is full rank and let $i\in\Qc^\star$.
  If ERC-OMP($\Ab,\Qc^\star,\stdacc{i}$) does not hold, then there
  exists $\yb\in\spansub{\Ab_{\Qc^\star}}$ for which OMP selects
  $\ab_i$ and then a wrong atom in the first two iterations.
  \label{cor:necess_omp1a}
\end{corollary}

\subsubsection{Discrimination between OMP and OLS at the $k$-th iteration}
We provide an element of discrimination between OMP and OLS when their
first $k-1$ iterations have selected true atoms, so that there is one
remaining true atom which has not been chosen. 
\begin{theorem}
  {\textbf{[Guaranteed success of the $k$-th iteration of OLS]}} If
  $\stdcro{\Ab_{\Qc^\star},\ab_j}$ is full rank for any
  $j\notin\Qc^\star$, then ERC-OLS($\Ab,\Qc^\star,k-1$) is
  true. Thus, if the first $k-1$ iterations of OLS select true atoms,
  the last true atom is necessarily selected in the $k$-th iteration.
  \label{th:erc_ols_last_is_true}
\end{theorem}
This result is straightforward from the definition of OLS in the
optimization viewpoint: ``OLS selects the new atom yielding the least
possible residual'' and because in the $k$-th iteration, the last true
atom yields a zero valued residual. Another (analytical) proof of
Theorem~\ref{th:erc_ols_last_is_true}, given below, is based on the
definition of ERC-OLS($\Ab,\Qc^\star,k-1$). It will enable us to
understand why the statement of Theorem~\ref{th:erc_ols_last_is_true}
is not valid for OMP.
\begin{IEEEproof} %
  Assume that OLS yields a subset $\Qc\subsetneq\Qc^\star$ after $k-1$
  iterations. Let $\ab_\mathrm{last}$ denote the last true atom so
  that $\Ab_{\Qc^\star}=\stdcro{\Ab_\Qc,\ab_\mathrm{last}}$ up to some
  column permutation. Since $\tilde{\Bb}_{\Qc^\star\backslash\Qc}$
  reduces to $\tilde{\bb}^\Qc_\mathrm{last}$ which is of unit norm,
  the pseudo-inverse $\tilde{\Bb}_{\Qc^\star\backslash\Qc}^\dag$ takes
  the form $\bigcro{\tilde{\bb}^\Qc_\mathrm{last}}^t$.
  Finally,~\eqref{eq:Ferc_ols_bis} simplifies to:
  \begin{equation}
    F_{\Qc^\star,\Qc}^{\mathrm{OLS}}(\ab_j)=\stdbars{
      \stdscal{\tilde{\bb}^\Qc_\mathrm{last},
        \tilde{\bb}^\Qc_{j}}}\leqslant 1
  \label{eq:Ferc_ols_last}
  \end{equation}
  since both vectors in the inner product are either of unit norm or
  equal to \zerob. Apply Lemma~\ref{lem:fullrank} in
  Appendix~\ref{app:reexpress_erc}: since for $j\notin\Qc^\star$,
  $\stdcro{\Ab_{\Qc^\star},\ab_j}$ is full rank,
  $\bigcro{\tilde{\bb}_{\mathrm{last}}^\Qc,\tilde{\bb}_{j}^\Qc}$ is
  full rank, thus~\eqref{eq:Ferc_ols_last} is a strict inequality.
\end{IEEEproof}
Similar to the calculation in the proof above, we rewrite
$F_{\Qc^\star,\Qc}^{\mathrm{OMP}}(\ab_j)$ defined
in~\eqref{eq:Ferc_omp_bis}:
\begin{equation}
  F_{\Qc^\star,\Qc}^{\mathrm{OMP}}(\ab_j)=
  \frac{\stdbars{
      \stdscal{\tilde{\ab}^\Qc_\mathrm{last},\tilde{\ab}^\Qc_{j}}}}
  {\stdnorm{\tilde{\ab}^\Qc_\mathrm{last}}^2}.
  \label{eq:Ferc_omp_last}
\end{equation}
However, we cannot ensure that
$F_{\Qc^\star,\Qc}^{\mathrm{OMP}}(\ab_j)\leqslant 1$ since
$\tilde{\ab}^\Qc_{j}$ and $\tilde{\ab}^\Qc_\mathrm{last}$ are not unit
norm vectors. We refer the reader to subsection~\ref{sec:brc_omp} for
a simple example with four atoms and two true atoms in which OMP is
not guaranteed to select the second true atom when the first has
already been chosen.

To further distinguish OMP and OLS, we elaborate a ``bad recovery
condition'' under which OMP is guaranteed to fail in the sense
that $\Qc^\star$ is not reachable.
\begin{theorem}
  {\textbf{[Sufficient condition for bad recovery with OMP]}}
  Assume that $\Ab_{\Qc^\star}$ is full rank. If
  \begin{equation}
    \min_{
      \substack{\Qc\subsetneq\Qc^\star\\\Card{\Qc}=k-1}}
    \Bigcro{\max_{j\notin\Qc^\star}
      F_{\Qc^\star,\Qc}^{\mathrm{OMP}}(\ab_j)}
    \geqslant 1,
    \tag*{BRC-OMP($\Ab,\Qc^\star$)}  
    \label{eq:brc_omp}
  \end{equation}
  then $\Qc^\star$ cannot be reached by OMP using any input in
  $\spansub{\Ab_{\Qc^\star}}$.
  \label{th:brc_omp}
\end{theorem}
Specifically, BRC-OMP($\Ab,\Qc^\star$) guarantees that a wrong
selection occurs at the $k$-th iteration when the previous iterations
have succeeded.
\begin{IEEEproof}%
  Assume that for some $\yb\in\spansub{\Ab_{\Qc^\star}}$, the first
  $k-1$ iterations of OMP succeed, \ie they select
  $\Qc\subsetneq\Qc^\star$ of cardinality $k-1$. Let
  $\ab_\mathrm{last}$ denote the last true atom
  ($\Ab_{\Qc^\star}=\stdcro{\Ab_{\Qc},\ab_\mathrm{last}}$ up to some
  permutation of columns). The residual $\rb_\Qc$ yielded by OMP after
  $k-1$ iterations is obviously proportional to
  $\tilde{\ab}^\Qc_\mathrm{last}$.

  BRC-OMP($\Ab,\Qc^\star$) implies that ERC-OMP($\Ab,\Qc^\star,\Qc$)
  is false, thus there exists $j\notin\Qc^\star$ such that
  $F_{\Qc^\star,\Qc}^{\mathrm{OMP}}(\ab_j)\geqslant 1$. According
  to~\eqref{eq:Ferc_omp_last},
  $\stdbars{\stdscal{\tilde{\ab}^\Qc_\mathrm{last},\tilde{\ab}^\Qc_{j}}}
  \geqslant\stdnorm{\tilde{\ab}^\Qc_\mathrm{last}}^2$ thus
  $\stdbars{\stdscal{\rb_\Qc,\tilde{\ab}^\Qc_{j}}}
  \geqslant\stdbars{\stdscal{\rb_\Qc,\tilde{\ab}^\Qc_{\mathrm{last}}}}$.
  We conclude that $\ab_{\mathrm{last}}$ cannot be chosen in the
  $k$-th iteration of OMP.
\end{IEEEproof}
Although BRC-OMP($\Ab,\Qc^\star$) may appear restrictive (as a minimum
is involved in the left-hand side), we will see in
Section~\ref{sec:empirical} that it may be frequently met, especially
when the atoms of \Ab are strongly correlated.

\section{Empirical evaluation of the OMP and OLS recovery conditions}
\label{sec:empirical}
The purpose of this section is twofold. In
subsection~\ref{sec:eval_ERC}, we evaluate and compare the ERC-OMP and
ERC-OLS conditions for several kinds of dictionaries. In particular,
we study the dependence of
$F_{\Qc^\star,\Qc}^{\mathrm{Oxx}}\triangleq\max_{j\notin\Qc^\star}
F_{\Qc^\star,\Qc}^{\mathrm{Oxx}}(\ab_j)$ with respect to the
dimensions $m,n$ of the dictionary and the subset cardinalities
$k=\Card{\Qc^\star}$ and $q=\Card{\Qc}$. This allows us to analyze,
for random and deterministic dictionaries, from which iteration $q$
the ERC-Oxx($\Ab,\Qc^\star,\Qc$) condition may be met, \ie
$F_{\Qc^\star,\Qc}^{\mathrm{Oxx}}<1$. In subsection~\ref{sec:brc_omp},
we emphasize the distinction between OMP and OLS by showing that the
bad recovery condition for OMP may be frequently met, especially when
some dictionary atoms are strongly correlated.

\subsection{Dictionaries under consideration}
\label{sec:dicos}
Our recovery conditions will be evaluated for three kinds of
dictionaries.

We consider first randomly Gaussian dictionaries whose entries obey
the standard Gaussian distribution. Once the dictionary elements are
drawn, we normalize each atom in such a way that $\stdnorm{\ab_i}=1$.

``Hybrid'' dictionaries are also studied, whose atoms result from an
additive mixture of a deterministic (constant) and a random component.
Specifically, we set $\ab_i=\alpha_i(\gb_i+t_i\unb)$ where $\gb_i$ is
drawn according to the standard Gaussian distribution, $\unb$ is the
(deterministic) vector whose entries are all equal to 1, and the
scalar $t_i$ obeys the uniform distribution on $[0,T]$, with $T>0$.
Once $\gb_i$ and $t_i$ are drawn, $\alpha_i$ is set in such a way that
$\stdnorm{\ab_i}=1$. In this simulation, the mutual coherence is
increased in comparison to the case $T=0$ (\ie for randomly Gaussian
dictionaries). The random vector $\gb_i$ plays the role of a noise
process added to the deterministic signal $t_i\unb$. When $T$ is
large, the atom normalization makes the noise level very low in
comparison with the deterministic component, thus the atoms are almost
deterministic, and look alike each other.

Finally, we consider a sparse spike train deconvolution problem of the
form $\yb=\hb\ast\xb$, where \hb is a Gaussian impulse response of
variance $\sigma^2$ (for simplicity, the smallest values in \hb are
thresholded so that \hb has a finite support of width $\lceil
6\sigma\rceil$). This is a typical inverse problem in which the
dictionary coherence is large. This problem is known to be a
challenging one since both OMP and OLS are likely to yield false
support recovery in practice~\cite{Lorenz09,Soussen11c,Bourguignon11}.
This is also true for basis pursuit~\cite{Dossal05a}. The problem can
be reformulated as $\yb=\Ab\xb$ where the dictionary \Ab gathers
shifted versions of the impulse response \hb. To be more specific, we
first consider a convolution operator with the same sampling rate for
the input and output signals $\xb$ and $\yb$, and we set boundary
conditions so that the convoluted signal $\hb\ast\xb$ resulting from
\xb can be fully observed without truncation. Thus, \Ab is a slightly
undercomplete ($m>n$ with $m\approx n$) Toeplitz matrix. Alternately,
we perform simulations in which the sampling rate of the input signal
\xb is higher than that of \yb (\ie \yb results from a down-sampling
of $\hb\ast\xb$), leading to an overcomplete dictionary \Ab which does
not have a Toeplitz structure anymore.

Regarding the last two problems, we found that the ERC factor
$F_{\Qc^\star}\triangleq F_{\Qc^\star,\emptyset}^{\mathrm{Oxx}}$ which
is the left hand-side in the ERC($\Ab,\Qc^\star$) condition can become
huge when $T$ (respectively, $\sigma$) is increased. For instance,
when $T$ is equal to 10, 100 and 1000, the average value of
$F_{\Qc^\star}$ is equal to 7, 54 and 322, respectively, for a
dictionary of size $100\times 1000$ and for $k=10$.

\subsection{Evaluation of the ERC-Oxx conditions}
\label{sec:eval_ERC}
We first show that for randomly Gaussian dictionaries, there is no
systematic implication between the ERC-OMP($\Ab,\Qc^\star,\Qc$) and
ERC-OLS($\Ab,\Qc^\star,\Qc$) conditions, nor between
ERC-OMP($\Ab,\Qc^\star,q$) and ERC-OLS($\Ab,\Qc^\star,q$). Then, we
perform more complete numerical simulations to assess the dependence
of $F_{\Qc^\star,\Qc}^{\mathrm{Oxx}}$ with respect to the size $(m,n)$
of the dictionary and the subset cardinalities $k$ and $q$ for the
three kinds of dictionaries. We will build ``phase transition
diagrams'' (in a sense to be defined below) to compare the OMP and OLS
recovery conditions. The general principle of our simulations is 1)~to
draw the dictionary \Ab and the subset $\Qc^\star$; and 2)~to
gradually increase $\Qc\subsetneq\Qc^\star$ by one element until
ERC-Oxx($\Ab,\Qc^\star,\Qc$) is met.
\begin{figure}[!t]
  \centering
  \DRAFT{
    \setlength{\tabcolsep}{0.1cm}
    \begin{tabular}{cc}
      \includegraphics*[width=75mm]{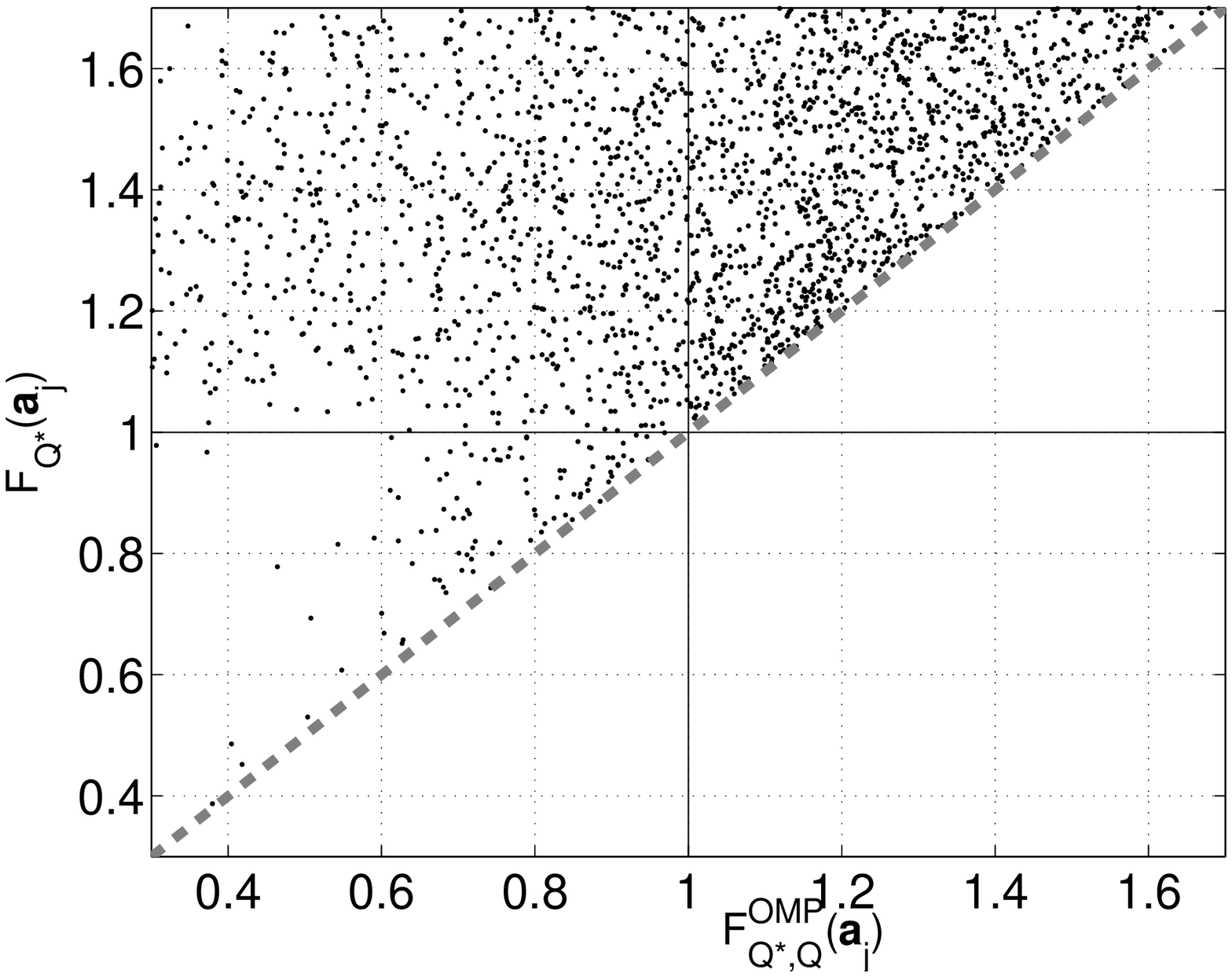}&
      \includegraphics*[width=75mm]{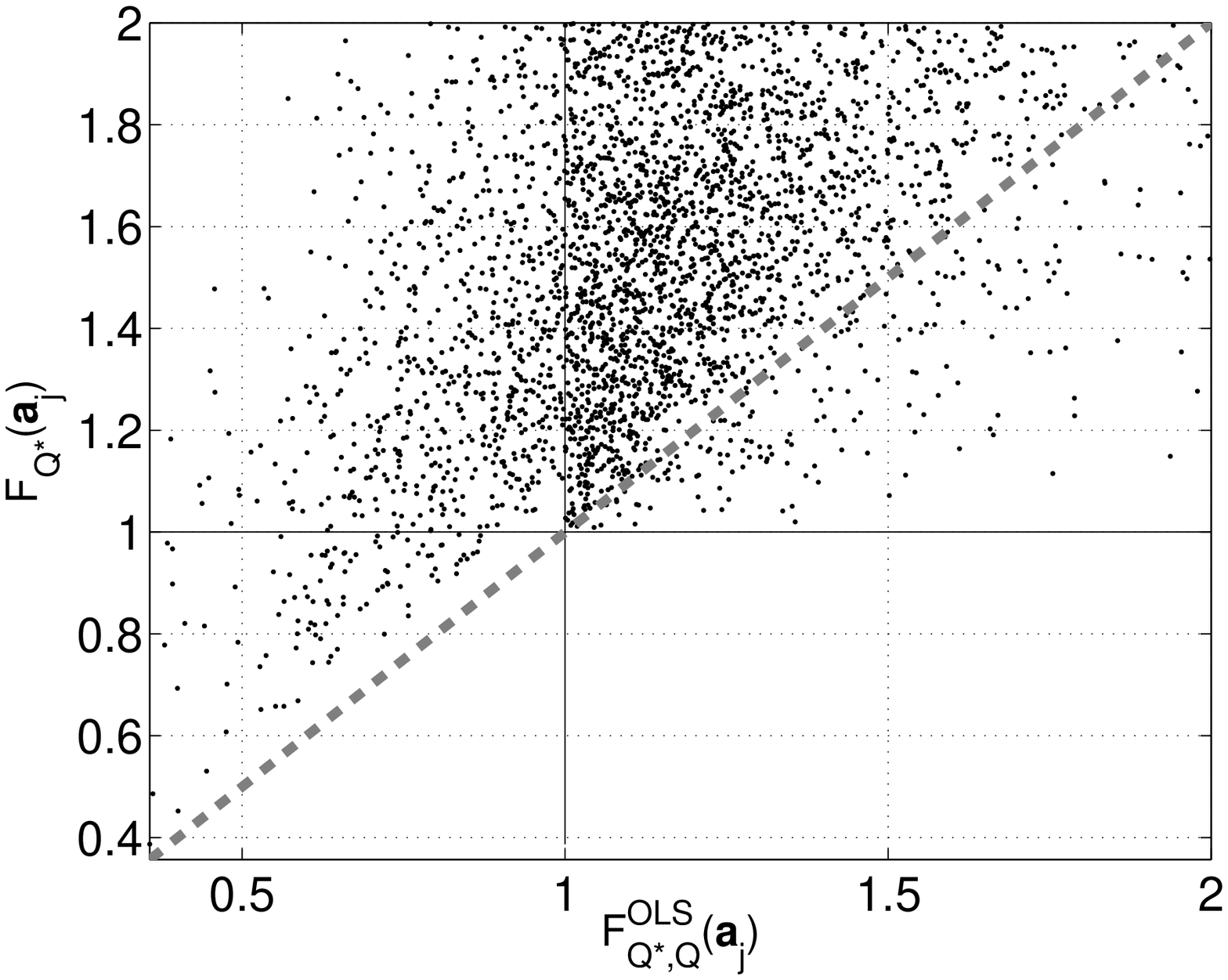}\\
      {\small(a)~$F_{\Qc^\star}(\ab_j)$
        \vs $F_{\Qc^\star,\Qc}^{\mathrm{OMP}}(\ab_j)$.
      } & 
      {\small(b)~$F_{\Qc^\star}(\ab_j)$
        \vs $F_{\Qc^\star,\Qc}^{\mathrm{OLS}}(\ab_j)$.
      }\\
      \includegraphics*[width=75mm]{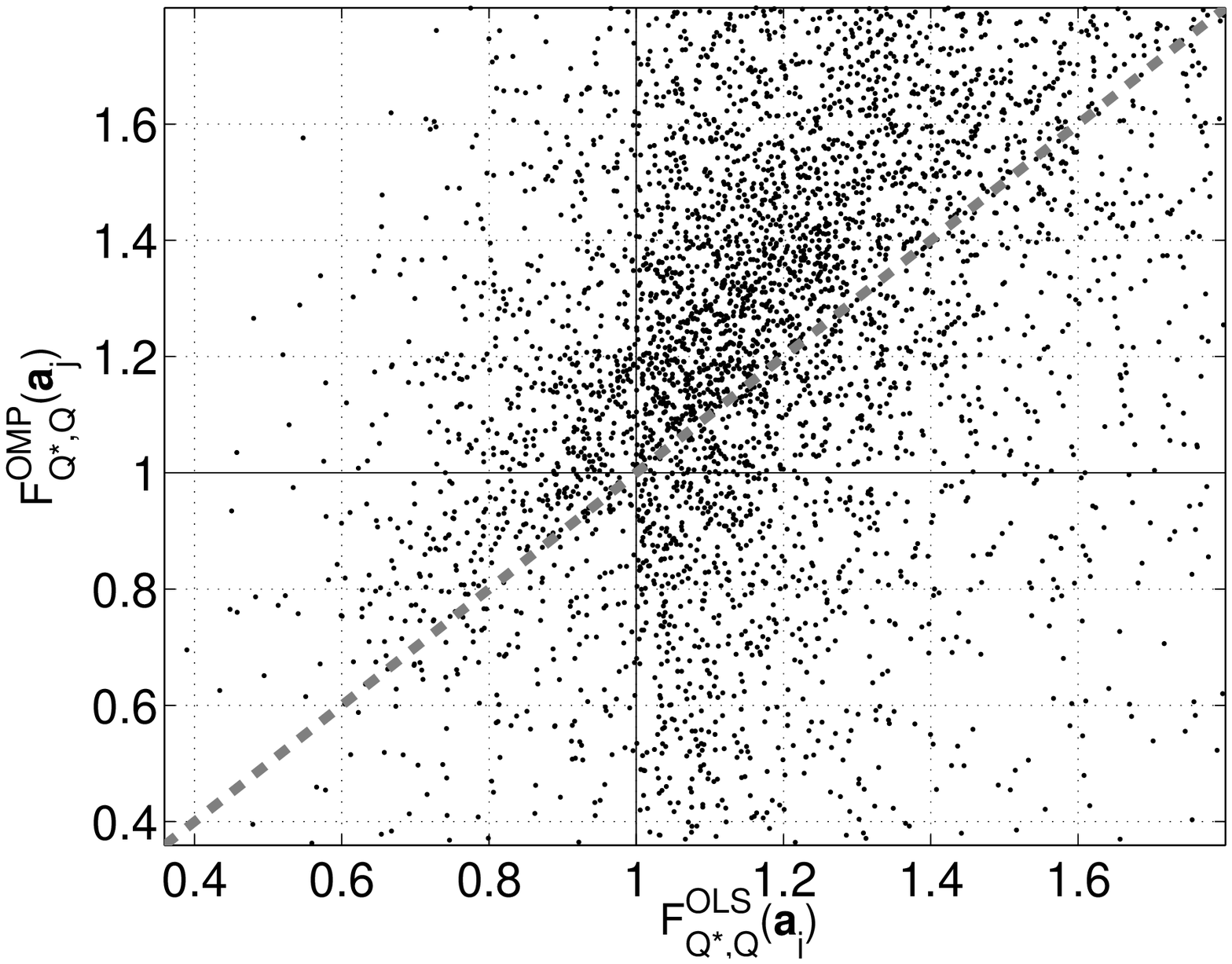}& \\
      {\small(c)~$F_{\Qc^\star,\Qc}^{\mathrm{OMP}}(\ab_j)$
        \vs $F_{\Qc^\star,\Qc}^{\mathrm{OLS}}(\ab_j)$.
      }
    }
    \FINAL{
      \setlength{\tabcolsep}{0.1cm}
      \begin{tabular}{cc}
        (a)&\begin{tabular}{c}
          \includegraphics*[width=70mm]{fig2b}
        \end{tabular}\\
        (b)&\begin{tabular}{c}
          \includegraphics*[width=70mm]{fig1b}
        \end{tabular}\\
        (c)&\begin{tabular}{c}
          \includegraphics*[width=70mm]{fig3b}
        \end{tabular}
      }
    \end{tabular}
    \caption{Comparison of the OMP and OLS exact recovery conditions.
      We draw 10.000 Gaussian dictionaries of size $100\times 11$ and 
      set $k=10$ so that there is only one wrong atom
      $\ab_j$. \Qc is always set to the first atom 
      ($\Card{\Qc}=1$). Plot of 
      (a)~$F_{\Qc^\star}(\ab_j)$
      \vs $F_{\Qc^\star,\Qc}^{\mathrm{OMP}}(\ab_j)$;
      (b)~$F_{\Qc^\star}(\ab_j)$
      \vs $F_{\Qc^\star,\Qc}^{\mathrm{OLS}}(\ab_j)$;
      (c)~$F_{\Qc^\star,\Qc}^{\mathrm{OMP}}(\ab_j)$
      \vs $F_{\Qc^\star,\Qc}^{\mathrm{OLS}}(\ab_j)$.
      For the last subfigure, we keep the trials for which 
      $F_{\Qc^\star}(\ab_j)\geqslant 1$. 
    }
\label{fig:1}
\end{figure}

\subsubsection{There is no logical implication between the ERC-OMP and
  ERC-OLS conditions}
\label{par:1}
We first investigate what is going on after the first iteration
($q=1$). We compare ERC-OMP($\Ab,\Qc^\star,\Qc$) and
ERC-OLS($\Ab,\Qc^\star,\Qc$) for a common dictionary and given subsets
$\Qc\subsetneq\Qc^\star$ with $q=1$. As the recovery conditions take
the form ``for all $j\notin\Qc^\star$,
$F_{\Qc^\star,\Qc}^{\mathrm{Oxx}}(\ab_j)<1$'', it is sufficient to
just consider the case where there is one wrong atom $\ab_j$ to study
the logical implication between the ERC-OMP and ERC-OLS conditions.
Therefore, in this paragraph, we consider undercomplete dictionaries
\Ab with $k+1$ atoms. Testing ERC($\Ab,\Qc^\star$),
ERC-OMP($\Ab,\Qc^\star,\Qc$) and ERC-OLS($\Ab,\Qc^\star,\Qc$) amounts
to evaluating $F_{\Qc^\star}(\ab_j)$,
$F_{\Qc^\star,\Qc}^{\mathrm{OMP}}(\ab_j)$ and
$F_{\Qc^\star,\Qc}^{\mathrm{OLS}}(\ab_j)$ and comparing them to 1.

Fig.~\ref{fig:1} is a scatter plot of the three criteria for 10.000
randomly Gaussian dictionaries \Ab of size $100\times 11$. The subset
$\Qc=\stdacc{1}$ is systematically chosen as the first atom of \Ab.
Figs.~\ref{fig:1}(a,b) are in good agreement with
Lemma~\ref{lem:erc_decrease}: we verify that
$F_{\Qc^\star,\Qc}^{\mathrm{OMP}}(\ab_j)\leqslant
F_{\Qc^\star}(\ab_j)$ whether the ERC holds or not, and that
$F_{\Qc^\star,\Qc}^{\mathrm{OLS}}(\ab_j)\leqslant
F_{\Qc^\star}(\ab_j)$ systematically occurs only when
$F_{\Qc^\star}(\ab_j)<1$. On Fig.~\ref{fig:1}(c) displaying
$F_{\Qc^\star,\Qc}^{\mathrm{OMP}}(\ab_j)$ versus
$F_{\Qc^\star,\Qc}^{\mathrm{OLS}}(\ab_j)$, we only keep the trials for
which $F_{\Qc^\star}(\ab_{j})\geqslant 1$, \ie ERC($\Ab,\Qc^\star$)
does not hold. Since both south-east and north-west quarter planes are
populated, we conclude that neither OMP nor OLS is uniformly better.
To be more specific, when ERC-OMP($\Ab,\Qc^\star,\Qc$) holds but
ERC-OLS($\Ab,\Qc^\star,\Qc$) does not, there exists an input
$\yb\in\spansub{\Ab_{\Qc^\star}}$ for which OLS selects
$\Qc=\stdacc{1}$ and then a wrong atom in the second iteration
(Theorem~\ref{th:necess_ols}). On the contrary, OMP is guaranteed to
exactly recover this input according to
Theorem~\ref{th:suffic_omp_ols}. The same situation can occur when
inverting the roles of OMP and OLS according to
Corollary~\ref{cor:necess_omp1a} and Theorem~\ref{th:suffic_omp_ols}
(note that this analysis becomes more complex when
$\Card{\Qc}\geqslant 2$ since ERC-OMP($\Ab,\Qc^\star,\Qc$) alone is
not a necessary condition for OMP anymore; Theorem~\ref{th:necess_omp}
also involves the assumption that \Qc is reachable).

We have compared ERC-OMP($\Ab,\Qc^\star,1$) and
ERC-OLS($\Ab,\Qc^\star,1$), which take into account all the possible
subsets of $\Qc^\star$ of cardinality $1$. Again, we found that when
ERC($\Ab,\Qc^\star$) is not met, it can occur that
ERC-OMP($\Ab,\Qc^\star,1$) holds while ERC-OLS($\Ab,\Qc^\star,1$) does
not and \emph{vice versa}.

\subsubsection{Phase transition analysis for overcomplete random  dictionaries}
We now address the case of overcomplete dictionaries. Moreover, we
study the dependence of the ERC-Oxx conditions with respect to the
cardinalities $k$ and $q$ for $k>q\geqslant 2$ and we compare them for
common problems ($\Ab,\Qc^\star,\Qc$).

Let us start with simple preliminary remarks. Because the
ERC-Oxx($\Ab,\Qc^\star,\Qc$) conditions take the form ``for all
$j\notin\Qc^\star$, $F_{\Qc^\star,\Qc}^{\mathrm{Oxx}}(\ab_j)<1$'',
they are more often met when the dictionary is undercomplete (or when
$m\approx n$) than in the overcomplete case: when the submatrix
$\Ab_{\Qc^\star}$ gathering the true atoms is given,
$\max_{j\notin\Qc^\star} F_{\Qc^\star,\Qc}^{\mathrm{Oxx}}(\ab_j)$ is
obviously increasing when additional wrong atoms $\ab_j$ are
incorporated, \ie when $n$ is increasing. Additionally, we notice that
for given \Ab and $\Qc^\star$, $F_{\Qc^\star,\Qc}^{\mathrm{OMP}}$
always decreases when $\Qc$ is growing by definition of
$F_{\Qc^\star,\Qc}^{\mathrm{OMP}}$. This might not be the case of
$F_{\Qc^\star,\Qc}^{\mathrm{OLS}}$ for specific settings but it
happens to be true in average for random dictionaries.
\begin{figure}[t]
  \centering
  {
    \DRAFT{\begin{tabular}{cc}}
      \FINAL{\begin{tabular}{c}}
        \FINAL{\includegraphics*[width=70mm]{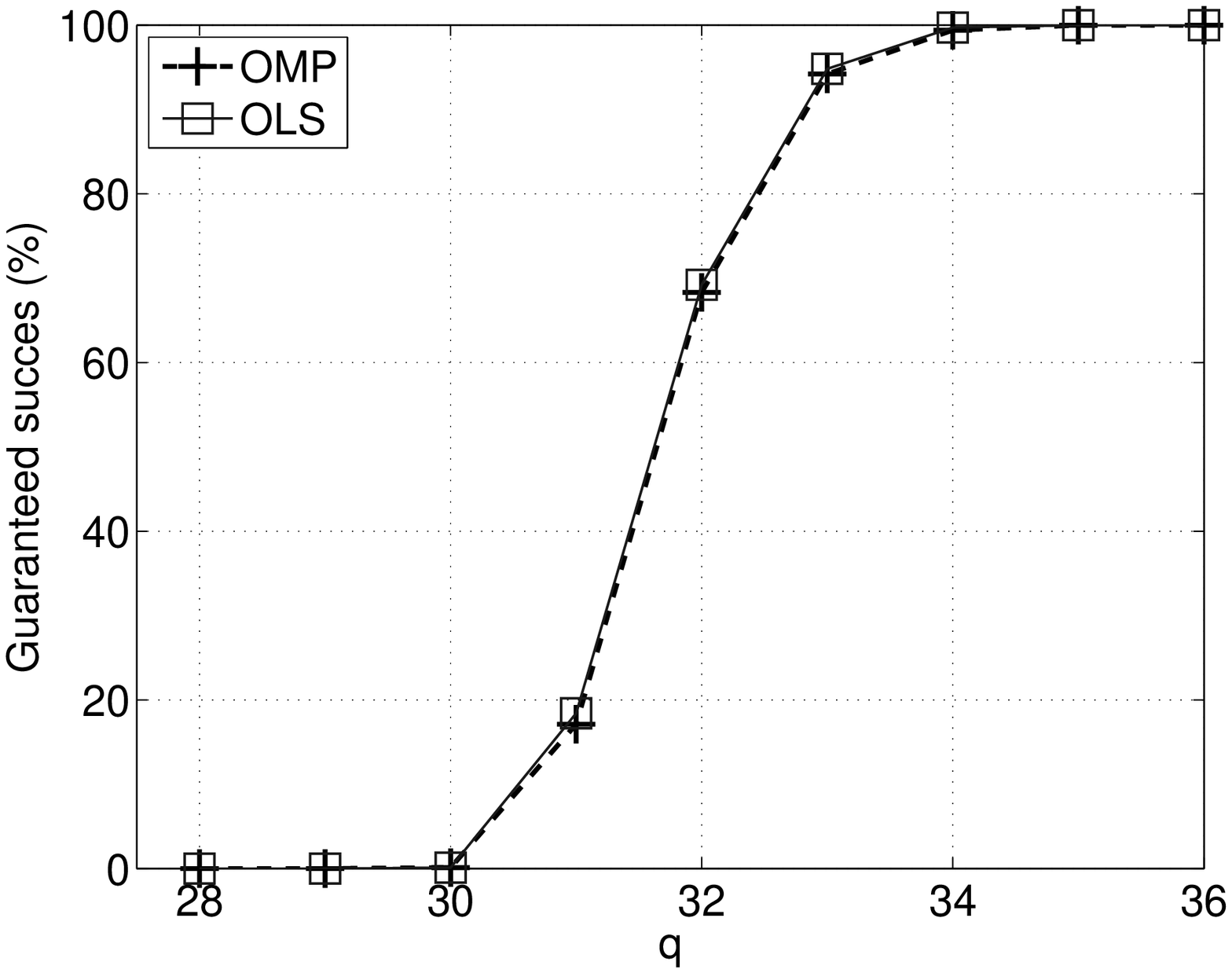}}
        \DRAFT{\includegraphics*[width=75mm]{phase_transG}}
        \DRAFT{&}
        \FINAL{\\ {\small{(a)~Random dictionaries ($T=0$)}}\\}
      \FINAL{\includegraphics*[width=70mm]{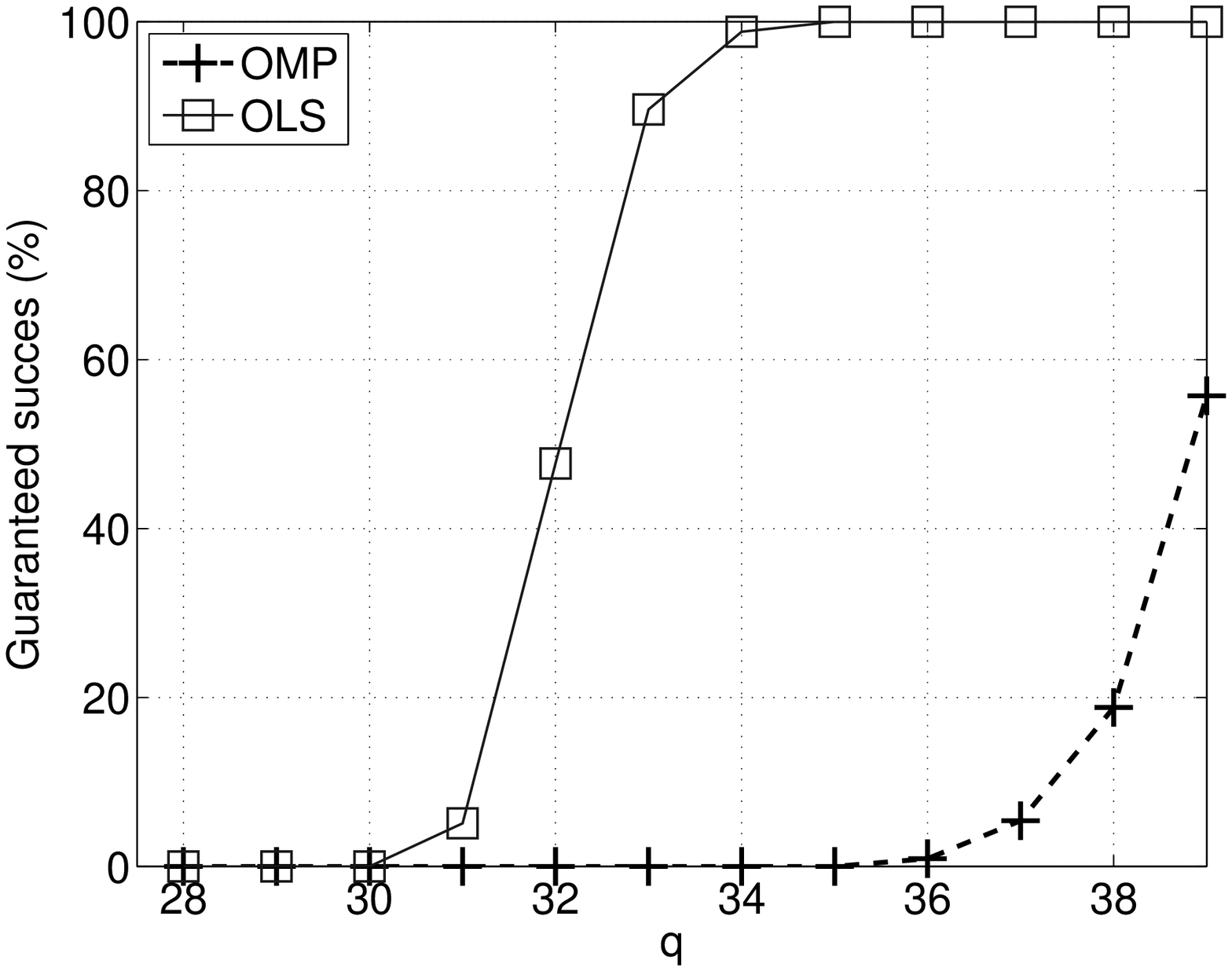}}
      \DRAFT{\includegraphics*[width=75mm]{phase_transGT}}\\
      \DRAFT{{\small{(a)~Random dictionaries ($T=0$)}} &}
      {\small (b)~Hybrid dictionaries ($T=10$)}
      \end{tabular}
  }
  \caption{Phase transition curves: for each $q<k$, we count the rate 
    of trials where ERC-Oxx($\Ab,\Qc^\star,\Qc)$ is true, 
    with $\Card{\Qc}=q$. The dictionaries are of size $200\times 600$, 
    $k$ is set to 40 and 1,000 Monte Carlo trials are performed.\; 
    (a)~Randomly Gaussian dictionaries;\, 
    (b)~Hybrid dictionaries with $T=10$.  }
  \label{fig:phase_transition}
\end{figure}

In the following experiments, $\Qc\subsetneq\Qc^\star$ is gradually
increased for fixed \Ab and $\Qc^\star$, and we search for the first
cardinality $q=\Card{\Qc}$ for which ERC-Oxx($\Ab,\Qc^\star,\Qc$) is
met. This allows us to define a ``phase transition
curve''~\cite{Tropp07,Donoho09} which separates the $q$-values for
which ERC-Oxx($\Ab,\Qc^\star,\Qc$) is never met, and is always met.
Examples of phase transition curves are given on
Fig.~\ref{fig:phase_transition} for random ($T=0$) and hybrid
dictionaries ($T=10$). Fig.~\ref{fig:phase_transition}(a) shows that
for $T=0$, the phase transition regime occurs in the same interval
$q\in\stdacc{30,\ldots,34}$ for both OMP and OLS and that the OMP and
OLS curves are superimposed. On the contrary, for hybrid dictionaries
(Fig.~\ref{fig:phase_transition}(b)), the mutual coherence increases
and the OLS curve is significantly above the OMP curve. Thus, the
guaranteed success for OLS occurs (in average) for an earlier
iteration than for OMP. For larger values of $T$ (\eg for $T=100$),
the ERC-OMP condition is never met before $q=k-1$, and even for
$q=k-1$, it is met for only $4~\%$ of trials.
\begin{figure*}[t]
  \centering
  {
    \setlength{\tabcolsep}{0pt}
    \begin{tabular}{@{}cccc@{}}
      \DRAFT{
      \includegraphics*[height=65mm]{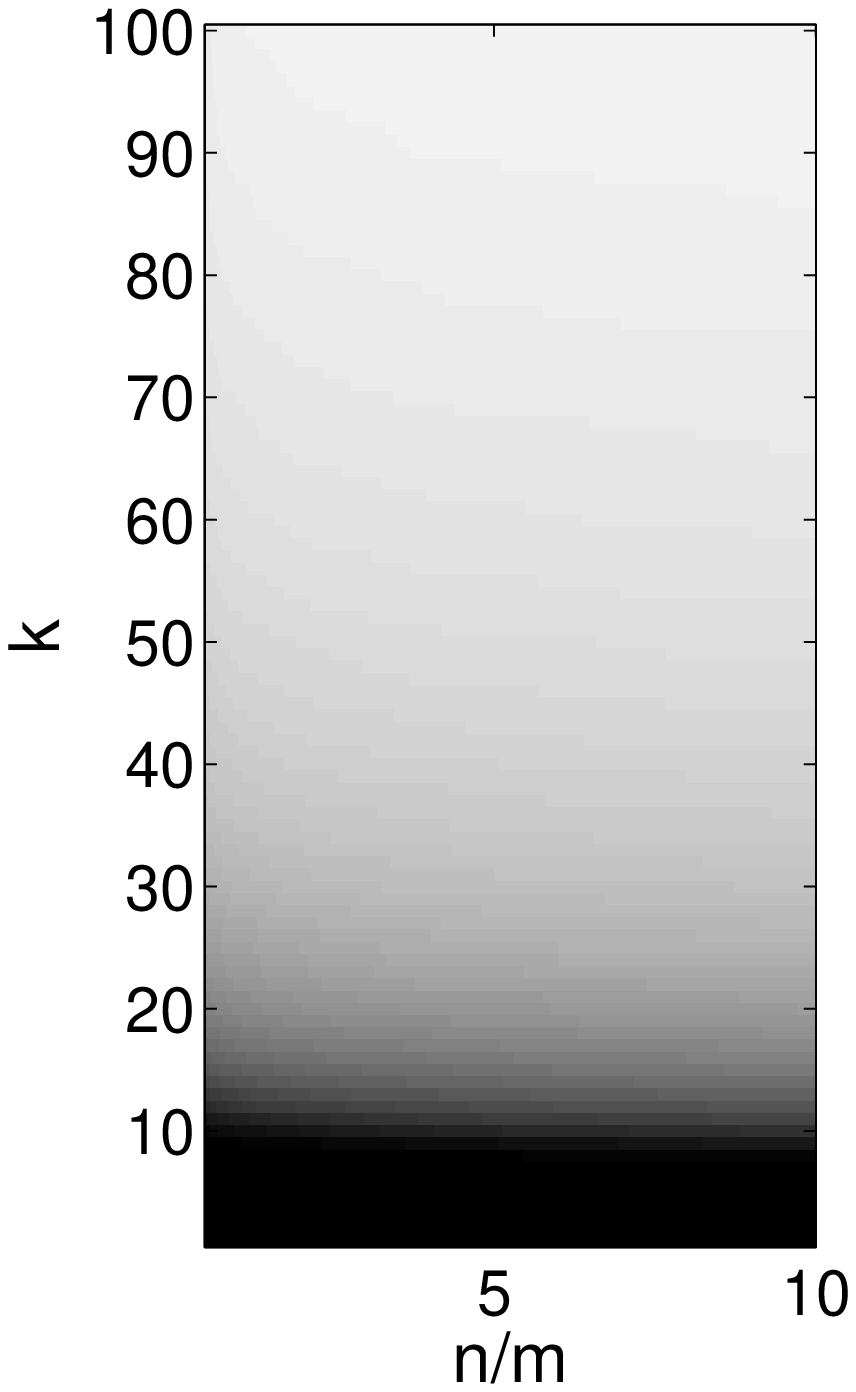}&
      \includegraphics*[height=65mm]{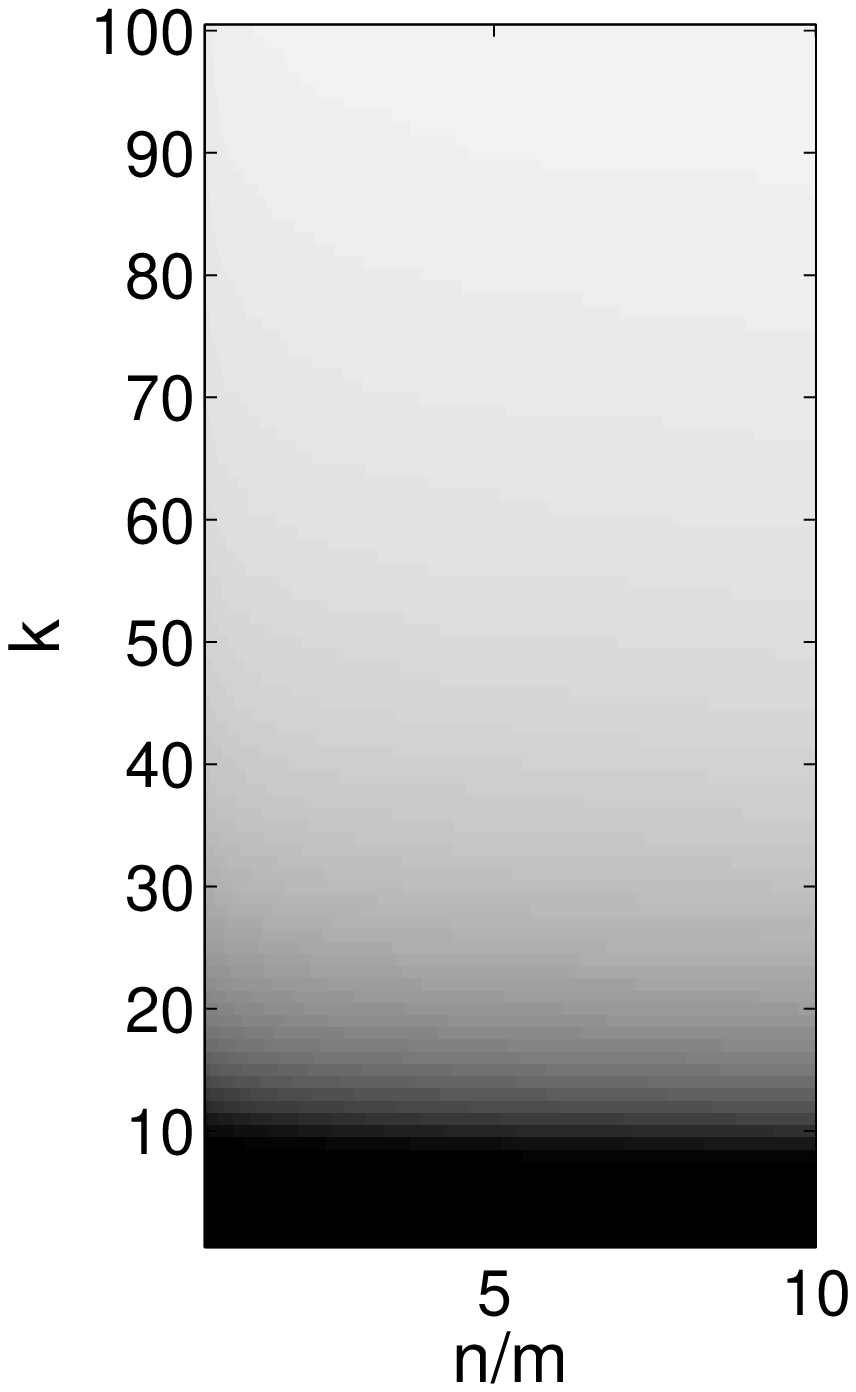}&
      \includegraphics*[height=65mm]{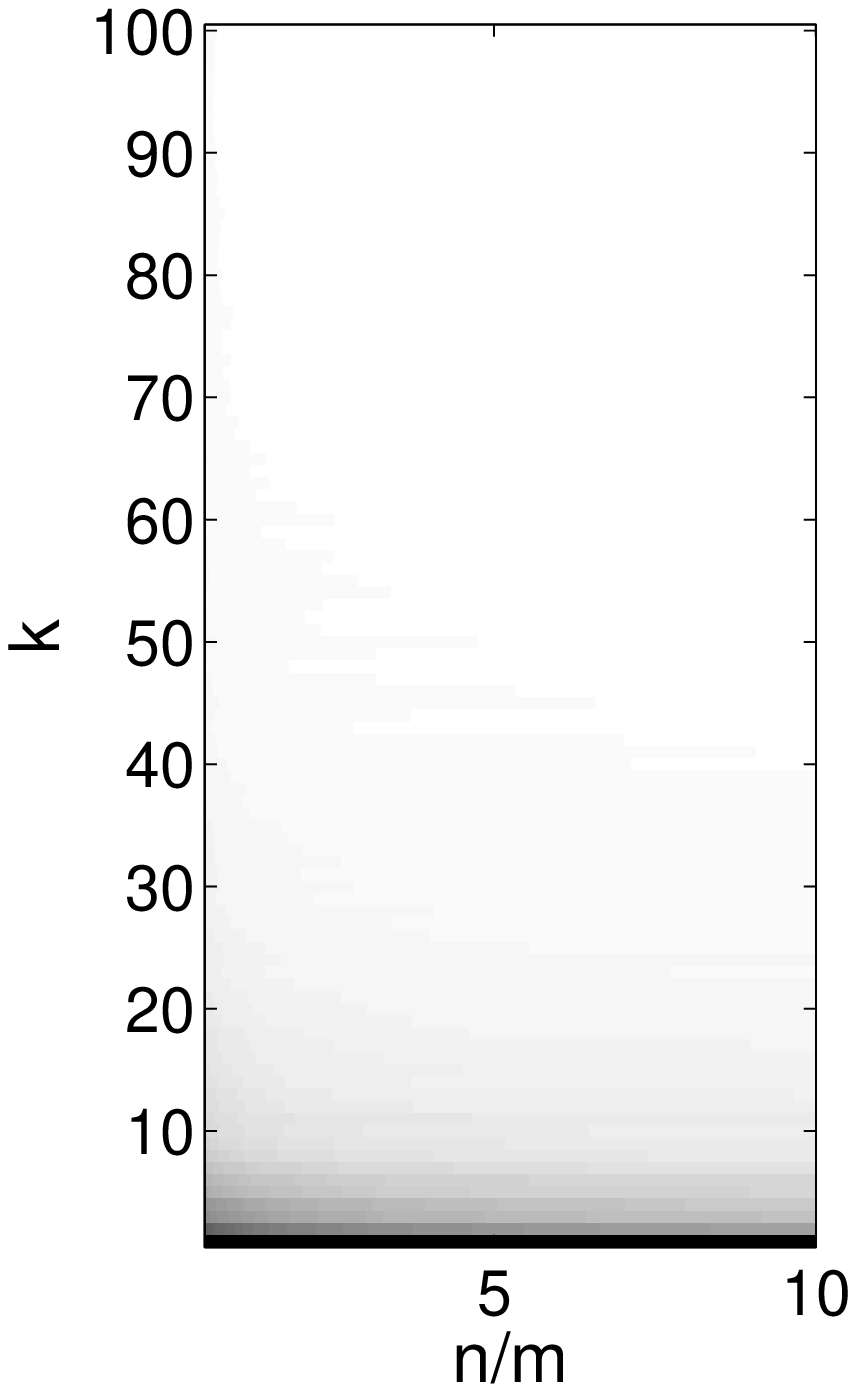}&
      \includegraphics*[height=65mm]{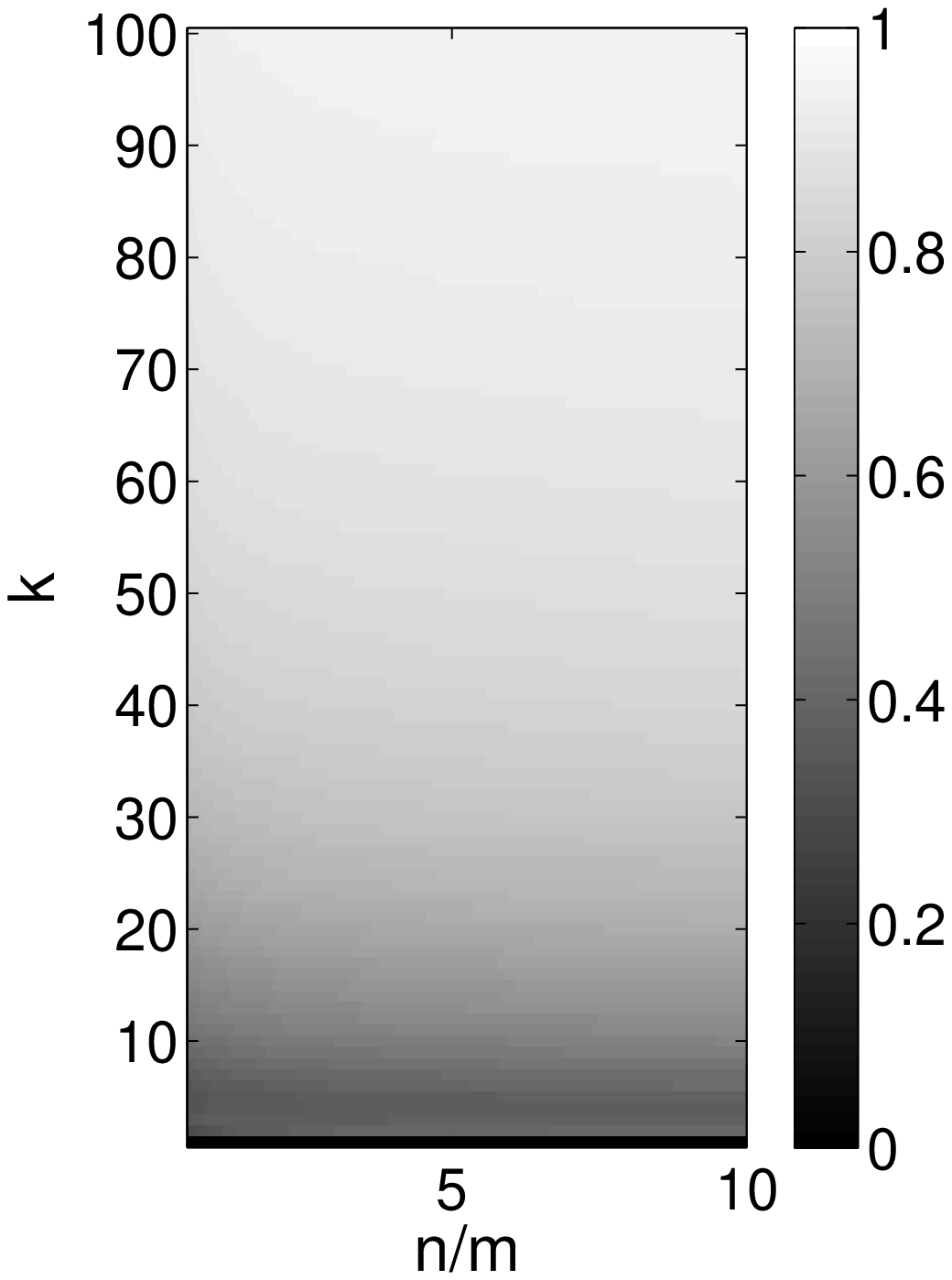}\\}
      \FINAL{
      \includegraphics*[height=70mm]{image_transG0omp_ji}&
      \includegraphics*[height=70mm]{image_transG0ols_ji}&
      \includegraphics*[height=70mm]{image_transG10omp_ji}&
      \includegraphics*[height=70mm]{image_transG10ols_ji}\\}
    {\small (a)~OMP, $T=0$} & 
    {\small (b)~OLS, $T=0$} &
    {\small (c)~OMP, $T=10$} & 
    {\small (d)~OLS, $T=10$}
    \end{tabular}
  }
  \caption{Phase transition diagrams for the
    ERC-Oxx($\Ab,\Qc^\star,\Qc$) condition. The gray levels represent
    the ratio $[q]^{\mathrm{Oxx}}(m,n,k)/k\in[0,1]$. Averaging is done
    over 200 draws of dictionary \Ab and subset $\Qc^\star$. \;
    (a,b)~Randomly Gaussian dictionaries of size $200\times n$ with
    $n\leqslant 2000$; (c,d)~Hybrid dictionaries of same size, with
    $T=10$.  }
  \label{fig:image_transition}
\end{figure*}

The experiment of Fig.~\ref{fig:phase_transition} is repeated for many
values of $k$ and dictionary sizes $m\times n$. For given \Ab and
$\Qc^\star$, let $q^{\mathrm{Oxx}}(m,n,k)$ denote the lowest value of
$q=\Card{\Qc}$ for which ERC-Oxx($\Ab,\Qc^\star,\Qc)$ is true. For
random and hybrid dictionaries, we perform 200 Monte Carlo simulations
in which random matrices \Ab and subsets $(\Qc^\star,\Qc)$ are drawn
and we compute the average values of $q^{\mathrm{Oxx}}$, denoted by
$[q]^{\mathrm{Oxx}}(m,n,k)$. This yields a phase transition
diagram~\cite{Donoho08,Donoho12} with the dictionary size (\eg $n/m$)
and the sparsity level $k$ in $x$- and $y$-axes, respectively. In this
image, the gray levels represent the ratio
$[q]^{\mathrm{Oxx}}(m,n,k)/k$ (see Fig.~\ref{fig:image_transition}).
Note that our phase transition diagrams are related to worst case
recovery conditions, so better performance may be achieved by actually
running Oxx for some simulated data $(\yb,\Ab)$ and testing whether
the support $\Qc^\star$ is found, where $\yb=\Ab\xb^\star$ and the
unknown nonzero amplitudes in $\xb^\star$ are drawn according to a
specific distribution.

A general comment regarding the results of
Fig.~\ref{fig:image_transition} is that the ERC-Oxx conditions are
satisfied early (for low values of $q/k$) when the unknown signal is
highly sparse ($k$ is low) or when $n/m$ is low, \ie when the
dictionary is not highly overcomplete. The ratio
$[q]^{\mathrm{Oxx}}(m,n,k)/k$ gradually grows with $k$ and $n/m$.
Regarding the OMP \vs OLS comparison, the phase diagrams obtained for
OMP and OLS look very much alike for Gaussian dictionaries ($T=0$). On
the contrary, we observe drastic differences in favor of OLS for
hybrid dictionaries (Fig.~\ref{fig:image_transition}(c,d)):
$F_{\Qc^\star,\Qc}^{\mathrm{OLS}}$ is significantly lower than
$F_{\Qc^\star,\Qc}^{\mathrm{OMP}}$.

We have performed similar tests for randomly uniform dictionaries (and
hybrid dictionaries based on a randomly uniform process) and we draw
conclusions similar to the Gaussian case. We have not encountered any
situation where $F_{\Qc^\star,\Qc}^{\mathrm{OMP}}$ is (in average)
significantly lower than $F_{\Qc^\star,\Qc}^{\mathrm{OLS}}$.

\subsubsection{ERC-Oxx evaluation for sparse spike train deconvolution dictionaries}
\label{par:3}
\begin{figure}[!t]
  \centering
  {
    \FINAL{\setlength{\tabcolsep}{0.1cm}}
    \begin{tabular}{cc}
      \FINAL{
        \begin{tabular}{c}(a)\end{tabular} & 
        \begin{tabular}{c}\includegraphics*[width=70mm]{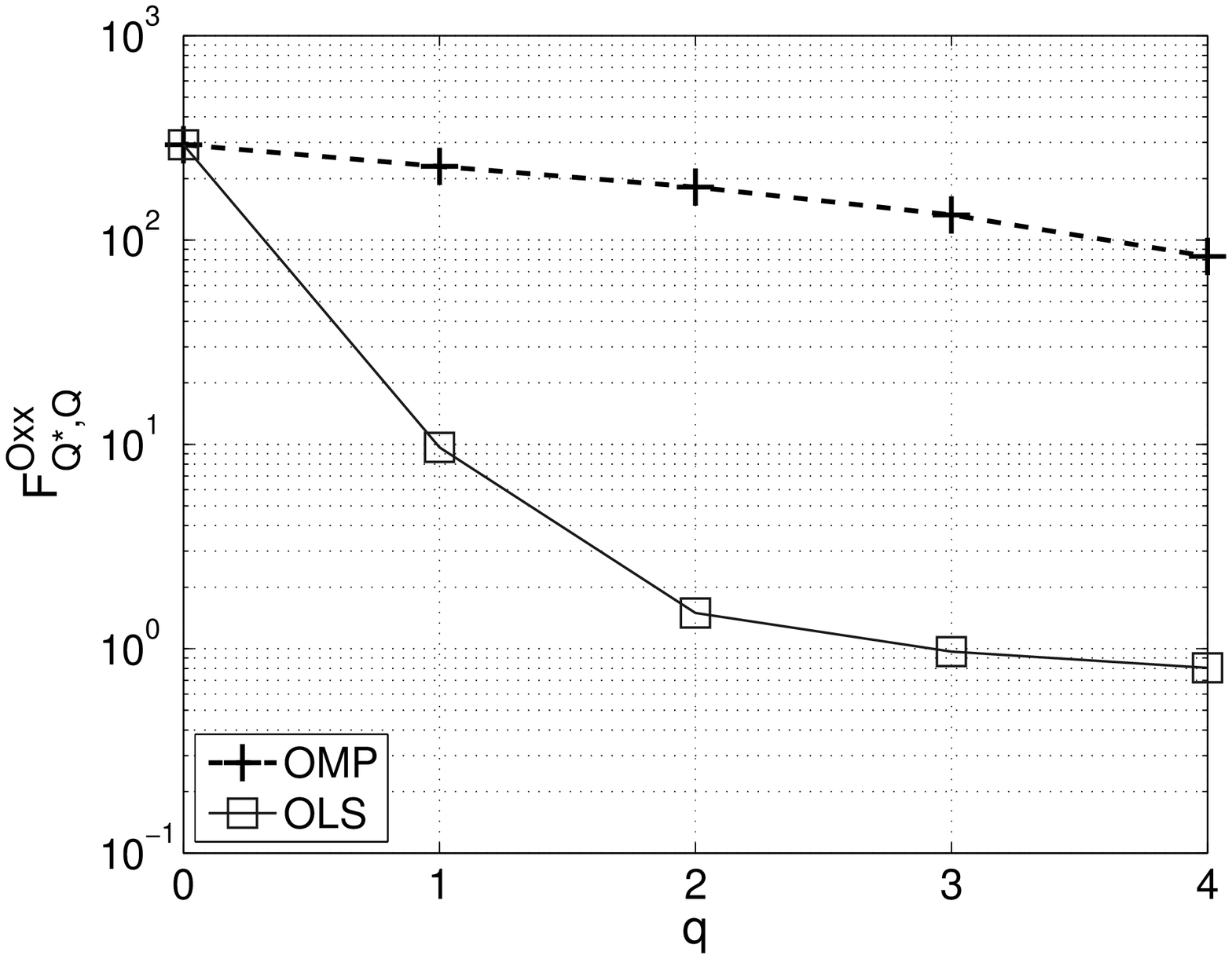}
          \end{tabular}
        }
        \DRAFT{\includegraphics*[width=75mm]{toeplitz_curveb}}
        \DRAFT{&}
        \FINAL{\\}
        \FINAL{
          \begin{tabular}{c}(b)\end{tabular} &
          \begin{tabular}{c}\includegraphics*[width=70mm]{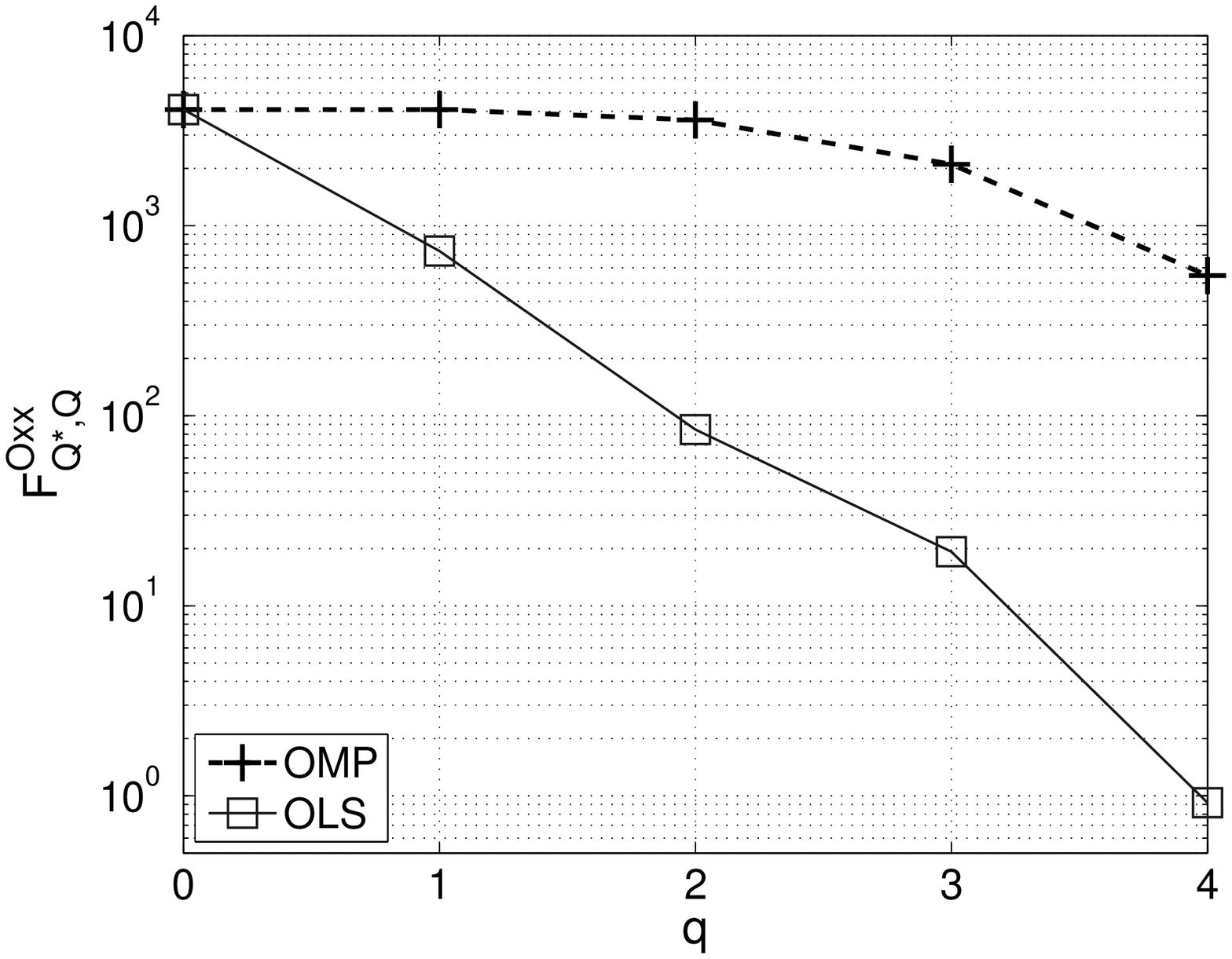}
          \end{tabular}
        }
        \DRAFT{\includegraphics*[width=75mm]{toeplitz_curvea}}\\
        \DRAFT{(a) & (b)}
    \end{tabular}
  }
  \caption{Curve representing $F_{\Qc^\star,\Qc}^{\mathrm{Oxx}}$ as a
    function of $q=\Card{\Qc}$ for the Gaussian deconvolution problem.
    $\Qc^\star$ is fixed and $\Qc\subsetneq\Qc^\star$ is gradually
    growing. $\Qc^\star$ and $\Qc$ are formed of the first $k=5$ and
    the first $q$ atoms, respectively with $q<k$.\; (a)~The Gaussian
    impulse response is of width $\sigma=50$ and the dictionary is of
    size $3000\times 2710$. \; (b)~$\sigma$ is set to 10, and the
    dictionary is of size $1000\times 4940$.  }
  \label{fig:curve_erc_toeplitz}
\end{figure}

We reproduced the above experiments for the convolutive dictionary
introduced in subsection~\ref{sec:dicos}. Since the dictionary is
deterministic, only one trial is performed per cardinality ($m,n,k$).
In each of the simulations hereafter, we set $\Qc$ and $\Qc^\star$ to
contiguous atoms. This is the worst situation because contiguous atoms
are the most highly correlated and exact support recovery may be more
easily achieved if we impose a minimum distance between true
atoms~\cite{Dossal05b,Lorenz09}. The curves of
Fig.~\ref{fig:curve_erc_toeplitz} represent
$F_{\Qc^\star,\Qc}^{\mathrm{Oxx}}$ with respect to $q$ for some given
$(\Ab,\Qc^\star)$. It is noticeable that the OLS curve decreases much
faster than the OMP curve, and that $F_{\Qc^\star,\Qc}^{\mathrm{OMP}}$
remains huge even after a number of iterations. For all our trials
where the true atoms strongly overlap, the
ERC-OMP($\Ab,\Qc^\star,\Qc$) condition is not met while
ERC-OLS($\Ab,\Qc^\star,\Qc$) may be fulfilled after a number of
iterations which is, however, close to $k$. Moreover, we found that
when $\sigma$ is large enough, $F_{\Qc^\star,\Qc}^{\mathrm{OMP}}$
remains larger than 1 even for $q=k-1$, whereas the ERC-OLS condition
is always met for $q=k-1$.

Empirical evaluations of the ERC condition for sparse spike train
deconvolution was already done in~\cite{Dossal05a}.
In~\cite{Dossal05a,Dossal05b,Lorenz09}, a stronger sufficient
condition than the ERC is evaluated for convolutive dictionaries. It
is a sufficient (but not necessary) exact recovery condition that is
easier to compute than the ERC because it does not require any matrix
inversion, and only relies on inner products between the dictionary
atoms (see~\cite[Lemma~3]{Gribonval08} for further details).
In~\cite{Dossal05a,Dossal05b}, it was pointed out that the ERC
condition is usually not fulfilled for convolutive dictionaries, but
when the true atoms are enough spaced, successful recovery is
guaranteed to occur. Our study can be seen as an alternative analysis
to~\cite{Dossal05a,Dossal05b,Lorenz09} in which no minimal distance
constraint is imposed.

\subsection{Examples where the bad recovery condition of OMP is met}
\label{sec:brc_omp}
We exhibit several situations in which the BRC-OMP($\Ab,\Qc^\star$)
condition may be fulfilled. This allows us to distinguish OMP from OLS
as we know that under regular conditions, any subset $\Qc^\star$ is
reachable using OLS at least for some input in
$\spansub{\Ab_{\Qc^\star}}$ (Lemma~\ref{lem:ols_reach}). The first
situation is a simple dictionary with four atoms, some of which being
strongly correlated. For this example, we show a stronger result than
the BRC: there exists a subset $\Qc^\star$ which is not reachable for
any $\yb\in\spansub{\Ab_{\Qc^\star}}$, but not even for any
$\yb\in\Rbb^m$. The other examples involve the random, hybrid and
deterministic dictionaries introduced in subsection~\ref{sec:dicos}.

\subsubsection{Example with four atoms}
\begin{example}
  Consider the simple dictionary
  \begin{align*}
    \Ab&=
    \left[
      {
        \setlength{\arraycolsep}{0.25cm}
        \begin{array}{rrrr}
          \cos\theta_1  &  \cos\theta_1&  0            &  0\\
          -\sin\theta_1 &  \sin\theta_1&  \cos\theta_2 & \cos\theta_2  \\
          0             &  0           &  \sin\theta_2 & -\sin\theta_2
        \end{array}
      }
    \right] 
  \end{align*}
  with $\Qc^\star=\stdacc{1,2}$. Set $\theta_2$ to an arbitrary value
  in $(0,\pi/2)$. When $\theta_1\ne 0$ is close enough to 0,
  BRC-OMP($\Ab,\Qc^\star$) is met. Moreover, OMP cannot reach
  $\Qc^\star$ in two iterations for any $\yb\in\Rbb^3$ (specifically,
  when $\yb\in\Rbb^3$ is proportional to neither $\ab_1$ nor $\ab_2$,
  $\ab_3$ or $\ab_4$ is selected in the first two iterations).
  \label{ex:omp_unreach}
\end{example}
\begin{IEEEproof}[Proof of Example~\ref{ex:omp_unreach}]
  We first prove that the BRC condition is met by calculating the
  factors $F_{\Qc^\star,\stdacc{1}}^{\mathrm{OMP}}(\ab_j)$ and
  $F_{\Qc^\star,\stdacc{2}}^{\mathrm{OMP}}(\ab_j)$ for
  $j\in\stdacc{3,4}$. Let us start with
  $F_{\Qc^\star,\stdacc{1}}^{\mathrm{OMP}}(\ab_j)$.

  The simple projection calculation
  $\tilde{\ab}_i=\ab_i-\stdscal{\ab_i,\ab_1}\ab_1$ (the tilde notation
  implicitly refers to $\Qc=\stdacc{1}$) leads to:
  \begin{align*}
    \tilde{\ab}_2=\sin(2\theta_1)\FINAL{&}\left[
      \begin{array}{@{}c@{}}
        \sin\theta_1\\
        \cos\theta_1\\
        0
      \end{array}
    \right ],\;\;\tilde{\ab}_3=\left[
      \begin{array}{@{}c@{}}
        \sin\theta_1\cos\theta_1\cos\theta_2\\
        \cos^2\theta_1\cos\theta_2\\
        \sin\theta_2
      \end{array}
    \right ]\FINAL{\\}
    \;\;\textrm{and}\FINAL{&}\;\;
    \tilde{\ab}_4=\left[
      \begin{array}{@{}c@{}}
        \sin\theta_1\cos\theta_1\cos\theta_2\\
        \cos^2\theta_1\cos\theta_2\\
        -\sin\theta_2
      \end{array}
    \right ].
  \end{align*}
  According to~\eqref{eq:Ferc_omp_last}, the OMP recovery
  factor reads for $j\in\stdacc{3,4}$:
  \begin{align}
    F_{\Qc^\star,\stdacc{1}}^{\mathrm{OMP}}(\ab_j)&=
    \frac{\stdbars{
        \stdscal{\tilde{\ab}_2,\tilde{\ab}_j}}}{\stdnorm{\tilde{\ab}_2}^2}
    =\frac{\stdbars{\cos\theta_1\cos\theta_2}}{\stdbars{\sin(2\theta_1)}}
    \label{eq:brc_4atomes}
  \end{align}
  given that $\stdnorm{\tilde{\ab}_2}=\stdbars{\sin(2\theta_1)}$ and
  $\stdbars{\stdscal{\tilde{\ab}_2,\tilde{\ab}_3}}=
  \stdbars{\stdscal{\tilde{\ab}_2,\tilde{\ab}_4}}=\stdnorm{\tilde{\ab}_2}\,
  \stdbars{\cos\theta_1\cos\theta_2}$.
  $F_{\Qc^\star,\stdacc{2}}^{\mathrm{OMP}}(\ab_j)$ can be obtained
  symmetrically by replacing $\theta_1$ by $-\theta_1$
  in~\eqref{eq:brc_4atomes}.  Thus, we have
  $F_{\Qc^\star,\stdacc{2}}^{\mathrm{OMP}}(\ab_j)=
  F_{\Qc^\star,\stdacc{1}}^{\mathrm{OMP}}(\ab_j)$. It follows that the
  left hand-side of the BRC-OMP($\Ab,\Qc^\star$) condition
  reads~\eqref{eq:brc_4atomes} and tends towards $+\infty$ when
  $\theta_1$ tends towards 0. Therefore, BRC-OMP($\Ab,\Qc^\star$) is
  met when $\stdbars{\theta_1}$ is small enough.
  \begin{figure}[t]
    \centering
    {
      \setlength{\unitlength}{0.00034996in}
\begingroup\makeatletter\ifx\SetFigFont\undefined%
\gdef\SetFigFont#1#2#3#4#5{%
  \reset@font\fontsize{#1}{#2pt}%
  \fontfamily{#3}\fontseries{#4}\fontshape{#5}%
  \selectfont}%
\fi\endgroup%
{\renewcommand{\dashlinestretch}{30}
\begin{picture}(9105,8229)(0,-10)
\thicklines
\put(1722,7482){\blacken\ellipse{134}{134}}
\put(1722,7482){\ellipse{134}{134}}
\put(2127,7347){\blacken\ellipse{134}{134}}
\put(2127,7347){\ellipse{134}{134}}
\put(2262,6987){\blacken\ellipse{134}{134}}
\put(2262,6987){\ellipse{134}{134}}
\put(2622,6222){\blacken\ellipse{134}{134}}
\put(2622,6222){\ellipse{134}{134}}
\put(2442,6582){\blacken\ellipse{134}{134}}
\put(2442,6582){\ellipse{134}{134}}
\put(2037,6897){\blacken\ellipse{134}{134}}
\put(2037,6897){\ellipse{134}{134}}
\put(2892,5502){\blacken\ellipse{134}{134}}
\put(2892,5502){\ellipse{134}{134}}
\put(2712,5772){\blacken\ellipse{134}{134}}
\put(2712,5772){\ellipse{134}{134}}
\put(4377,2667){\blacken\ellipse{134}{134}}
\put(4377,2667){\ellipse{134}{134}}
\put(4692,2037){\blacken\ellipse{134}{134}}
\put(4692,2037){\ellipse{134}{134}}
\put(4692,1767){\blacken\ellipse{134}{134}}
\put(4692,1767){\ellipse{134}{134}}
\put(5007,1677){\blacken\ellipse{134}{134}}
\put(5007,1677){\ellipse{134}{134}}
\put(5007,1272){\blacken\ellipse{134}{134}}
\put(5007,1272){\ellipse{134}{134}}
\put(5412,1047){\blacken\ellipse{134}{134}}
\put(5412,1047){\ellipse{134}{134}}
\put(5277,552){\blacken\ellipse{134}{134}}
\put(5277,552){\ellipse{134}{134}}
\path(3612,4062)(7662,6087)
\blacken\path(7547.961,5979.669)(7662.000,6087.000)(7507.711,6060.167)(7547.961,5979.669)
\path(3612,4062)(4377,4062)
\blacken\path(4227.000,4017.000)(4377.000,4062.000)(4227.000,4107.000)(4227.000,4017.000)
\thinlines
\dashline{60.000}(1677,7932)(5592,147)
\thicklines
%
	\path(2100,8000)(5175,102)
	\path(1272,7842)(5997,282)
\blacken\path(6748.650,5969.340)(6852.000,6087.000)(6700.950,6045.660)(6748.650,5969.340)
\path(6852,6087)(3612,4062)
\thinlines
\dashline{70.000}(7662,6087)(8427,6087)
\dashline{70.000}(6852,6087)(7617,6087)
\thicklines
\blacken\path(7507.711,2063.833)(7662.000,2037.000)(7547.961,2144.331)(7507.711,2063.833)
\path(7662,2037)(3612,4062)
\thinlines
\dashline{90.000}(5367,7617)(1857,462)
\thicklines
%
	\path(5000,7818)(2230,292)
	\path(5800,7506)(1500,687)
\thinlines
\dashline{60.000}(9012,4062)(112,4062)
\dashline{60.000}(3612,8202)(3612,12)
\thicklines
\path(3657,4062)(8472,6087)
\blacken\path(8351.176,5987.368)(8472.000,6087.000)(8316.285,6070.330)(8351.176,5987.368)
\path(5017,7662)(5107,7662)(5107,7572)
	(5017,7572)(5017,7662)
\path(5277,7392)(5367,7392)(5367,7302)
	(5277,7302)(5277,7392)
\path(5187,6987)(5277,6987)(5277,6897)
	(5187,6897)(5187,6987)
\path(4737,6942)(4827,6942)(4827,6852)
	(4737,6852)(4737,6942)
\path(4827,7212)(4917,7212)(4917,7122)
	(4827,7122)(4827,7212)
\path(4917,6627)(5007,6627)(5007,6537)
	(4917,6537)(4917,6627)
\path(4647,6492)(4737,6492)(4737,6402)
	(4647,6402)(4647,6492)
\path(4602,6087)(4692,6087)(4692,5997)
	(4602,5997)(4602,6087)
\path(4242,5637)(4332,5637)(4332,5547)
	(4242,5547)(4242,5637)
\path(4197,5277)(4287,5277)(4287,5187)
	(4197,5187)(4197,5277)
\path(3017,2757)(3107,2757)(3107,2667)
	(3017,2667)(3017,2757)
\path(2622,2037)(2712,2037)(2712,1947)
	(2622,1947)(2622,2037)
\path(2127,1452)(2217,1452)(2217,1362)
	(2127,1362)(2127,1452)
\path(2262,1047)(2352,1047)(2352,957)
	(2262,957)(2262,1047)
\path(1992,687)(2082,687)(2082,597)
	(1992,597)(1992,687)
\path(1812,867)(1902,867)(1902,777)
	(1812,777)(1812,867)
\path(1857,1227)(1947,1227)(1947,1137)
	(1857,1137)(1857,1227)
\path(2567,1632)(2657,1632)(2657,1542)
	(2567,1542)(2567,1632)
\put(7800,6342){\makebox(0,0)[lb]{\smash{{\SetFigFont{10}{10.0}{\rmdefault}{\mddefault}{\updefault}$\tilde{\ab}_3$}}}}
\put(8137,5517){\makebox(0,0)[lb]{\smash{{\SetFigFont{10}{10.0}{\rmdefault}{\mddefault}{\updefault}$\tilde{\ab}_3+\tilde{\ab}_2$}}}}
\put(5907,6342){\makebox(0,0)[lb]{\smash{{\SetFigFont{10}{10.0}{\rmdefault}{\mddefault}{\updefault}$\tilde{\ab}_3-\tilde{\ab}_2$}}}}
\put(7497,2347){\makebox(0,0)[lb]{\smash{{\SetFigFont{10}{10.0}{\rmdefault}{\mddefault}{\updefault}$\tilde{\ab}_4$}}}}
\put(4467,3737){\makebox(0,0)[lb]{\smash{{\SetFigFont{10}{10.0}{\rmdefault}{\mddefault}{\updefault}$\tilde{\ab}_2$}}}}
\put(3162,4007){\makebox(0,0)[lb]{\smash{{\SetFigFont{10}{10.0}{\rmdefault}{\mddefault}{\updefault}$\zerob$}}}}
\end{picture}
}
    }
    \caption{Example~\ref{ex:omp_unreach}: drawing of the plane
      $\spansub{\ab_1}^\perp$. The tilde notation refers to the subset
      $\Qc=\stdacc{1}$. When $\theta_1$ is close to 0, $\tilde{\ab}_2$
      is of very small norm since $\ab_2$ is almost equal to $\ab_1$,
      while $\ab_3$ and $\ab_4$, which are almost orthogonal to
      $\ab_1$, yield projections $\tilde{\ab}_3$ and $\tilde{\ab}_4$
      that are almost of unit norm. The angles
      $(\tilde{\ab}_2,\tilde{\ab}_3)$ and
      $(\tilde{\ab}_2,\tilde{\ab}_4)$ tend to $\theta_2$ and
      $-\theta_2$ when $\theta_1\rightarrow 0$. The bullet and square
      points correspond to positions \rb satisfying
      $\stdbars{\stdscal{\rb,\tilde{\ab}_2}}
      \geqslant\stdbars{\stdscal{\rb,\tilde{\ab}_3}}$ and
      $\stdbars{\stdscal{\rb,\tilde{\ab}_2}}
      \geqslant\stdbars{\stdscal{\rb,\tilde{\ab}_4}}$, respectively.
      The central directions of these two cones are orthogonal to
      $\tilde{\ab}_3$ and $\tilde{\ab}_4$, respectively (dashed
      lines). Both cones only intersect at $\rb=\zerob$, therefore OMP
      cannot successively select $\ab_1$ and $\ab_2$ in the first two
      iterations.  }
    \label{fig:cone}
  \end{figure}

  To show that $\Qc^\star$ is not reachable for any $\yb\in\Rbb^3$,
  let us assume that OMP selects a true atom in the first iteration.
  Because there is a symmetry between $\ab_1$ and $\ab_2$, we can
  assume without loss of generality that $\ab_1$ is selected. Then,
  the data residual $\rb$ after the first iteration lies in
  $\spansub{\ab_1}^\perp$ which is of dimension 2. We show using
  geometrical arguments, that $\ab_2$ cannot be selected in the second
  iteration for any
  $\rb\in\spansub{\ab_1}^\perp\backslash\stdacc{\zerob}$. We refer the
  reader to Fig.~\ref{fig:cone} for a 2D display of the projected
  atoms in the plane $\spansub{\ab_1}^\perp$.

  Let \Cc denote the set of points $\rb\in\Rbb^2$ satisfying
  $\stdbars{\stdscal{\rb,\tilde{\ab}_2}}
  \geqslant\stdbars{\stdscal{\rb,\tilde{\ab}_3}}$. $\rb\in\Cc$ \IFF
  there exist $(\varepsilon_2,\varepsilon_3)\in\stdacc{-1,1}^2$ such
  that $\varepsilon_2\stdscal{\rb,\tilde{\ab}_2}
  \geqslant\varepsilon_3\stdscal{\rb,\tilde{\ab}_3}\geqslant 0$, \ie
    \begin{align}
      \stdscal{\rb,\varepsilon_2\tilde{\ab}_2-\varepsilon_3\tilde{\ab}_3}
      \geqslant 0 \;\;\mathrm{and}\;\;
      \stdscal{\rb,\varepsilon_3\tilde{\ab}_3}\geqslant 0.
      \label{eq:cone}
    \end{align}
    For each sign pattern
    $(\varepsilon_2,\varepsilon_3)$,~\eqref{eq:cone} yields a 2D half
    cone defined as the intersection of two half-planes delimited by
    the directions which are orthogonal to $\tilde{\ab}_3$ and
    $\varepsilon_2\tilde{\ab}_2-\varepsilon_3\tilde{\ab}_3$.
    Moreover, the opposite sign pattern
    $(-\varepsilon_2,-\varepsilon_3)$ yields the remaining part of the
    same 2D cone. Consequently, the four possible sign patterns
    $(\varepsilon_2,\varepsilon_3)\in\stdacc{-1,1}^2$ yield both cones
    delimited by the orthogonal directions to $\tilde{\ab}_3$ and
    $\tilde{\ab}_2+\tilde{\ab}_3$, and to $\tilde{\ab}_3$ and
    $-\tilde{\ab}_2+\tilde{\ab}_3$, respectively.  Because these cones
    are adjacent, their union \Cc is the cone delimited by the
    orthogonal directions to $\tilde{\ab}_3+\tilde{\ab}_2$ and
    $\tilde{\ab}_3-\tilde{\ab}_2$ (plain lines in the south-east and
    north-west directions in Fig.~\ref{fig:cone}). Similarly, the
    condition $\stdbars{\stdscal{\rb,\tilde{\ab}_2}}
    \geqslant\stdbars{\stdscal{\rb,\tilde{\ab}_4}}$ yields another 2D
    cone whose central direction is orthogonal to $\tilde{\ab}_4$.
    When $\theta_1$ is close to 0, both cones only intersect at
    $\rb=\zerob$ (since their inner angle tends towards 0), thus
    \begin{equation*}
      \forall\rb\in\Rbb^2\backslash\stdacc{\zerob},\;
      \stdbars{\stdscal{\rb,\tilde{\ab}_2}}
      <\max(\stdbars{\stdscal{\rb,\tilde{\ab}_3}},
      \stdbars{\stdscal{\rb,\tilde{\ab}_4}}).
    \end{equation*}
    We conclude that $\ab_2$ cannot be selected in the second iteration
    according to the OMP rule~\eqref{eq:omp_rule}.
\end{IEEEproof}

\subsubsection{Numerical simulation of the BRC condition}
\begin{figure}[t]
  \centering
  \DRAFT{\begin{tabular}{cc}}
    \FINAL{\begin{tabular}{c}}
      \begin{tabular}{c}\includegraphics*[width=75mm]{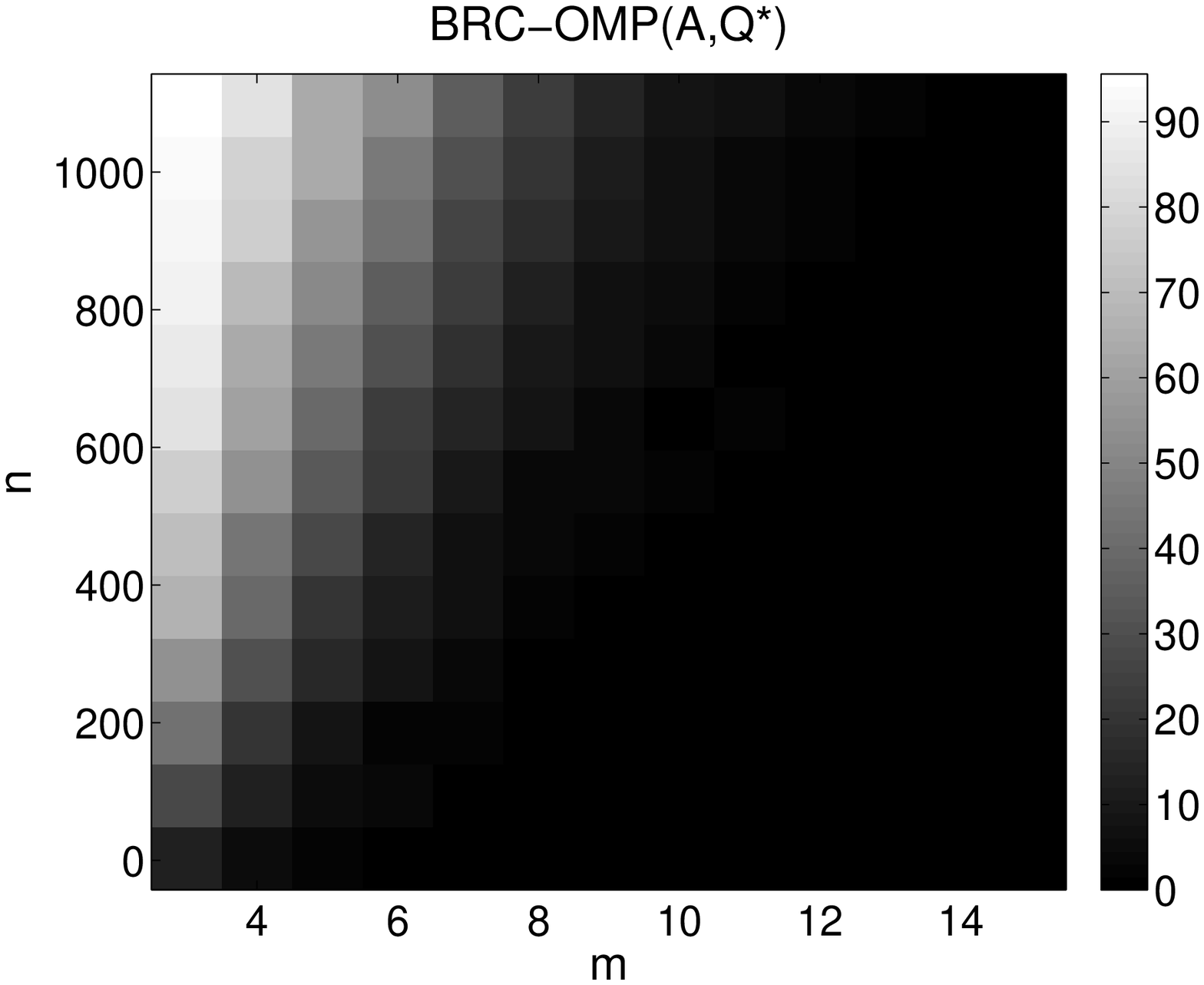}\end{tabular}
      \FINAL{\\{\small (a)~Gaussian dictionaries}\\}\DRAFT{&}
      \begin{tabular}{c}\includegraphics*[width=75mm]{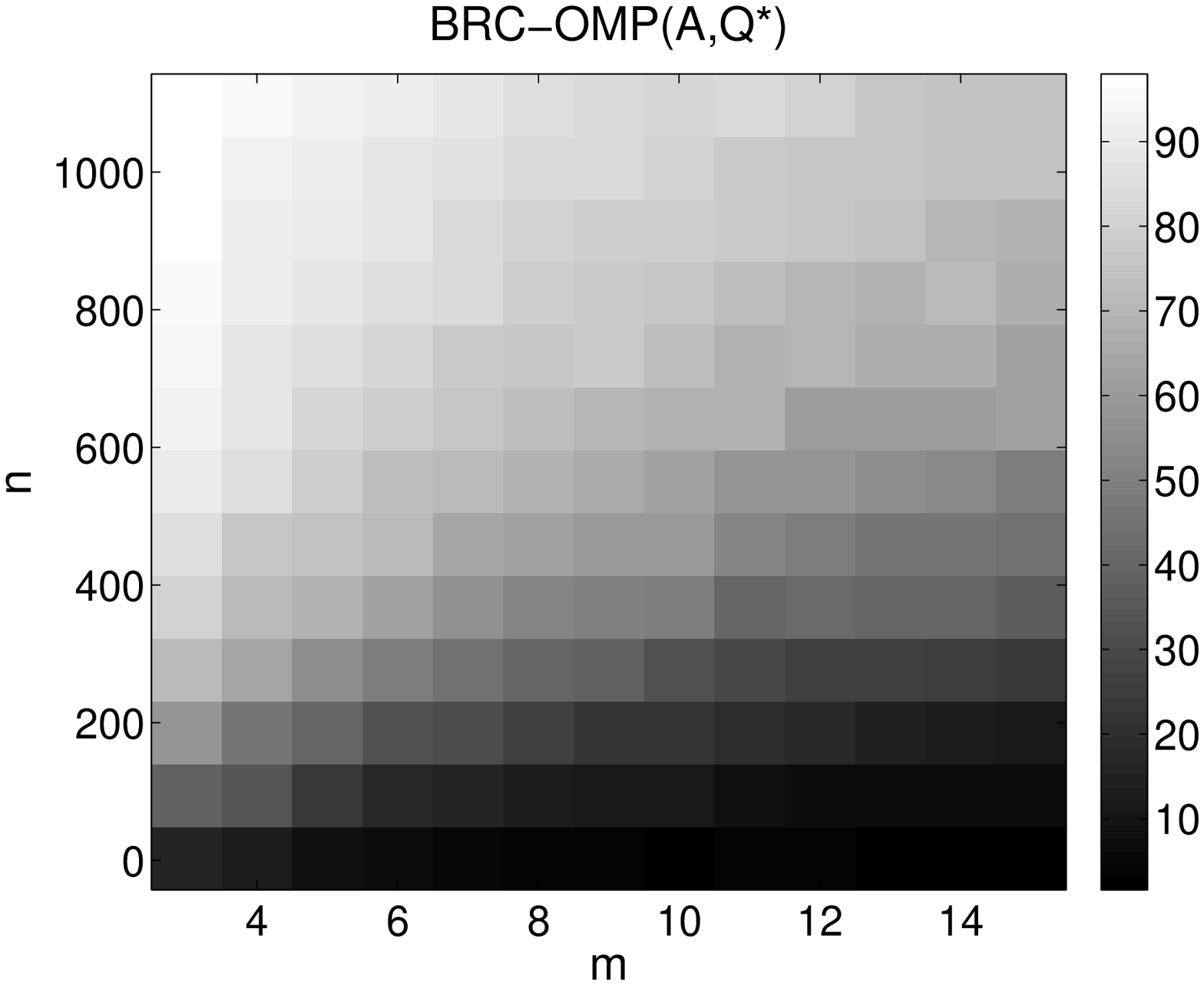}\end{tabular}\\
      \DRAFT{\small (a)~Gaussian dictionaries &}
      {\small (b)~Hybrid dictionaries ($T=10$)}
    \end{tabular}
    \caption{ Evaluation of the bad recovery condition
      BRC-OMP($\Ab,\Qc^\star$) for randomly Gaussian~(a) and
      hybrid~(b) dictionaries of various sizes $(m,n)$. 1,000 trials
      are performed per dictionary size, and $\Qc^\star$ is always set
      to the first two atoms ($k=2$). The gray levels correspond to
      the rate of guaranteed failure, \ie the proportion of trials
      where BRC-OMP($\Ab,\Qc^\star$) holds. }
 \label{fig:brc_omp}
\end{figure}
We test the BRC-OMP condition for various dictionary sizes ($m,n$) for
the random, hybrid and convolutive dictionaries introduced in
subsection~\ref{sec:dicos}. The average results related to the random
and hybrid dictionaries are gathered in Fig.~\ref{fig:brc_omp} in the
case $k=2$. For randomly Gaussian dictionaries, we observe that the
BRC-OMP condition may be met for strongly overcomplete dictionaries,
\ie when $n\gg m$ (Fig.~\ref{fig:brc_omp}~(a)). In the special case
$k=2$, it is noticeable that OLS performs at least as well as OMP
whether the BRC condition if fulfilled or not: when the first
iteration (common to both algorithms) has succeeded, OLS cannot fail
according to Theorem~\ref{th:erc_ols_last_is_true} while OMP is
guaranteed to fail in cases where the BRC holds. For the hybrid
dictionaries, the BRC condition is more frequently met when the
dictionary is moderately overcomplete, \ie for large values of $m/n$.
This result is in coherence with our evaluations of the ERC-Oxx
condition (see, \eg Fig.~\ref{fig:image_transition}(c)) which are more
rarely met for hybrid dictionary than for random dictionaries.

We performed similar tests for the sparse spike train deconvolution
problem with a Gaussian impulse response of width $\sigma$, and with
$k=2$ (the true atoms are contiguous, thus they are strongly
correlated). We repeated the simulation of Fig.~\ref{fig:brc_omp} for
various sizes $m\approx n$ and various widths $\sigma$, and we found
that whatever $(m,n)$, the BRC condition is always met for
$\sigma\geqslant 1.5$ and never met when $\sigma\leqslant 1.4$. The
images of Fig.~\ref{fig:brc_omp} thus become uniformly white and
uniformly black, respectively. To be more specific, the value of the
left hand-side of the BRC-OMP($\Ab,\Qc^\star$) condition gradually
increases with $\sigma$, \eg this value reaches 10, 35 and 48 for
$\sigma=10$, 20 and 50, respectively for dictionaries of size
$m\approx n$, with $m=3000$. This result is in coherence with that of
Fig.~\ref{fig:curve_erc_toeplitz} which already indicated that the
$F_{\Qc^\star,\Qc}^{\mathrm{OMP}}$ factor becomes huge for convolutive
problems with strongly correlated atoms.

Note that when $\Qc^\star$ does not involve contiguous atoms but
``spaced atoms'' which are less correlated, the bad recovery condition
are met for larger values of $\sigma$: denoting by $\Delta$ the
minimum distance between two true atoms, the lowest $\Delta$ value for
which the BRC is met turns out to be an increasing affine function of
$\sigma$. Similar empirical studies were done in~\cite{Dossal05a} for
the exact recovery condition for spaced atoms, and
in~\cite{Dossal05a,Lorenz09} for the weak exact recovery condition
of~\cite[Lemma~3]{Gribonval08}. In particular, the numerical
simulations in~\cite{Lorenz09} for the Gaussian deconvolution problem
demonstrate that the latter condition is met for larger $\sigma$'s
when the minimum distance between true atoms is increased and the
limit $\Delta$ value corresponding to the phase transition is also an
affine function of $\sigma$. Our bad recovery condition results are
thus a complement to those of~\cite{Lorenz09}.

\section{Conclusions}
\label{sec:cl}
Our first contribution is an original analysis of OLS based on the
extension of the ERC condition. We showed that when the ERC holds, OLS
is guaranteed to yield an exact support recovery. Although OLS has
been acknowledged in several communities for two decades, such a
theoretical analysis was lacking. Our second contribution is a
parallel study of OMP and OLS when a number of iterations have been
performed and true atoms have been selected. We found that neither OMP
nor OLS is uniformly better. In particular, we showed using randomly
Gaussian dictionaries that when the ERC is not met but the first
iteration (which is common to OMP and OLS) selects a true atom, there
are counter-examples for which OMP is guaranteed to yield an exact
support recovery while OLS does not, and \emph{vice versa}.

Finally, several elements of analysis suggest that OLS behaves better
than OMP. First, any subset \Qc can be reached by OLS using some input
in $\spansub{\Ab_\Qc}$ while for some dictionaries, it may occur that
some subsets are never reached by OMP for any $\yb\in\Rbb^m$. In other
words, OLS has a stronger capability of exploration. Secondly, when
all true atoms except one have been found by OLS and no wrong
selection occurred, OLS is guaranteed to find the last true atom in
the following iteration while OMP may fail.

For problems in which the dictionary is far from orthogonal and some
dictionary atoms are strongly correlated, we found in our experiments
that the OLS recovery condition might be met after some iterations
while the OMP recovery condition is rarely met. We did not encounter
the opposite situation where the OMP recovery condition is frequently
met after fewer iterations than the OLS condition. Moreover,
guaranteed failure of OMP may occur more often when the dictionary
coherence is large. These results are in coherence with empirical
studies reporting that OLS usually outperforms OMP at the price of a
larger numerical cost~\cite{RebolloNeira02,Davies12}. In our
experience, OLS yields a residual error which may be by far lower than
that of OMP after the same number of iterations~\cite{Soussen11c}.
Moreover, it performs better support recoveries in terms of ratio
between the number of good detections and of false
alarms~\cite{Bourguignon11}.

\appendices

\section{Necessary and sufficient conditions of exact 
  recovery for OMP and OLS}
\label{sec:cns}
This appendix includes the complete analysis of our OMP and OLS
recovery conditions.

\subsection{Sufficient conditions}
\label{sec:sufficient}
We show that when Oxx happens to select true atoms during its early
iterations, it is guaranteed to recover the whole unknown support in
the subsequent iterations when the ERC-Oxx($\Ab,\Qc^\star,\Qc$)
condition is fulfilled. We establish Theorem~\ref{th:suffic_omp_ols}
whose direct consequence is Theorem~\ref{th:tropp_ols} stating that
when ERC($\Ab,\Qc^\star$) holds, OLS is guaranteed to succeed.
\FINAL{
  \begin{table*}[t]
  \normalsize
  \begin{align}
    F_{\Qc^\star,\Qc}^{\mathrm{OMP}}(\ab_j)&=
    F_{\Qc^\star,\Qc'}^{\mathrm{OMP}}(\ab_j)+
    \,\bigbars{\bigpth{\Ab_{\Qc^\star}^\dag\ab_j}(\ell)}
    \tag{\ref{eq:omp_erc_rec}}\\
    F_{\Qc^\star,\Qc}^{\mathrm{OLS}}(\ab_j)&= \Biggbars{
      \chi_{j}^{\Qc,\Qc'}-\eta_{j}^{\Qc,\Qc'}
      \sum_{i\in\Qc^\star\backslash\Qc'}
      \frac{\betab_j^{\Qc^\star\backslash\Qc'}(i)\chi_i^{\Qc,\Qc'}}
      {\eta_i^{\Qc,\Qc'}}}+
    \eta_{j}^{\Qc,\Qc'}\sum_{i\in\Qc^\star\backslash\Qc'}
    \frac{\bigbars{\betab_j^{\Qc^\star\backslash\Qc'}(i)}}
    {\eta_i^{\Qc,\Qc'}}
    \tag{\ref{eq:ols_erc_rec}}
\end{align}
\hrule
\end{table*}
}

\subsubsection{ERC-Oxx are sufficient recovery conditions at a given iteration}
\label{sec:tropp_extens}
We follow the analysis of~\cite[Theorem~3.1]{Tropp04} to extend
Tropp's exact recovery condition to a sufficient condition dedicated
to the $(q+1)$-th iteration of Oxx.
\begin{lemma}
  Assume that $\Ab_{\Qc^\star}$ is full rank. If Oxx with
  $\yb\in\spansub{\Ab_{\Qc^\star}}$ as input selects $q$ true atoms
  $\Qc\subsetneq\Qc^\star$ and ERC-Oxx($\Ab,\Qc^\star,\Qc$) holds,
  then the $(q+1)$-th iteration of Oxx selects a true atom.
  \label{lem:tropp_omp_ols}
\end{lemma}
\begin{IEEEproof} 
  According to the selection
  rule~\eqref{eq:omp_rule}-\eqref{eq:ols_rule}, Oxx selects a true
  atom at iteration $(q+1)$ \IFF
  \begin{equation}
  \phi(\rb_\Qc)\triangleq\frac{
    \max_{i\notin\Qc^\star}\stdbars{\stdscal{\rb_\Qc,\tilde{\cb}_i}}}
  {\max_{i\in\Qc^\star\backslash\Qc}\stdbars{\stdscal{\rb_\Qc,\tilde{\cb}_i}}}
  < 1.
  \label{eq:cdt1}
  \end{equation}
  Let us gather the vectors $\tilde{\cb}_i$ indexed by
  $i\notin\Qc^\star$ and $i\in\Qc^\star\backslash\Qc$ in two matrices
  $\tilde{\Cb}_{\bullet\backslash\Qc^\star}$ and
  $\tilde{\Cb}_{\Qc^\star\backslash\Qc}$ of dimensions $m\times(n-k)$
  and $m\times(k-q)$, respectively where the notation $\bullet$ stands
  for all indices $i\in\stdacc{1,\ldots,n}$. The
  condition~\eqref{eq:cdt1} rereads:
  \begin{equation*} 
    \phi(\rb_\Qc)=\frac{\stdnorm{\tilde{\Cb}_{\bullet\backslash\Qc^\star}^t\rb_\Qc}_\infty}{
      \stdnorm{\tilde{\Cb}_{\Qc^\star\backslash\Qc}^t\rb_\Qc}_\infty} < 1.
  \end{equation*}
  Following Tropp's analysis, we re-arrange the vector $\rb_\Qc$
  occurring in the numerator. Since $\rb_\Qc=\Pb_{\Qc}^{\perp}\yb$ and
  $\yb\in\spansub{\Ab_{\Qc^\star}}$, $\rb_\Qc\in
  \spansub{\tilde{\Ab}_{\Qc^\star\backslash\Qc}}=
  \spansub{\tilde{\Cb}_{\Qc^\star\backslash\Qc}}$. We rewrite
  $\rb_\Qc$ as $\tilde{\Pb}_{\Qc^\star\backslash\Qc}\rb_\Qc$ where
  $\tilde{\Pb}_{\Qc^\star\backslash\Qc}$ stands for the orthogonal
  projector on $\spansub{\tilde{\Cb}_{\Qc^\star\backslash\Qc}}$:
  $\tilde{\Pb}_{\Qc^\star\backslash\Qc}=\tilde{\Pb}_{\Qc^\star\backslash\Qc}^t=
  \bigpth{\tilde{\Cb}_{\Qc^\star\backslash\Qc}
    \tilde{\Cb}_{\Qc^\star\backslash\Qc}^\dag}^t$. $\phi(\rb_\Qc)$
  rereads
  \begin{equation*} 
    \phi(\rb_\Qc)=\frac{\stdnorm{
        \bigpth{\tilde{\Cb}_{\Qc^\star\backslash\Qc}^\dag
          \tilde{\Cb}_{\bullet\backslash\Qc^\star}}^t
        \tilde{\Cb}_{\Qc^\star\backslash\Qc}^t\rb_\Qc}_\infty}{
      \stdnorm{\tilde{\Cb}_{\Qc^\star\backslash\Qc}^t\rb_\Qc}_\infty}.
  \end{equation*}

  This expression can obviously be majorized using the matrix norm:
  \begin{equation} 
    \phi(\rb_\Qc)\leqslant\stdnorm{
      \bigpth{\tilde{\Cb}_{\Qc^\star\backslash\Qc}^\dag
        \tilde{\Cb}_{\bullet\backslash\Qc^\star}}^t}_{\infty,\infty}.
  \label{eq:suffcdt4}
  \end{equation}
  Since the $\ell_\infty$ norm of a matrix is equal to the $\ell_1$
  norm of its transpose and $\|\,.\,\|_{1,1}$ equals the maximum
  column sum of the absolute value of its
  argument~\cite[Theorem~3.1]{Tropp04}, the upper bound
  of~\eqref{eq:suffcdt4} rereads
  \begin{align*}
    \max_{j\notin\Qc^\star}
    \stdnorm{\tilde{\Cb}_{\Qc^\star\backslash\Qc}^\dag
      \tilde{\cb}_{j}}_1=
    \max_{j\notin\Qc^\star}F_{\Qc^\star,\Qc}^{\mathrm{Oxx}}(\ab_j)
  \end{align*}
  according to Lemma~\ref{lem:reexpress_erc}.
  By definition of ERC-Oxx($\Ab,\Qc^\star,\Qc$), this upper bound is
  lower than 1 thus $\phi(\rb_\Qc)<1$. 
\end{IEEEproof} 

\subsubsection{Recursive expression of the ERC-Oxx formulas}
\label{sec:rec_expr}
We elaborate recursive expressions of
$F_{\Qc^\star,\Qc}^{\mathrm{Oxx}}(\ab_j)$ when \Qc is increased by one
element resulting in the new subset $\Qc'\subsetneq\Qc^\star$ (here,
we do not consider the case where $\Qc'=\Qc^\star$ since
$F_{\Qc^\star,\Qc^\star}^{\mathrm{Oxx}}(\ab_j)$ is not properly
defined,~\eqref{eq:Ferc_omp} and~\eqref{eq:Ferc_ols} being empty
sums). We will use the notation $\Qc'=\Qc\cup\stdacc{\ell}$ where
$\ell\in\Qc^\star\backslash\Qc$. To avoid any confusion,
$\tilde{\ab}_i$ will be systematically replaced by
$\tilde{\ab}_i^{\Qc}$ and $\tilde{\ab}_i^{\Qc'}$ to express the
dependence upon \Qc and $\Qc'$, respectively. In the same way,
$\tilde{\bb}_i$ will be replaced by $\tilde{\bb}_i^{\Qc}$ or
$\tilde{\bb}_i^{\Qc'}$ but for simplicity, we will keep the matrix
notations $\tilde{\Bb}_{\Qc^\star\backslash\Qc}$ and
$\tilde{\Bb}_{\Qc^\star\backslash\Qc'}$ without
superscript,~$\tilde{}$~referring to \Qc and $\Qc'$, respectively.

Let us first link $\tilde{\bb}_i^{\Qc}$ to $\tilde{\bb}_i^{\Qc'}$ when
$\tilde{\ab}_i^{\Qc'}\ne\zerob$.
\begin{lemma}
  \label{lem:recurs_bi}
  Assume that $\Ab_{\Qc'}$ is full rank and
  $\Qc'=\Qc\cup\stdacc{\ell}\subsetneq\Qc^\star$. Then,
  $\spansub{\Ab_{\Qc}}^\perp$ is the orthogonal direct sum of the
  subspaces $\spansub{\Ab_{\Qc'}}^\perp$ and
  $\spansub{\tilde{\ab}_{\ell}^\Qc}$, and the normalized projection of
  any atom $\ab_i\notin\spansub{\Ab_{\Qc'}}$ takes the form:
  \begin{align}
    \tilde{\bb}_i^{\Qc}&=\eta_{i}^{\Qc,\Qc'}\tilde{\bb}_i^{\Qc'}+
    \chi_{i}^{\Qc,\Qc'}\tilde{\bb}_{\ell}^{\Qc}
  \label{eq:recurs_bi}
  \end{align}
  where
  \begin{align}
  \eta_i^{\Qc,\Qc'}&=\frac{\bignorm{\tilde{\ab}_i^{\Qc'}}}{
    \bignorm{\tilde{\ab}_i^{\Qc}}}\,\in(0,1],
  \label{eq:eta}\\
  \chi_i^{\Qc,\Qc'}&=\stdscal{\tilde{\bb}_i^{\Qc},\tilde{\bb}_{\ell}^{\Qc}},
  \label{eq:chi}\\
    \bigpth{\eta_i^{\Qc,\Qc'}}^2&+\bigpth{\chi_i^{\Qc,\Qc'}}^2=1.
    \label{eq:pythagore}
  \end{align}
\end{lemma}
\begin{IEEEproof} %
  Since $\Qc\subsetneq\Qc'$, we have
  $\spansub{\Ab_{\Qc'}}^\perp\subseteq\spansub{\Ab_{\Qc}}^\perp$.
  Because $\Ab_{\Qc'}$ is full rank, $\spansub{\Ab_{\Qc'}}^\perp$ and
  $\spansub{\Ab_{\Qc}}^\perp$ are of consecutive dimensions.
  Moreover, $\tilde{\ab}_{\ell}^\Qc=\ab_{\ell}-\Pb_{\Qc}\ab_{\ell}
  \in\spansub{\Ab_{\Qc'}}\cap\spansub{\Ab_{\Qc}}^\perp$, and
  $\tilde{\ab}_{\ell}^\Qc\ne\zerob$ since $\Ab_{\Qc'}$ is full rank.
  As a vector of $\spansub{\Ab_{\Qc'}}$, $\tilde{\ab}_{\ell}^\Qc$ is
  orthogonal to $\spansub{\Ab_{\Qc'}}^\perp$. It follows that
  $\spansub{\tilde{\ab}_{\ell}^\Qc}$ is the orthogonal complement of
  $\spansub{\Ab_{\Qc'}}^\perp$ in $\spansub{\Ab_{\Qc}}^\perp$.

  The orthogonal decomposition of $\tilde{\ab}_i=\Pb_{\Qc}^\perp\ab_i$
  reads:
  \begin{align*}
    \tilde{\ab}_i^{\Qc}&=\tilde{\ab}_i^{\Qc'}+
    \stdscal{\tilde{\ab}_i^{\Qc},\tilde{\bb}_{\ell}^\Qc}\tilde{\bb}_{\ell}^\Qc
  \end{align*}
  since $\tilde{\bb}_{\ell}^\Qc$ is of unit norm. Replacing
  $\tilde{\ab}_i^{\Qc}=\stdnorm{\tilde{\ab}_i^{\Qc}}\,\tilde{\bb}_i^{\Qc}$
  and
  $\tilde{\ab}_i^{\Qc'}=\stdnorm{\tilde{\ab}_i^{\Qc'}}\,\tilde{\bb}_i^{\Qc'}$
  yields~\eqref{eq:recurs_bi}-\eqref{eq:chi}. Pythagoras' theorem
  yields~\eqref{eq:pythagore}. The assumption
  $\ab_i\notin\spansub{\Ab_{\Qc'}}$ implies that
  $\tilde{\ab}_i^{\Qc'}\ne\zerob$, then $\eta_i^{\Qc,\Qc'}>0$.
\end{IEEEproof}
\begin{lemma}
  \label{lem:recurs_Bi} 
  Assume that $\Ab_{\Qc^\star}$ is full rank. Let
  $\Qc\subsetneq\Qc'\subsetneq\Qc^\star$ with
  $\Qc'=\Qc\cup\stdacc{\ell}$. Then,
  $\spansub{\tilde{\Bb}_{\Qc^\star\backslash\Qc}}$ is the orthogonal
  direct sum of $\spansub{\tilde{\Bb}_{\Qc^\star\backslash\Qc'}}$ and
  $\spansub{\tilde{\bb}_{\ell}^\Qc}$.
\end{lemma}
\begin{IEEEproof}
  According to Corollary~\ref{cor:fullrank} in
  Appendix~\ref{app:reexpress_erc},
  $\tilde{\Bb}_{\Qc^\star\backslash\Qc}$ and
  $\tilde{\Bb}_{\Qc^\star\backslash\Qc'}$ are full rank matrices, thus
  their column spans are of consecutive cardinalities.
  Lemma~\ref{lem:recurs_bi} states that $\tilde{\bb}_{\ell}^\Qc$ is
  orthogonal to $\spansub{\Ab_{\Qc'}}^\perp$, thus it is orthogonal to
  $\tilde{\bb}_{i}^{\Qc'}\in\spansub{\Ab_{\Qc'}}^\perp$ for all $i
  \in\Qc^\star\backslash\Qc'$.
\end{IEEEproof}

We finally establish a link between
$F_{\Qc^\star,\Qc}^{\mathrm{Oxx}}(\ab_j)$ and
$F_{\Qc^\star,\Qc'}^{\mathrm{Oxx}}(\ab_j)$. It is a simple recursive
relation in the case of OMP. For OLS, we cannot directly relate the
two quantities but we express
$F_{\Qc^\star,\Qc}^{\mathrm{OLS}}(\ab_j)=
\bignorm{\tilde{\Bb}_{\Qc^\star\backslash\Qc}^\dag
  \tilde{\bb}_{j}^\Qc}_1$ with respect to
$\tilde{\Bb}_{\Qc^\star\backslash\Qc'}^\dag\tilde{\bb}_{j}^{\Qc'}$.
\begin{lemma}
  \label{lem:erc_omp_ols_rec} 
  Assume that $\Ab_{\Qc^\star}$ is full rank. Let
  $\Qc\subsetneq\Qc'\subsetneq\Qc^\star$ with
  $\Qc'=\Qc\cup\stdacc{\ell}$ and let $j\notin\Qc^\star$. If
  $\ab_j\notin\spansub{\Ab_{\Qc'}}$, then
  \FINAL{$F_{\Qc^\star,\Qc'}^{\mathrm{Oxx}}(\ab_j)$ takes the
    forms~\eqref{eq:omp_erc_rec}
    \refstepcounter{equation}\label{eq:omp_erc_rec}
    and~\eqref{eq:ols_erc_rec}
    \refstepcounter{equation}\label{eq:ols_erc_rec}}
  \DRAFT{
    \begin{align}
      F_{\Qc^\star,\Qc}^{\mathrm{OMP}}(\ab_j)&=
      F_{\Qc^\star,\Qc'}^{\mathrm{OMP}}(\ab_j)+
      \,\bigbars{\bigpth{\Ab_{\Qc^\star}^\dag\ab_j}(\ell)}
      \label{eq:omp_erc_rec}\\
      F_{\Qc^\star,\Qc}^{\mathrm{OLS}}(\ab_j)&= \Biggbars{
        \chi_{j}^{\Qc,\Qc'}-\eta_{j}^{\Qc,\Qc'}
        \sum_{i\in\Qc^\star\backslash\Qc'}
        \frac{\betab_j^{\Qc^\star\backslash\Qc'}(i)\chi_i^{\Qc,\Qc'}}
        {\eta_i^{\Qc,\Qc'}}}+
      \eta_{j}^{\Qc,\Qc'}\sum_{i\in\Qc^\star\backslash\Qc'}
      \frac{\bigbars{\betab_j^{\Qc^\star\backslash\Qc'}(i)}}
      {\eta_i^{\Qc,\Qc'}}
      \label{eq:ols_erc_rec}
    \end{align}
  }
  where $\eta_i^{\Qc,\Qc'}$ and $\chi_i^{\Qc,\Qc'}$ are defined
  in~\eqref{eq:eta}-\eqref{eq:chi} and
  $\betab^{\Qc^\star\backslash\Qc'}_{j}\triangleq
  \tilde{\Bb}_{\Qc^\star\backslash\Qc'}^\dag\tilde{\bb}_{j}^{\Qc'}$.
\end{lemma}
\begin{IEEEproof} 
  \eqref{eq:omp_erc_rec} straightforwardly follows from the
  definition~\eqref{eq:Ferc_omp} of
  $F_{\Qc^\star,\Qc}^{\mathrm{OMP}}(\ab_j)$.

  Let us now establish~\eqref{eq:ols_erc_rec}. We denote by
  $\tilde{\Pb}_{\Qc^\star\backslash\Qc}$ and
  $\tilde{\Pb}_{\Qc^\star\backslash\Qc'}$ the orthogonal projectors on
  $\spansub{\tilde{\Bb}_{\Qc^\star\backslash\Qc}}$ and
  $\spansub{\tilde{\Bb}_{\Qc^\star\backslash\Qc'}}$. Because
  $\spansub{\tilde{\Bb}_{\Qc^\star\backslash\Qc}}$ is the orthogonal
  direct sum of $\spansub{\tilde{\Bb}_{\Qc^\star\backslash\Qc'}}$ and
  $\spansub{\tilde{\bb}_{\ell}^\Qc}$ (Lemma~\ref{lem:recurs_Bi}), we
  have the orthogonal decomposition:
  \begin{align*}
    \tilde{\Pb}_{\Qc^\star\backslash\Qc}\tilde{\bb}_{j}^{\Qc}&=
    \tilde{\Pb}_{\Qc^\star\backslash\Qc'}\tilde{\bb}_{j}^{\Qc}+
    \chi_{j}^{\Qc,\Qc'}\tilde{\bb}_{\ell}^{\Qc}.
  \end{align*}
  \eqref{eq:recurs_bi} yields
  \begin{align*}
    \tilde{\Pb}_{\Qc^\star\backslash\Qc}\tilde{\bb}_{j}^{\Qc}&=
    \eta_{j}^{\Qc,\Qc'}
    \tilde{\Pb}_{\Qc^\star\backslash\Qc'}\tilde{\bb}_{j}^{\Qc'}
    +\chi_{j}^{\Qc,\Qc'}\tilde{\bb}_{\ell}^{\Qc}
  \end{align*}
  ($\tilde{\Pb}_{\Qc^\star\backslash\Qc'}\tilde{\bb}_{\ell}^{\Qc}=\zerob$
  according to Lemma~\ref{lem:recurs_Bi}) and then
  \begin{align*}
    \tilde{\Pb}_{\Qc^\star\backslash\Qc}\tilde{\bb}_{j}^{\Qc}&=
    \eta_{j}^{\Qc,\Qc'}\sum_{i\in\Qc^\star\backslash\Qc'}
    \betab^{\Qc^\star\backslash\Qc'}_{j}(i)\tilde{\bb}_{i}^{\Qc'}+
    \chi_{j}^{\Qc,\Qc'}\tilde{\bb}_{\ell}^{\Qc}
  \end{align*}
  by definition of $\betab^{\Qc^\star\backslash\Qc'}_{j}$.
  In the latter equation, we re-express $\tilde{\bb}_{i}^{\Qc'}$ with
  respect to $\tilde{\bb}_{i}^{\Qc}$ using~\eqref{eq:recurs_bi}:
  \begin{align*}
    \tilde{\Pb}_{\Qc^\star\backslash\Qc}&\tilde{\bb}_{j}^{\Qc}
    =\eta_{j}^{\Qc,\Qc'}\sum_{i\in\Qc^\star\backslash\Qc'}
    \frac{\betab^{\Qc^\star\backslash\Qc'}_{j}(i)}{
      \eta_{i}^{\Qc,\Qc'}}\,\tilde{\bb}_{i}^{\Qc}+
    \FINAL{\nonumber\\&} 
    \Biggacc{\chi_{j}^{\Qc,\Qc'}-
      \eta_{j}^{\Qc,\Qc'}\sum_{i\in\Qc^\star\backslash\Qc'}
      \frac{\betab^{\Qc^\star\backslash\Qc'}_{j}(i)\chi_{i}^{\Qc,\Qc'}}
      {\eta_{i}^{\Qc,\Qc'}}} \tilde{\bb}_{\ell}^{\Qc}.
  \label{eq:decompos_bbad2}
  \end{align*}
  Thus, $F_{\Qc^\star,\Qc}^{\mathrm{OLS}}(\ab_j)=
  \bignorm{\tilde{\Bb}_{\Qc^\star\backslash\Qc}^\dag
    \tilde{\bb}_{j}^\Qc}_1$ reads~\eqref{eq:ols_erc_rec}.
\end{IEEEproof} 

\subsubsection{The ERC is a sufficient recovery condition for OLS}
\label{sec:erc_suffic}
The key result of Lemma~\ref{lem:erc_decrease} (see
Section~\ref{sec:overview_sufficient}) states that when
$j\notin\Qc^\star$, $F_{\Qc^\star,\Qc}^{\mathrm{OLS}}(\ab_j)$ is
decreasing when $\Qc\subsetneq\Qc^\star$ is growing provided that
$F_{\Qc^\star,\Qc}^{\mathrm{OLS}}(\ab_j)<1$, and that
$F_{\Qc^\star,\Qc}^{\mathrm{OMP}}(\ab_j)$ is always decreasing.
\begin{IEEEproof}[Proof of Lemma~\ref{lem:erc_decrease}]
  It is sufficient to prove the result when
  $\Card{\Qc'}=\Card{\Qc}+1$. The case $\Card{\Qc'}>\Card{\Qc}+1$
  obviously deduces from the former case by recursion.

  Let $\Qc\subsetneq\Qc'\subsetneq\Qc^\star$ with
  $\Card{\Qc'}=\Card{\Qc}+1$. The result is obvious when
  $\ab_j\in\spansub{\Ab_{\Qc'}}$: $\tilde{\ab}_{j}=\zerob$ then
  $F_{\Qc^\star,\Qc'}^{\mathrm{Oxx}}(\ab_j)=0$. When
  $\ab_j\notin\spansub{\Ab_{\Qc'}}$,~\eqref{eq:omp_decrease} obviously
  deduces from~\eqref{eq:omp_erc_rec}. The proof
  of~\eqref{eq:ols_decrease} relies on the study of function
  $\varphi(\eta)=\stdbars{\sqrt{1-\eta^2}-C\eta}+D\eta$ which is fully
  defined in~\eqref{eq:varphi},~\eqref{eq:C} and~\eqref{eq:D} in
  Appendix~\ref{app:etude_fct}. Because this study is rather
  technical, we place it in Appendix~\ref{app:etude_fct}.

  We notice that $F_{\Qc^\star,\Qc}^{\mathrm{OLS}}(\ab_j)$ given
  in~\eqref{eq:ols_erc_rec} takes the form
  $\varphi\bigpth{\eta_{j}^{\Qc,\Qc'}}$ where the variables occurring
  in $C$ and $D$ (see~\eqref{eq:C} and~\eqref{eq:D}) are set to
  $N\leftarrow\Card{\Qc^\star\backslash\Qc'}$,
  $\eta_i\leftarrow\eta_i^{\Qc,\Qc'}$,
  $\chi_i\leftarrow\chi_i^{\Qc,\Qc'}$, and
  $\betab\leftarrow\sgn\bigpth{\chi_{j}^{\Qc,\Qc'}}
  \betab_j^{\Qc^\star\backslash\Qc'}$. Now, we invoke
  Lemma~\ref{lem:1Dter} in Appendix~\ref{app:etude_fct}: as
  $F_{\Qc^\star,\Qc'}^{\mathrm{OLS}}(\ab_j)=
  \bignorm{\betab_j^{\Qc^\star\backslash\Qc'}}_1$ plays the role of
  $\stdnorm{\betab}_1$, $F_{\Qc^\star,\Qc}^{\mathrm{OLS}}(\ab_j)<1$
  implies that $F_{\Qc^\star,\Qc'}^{\mathrm{OLS}}(\ab_j)\leqslant
  F_{\Qc^\star,\Qc}^{\mathrm{OLS}}(\ab_j)$.
\end{IEEEproof} 
We deduce from Lemmas~\ref{lem:erc_decrease}
and~\ref{lem:tropp_omp_ols} that ERC-Oxx($\Ab,\Qc^\star,\Qc$) are
sufficient recovery conditions when $\Qc\subsetneq\Qc^\star$ has been
reached (Theorem~\ref{th:suffic_omp_ols}).
\begin{IEEEproof}[Proof of Theorem~\ref{th:suffic_omp_ols}]
  Apply Lemma~\ref{lem:tropp_omp_ols} at each iteration
  $q,\ldots,k-1$ until the increased subset $\Qc'$ matches
  $\Qc^\star$. The ERC-Oxx($\Ab,\Qc^\star,\,.\,$) assumption of
  Lemma~\ref{lem:tropp_omp_ols} is always fulfilled according to
  Lemma~\ref{lem:erc_decrease}.
\end{IEEEproof}
Finally, we prove that ERC($\Ab,\Qc^\star$) is a necessary and
sufficient condition of successful recovery for OLS
(Theorem~\ref{th:tropp_ols}).
\begin{IEEEproof}[Proof of Theorem~\ref{th:tropp_ols}]
  The sufficient condition is a special case of
  Theorem~\ref{th:suffic_omp_ols} for $\Qc=\emptyset$. The necessary
  condition identifies with that of Theorem~\ref{th:tropp_omp} since
  ERC-OLS($\Ab,\Qc^\star,\emptyset$) simplifies to
  ERC($\Ab,\Qc^\star$).
\end{IEEEproof} 

\subsection{Necessary conditions}
\label{sec:necessary}
We provide the technical analysis to prove that
ERC-Oxx($\Ab,\Qc^\star,\Qc$) is not only a sufficient condition of
exact recovery when $\Qc\subsetneq\Qc^\star$ has been reached, but
also a necessary condition in the worst case. We will prove
Theorems~\ref{th:necess_ols} and~\ref{th:necess_omp} (see
Section~\ref{sec:overview}) generalizing Tropp's necessary
condition~\cite[Theorem~3.10]{Tropp04} to any iteration of OMP and
OLS.

We will first assume that Oxx exactly recovers
$\Qc\subsetneq\Qc^\star$ in $q=\Card{\Qc}$ iterations with some input
vector in $\spansub{\Ab_\Qc}$. This reachability assumption allows us
to carry out a parallel analysis of OMP and OLS
(subsection~\ref{sec:stage1}) leading to the following proposition.
\begin{proposition}
  {\textbf{[Necessary condition for Oxx after $q$ iterations]}} Assume
  that $\Ab_{\Qc^\star}$ is full rank and $\Qc\subsetneq\Qc^\star$ is
  reachable from an input in $\spansub{\Ab_\Qc}$ by Oxx. If
  ERC-Oxx($\Ab,\Qc^\star,\Qc$) does not hold, then there exists
  $\yb\in\spansub{\Ab_{\Qc^\star}}$ for which Oxx selects \Qc in the
  first $q$ iterations and then a wrong atom at iteration $(q+1)$.
  \label{prop:necess_omp_ols}
\end{proposition}
This proposition coincides with Theorem~\ref{th:necess_omp} in the
case of OMP whereas for OLS, Theorem~\ref{th:necess_ols} does not
require the assumption that \Qc is reachable
(subsection~\ref{sec:stage2}).

\subsubsection{Parallel analysis of OMP and OLS}
\label{sec:stage1}
\begin{IEEEproof}[Proof of Proposition~\ref{prop:necess_omp_ols}]
  We proceed the proof of Lemma~\ref{lem:tropp_omp_ols} backwards.
  By assumption, the right hand-side of inequality~\eqref{eq:suffcdt4} 
  is equal to
  \begin{equation*}
    \stdnorm{\bigpth{\tilde{\Cb}_{\Qc^\star\backslash\Qc}^\dag
        \tilde{\Cb}_{\bullet\backslash\Qc^\star}}^t}_{\infty,\infty}= 
    \max_{j\notin\Qc^\star}F_{\Qc^\star,\Qc}^{\mathrm{Oxx}}(\ab_j)\geqslant 1.
  \end{equation*}
  By definition of induced norms, there exists a vector
  $\vb\in\Rbb^{k-q}$ satisfying $\vb\ne\zerob$ and
  \begin{align}
    \frac{\stdnorm{ \bigpth{\tilde{\Cb}_{\Qc^\star\backslash\Qc}^\dag
          \tilde{\Cb}_{\bullet\backslash\Qc^\star}}^t
        \vb}_\infty}{\stdnorm{\vb}_\infty } = \stdnorm{
      \bigpth{\tilde{\Cb}_{\Qc^\star\backslash\Qc}^\dag
        \tilde{\Cb}_{\bullet\backslash\Qc^\star}}^t}_{\infty,\infty}\geqslant
    1.
    \label{eq:egdemERC}
  \end{align}
  Define 
  \begin{align}
    \label{eq:expression_yt_CN}
    \hat{\yb} &= \Ab_{\Qc^\star\backslash\Qc}
    (\tilde{\Cb}_{\Qc^\star\backslash\Qc}^t 
    \tilde{\Ab}_{\Qc^\star\backslash\Qc})^{-1}\vb.
  \end{align}
  The matrix inversion in~\eqref{eq:expression_yt_CN} is well defined
  since $\tilde{\Ab}_{\Qc^\star\backslash\Qc}$ is full rank
  (Corollary~\ref{cor:fullrank} in Appendix~\ref{app:reexpress_erc})
  and
  $\tilde{\Cb}_{\Qc^\star\backslash\Qc}=\tilde{\Ab}_{\Qc^\star\backslash\Qc}$
  or $\tilde{\Bb}_{\Qc^\star\backslash\Qc}$ reads as the right product
  of $\tilde{\Ab}_{\Qc^\star\backslash\Qc}$ with a nondegenerate
  diagonal matrix.
  By taking into account that
  $\tilde{\Ab}_{\Qc^\star\backslash\Qc}=
  \Pb_{\Qc}^\perp{\Ab}_{\Qc^\star\backslash\Qc}$, we obtain that
  \begin{align}\label{eq:connec_it1}
    \vb = \tilde{\Cb}_{\Qc^\star\backslash\Qc}^t\Pb_{\Qc}^\perp\hat{\yb}.
  \end{align}

  Since the left hand-side of~\eqref{eq:egdemERC} identifies with
  $\phi(\Pb_{\Qc}^\perp\hat{\yb})$ where $\phi$ is defined
  in~\eqref{eq:cdt1},~\eqref{eq:egdemERC} yields:
  \begin{equation}
    \max_{j\notin\Qc^\star}\stdbars{\stdscal{\Pb_{\Qc}^\perp\hat{\yb},\tilde{\cb}_{j}}}
    \geqslant
    \max_{i\in\Qc^\star\backslash\Qc}\stdbars{\stdscal{\Pb_{\Qc}^\perp\hat{\yb},\tilde{\cb}_{i}}}.
    \label{eq:erc_omp_ols_failure}
  \end{equation}
  Moreover, we have $\Pb_{\Qc}^\perp\hat{\yb}\ne\zerob$ according
  to~\eqref{eq:connec_it1} and $\vb\ne\zerob$.

  Now, let $\zb\in\spansub{\Ab_\Qc}$ denote the input for which Oxx
  recovers \Qc. According to Lemma~\ref{lem:continuity} in
  Appendix~\ref{app:CN}, the first $q$ iterations of Oxx with the
  modified input $\yb=\zb+\varepsilon\hat{\yb}$ also select \Qc when
  $\varepsilon>0$ is sufficiently small. Because $\Pb_{\Qc}^\perp\yb
  =\varepsilon\Pb_{\Qc}^\perp\hat{\yb}$
  and~\eqref{eq:erc_omp_ols_failure} holds, the $(q+1)$-th iteration
  of Oxx necessarily selects a wrong atom.
\end{IEEEproof} 
At this point, we have proved Theorem~\ref{th:necess_omp} which is
relative to OMP. 

\subsubsection{OLS ability to reach any subset}
\label{sec:stage2}
In order to prove Theorem~\ref{th:necess_ols}, we establish that any
subset \Qc can be reached using OLS with some input
$\yb\in\spansub{\Ab_\Qc}$ (Lemma~\ref{lem:ols_reach}). To generate
\yb, we assign decreasing weight coefficients to the atoms
$\stdacc{\ab_i,\,i\in\Qc}$ with a rate of decrease which is high
enough.
\begin{IEEEproof}[Proof of Lemma~\ref{lem:ols_reach}]
  Without loss of generality, we assume that the elements of \Qc
  correspond to the first $q$ atoms. For arbitrary values of
  $\varepsilon_2,\ldots, \varepsilon_{q}>0$, we define the following
  recursive construction:
  \begin{itemize}
    \item $\yb_1=\ab_1$,
    \item $\yb_{p}=\yb_{p-1}+\varepsilon_{p}\ab_{p}$ for
      $p\in\stdacc{2,\ldots,q}$.
  \end{itemize}
  ($\yb_{p}$ implicitly depends on
  $\varepsilon_2,\ldots,\varepsilon_{p}$) and set
  $\yb\triangleq\yb_{q}$. We show by recursion that there
  exist $\varepsilon_2,\ldots,\varepsilon_{p}>0$ such that OLS
  with $\yb_{p}$ as input successively selects
  $\ab_1,\ldots,\ab_{p}$ during the first $p$
  iterations (in particular, the selection rule~\eqref{eq:ols_rule}
  always yields a unique maximum).

  The statement is obviously true for $\yb_1=\ab_1$. Assume that it is
  true for $\yb_{p-1}$ with some $\varepsilon_2,\ldots,
  \varepsilon_{p-1}>0$ (these parameters will remain fixed in the
  following). According to Lemma~\ref{lem:continuity} in
  Appendix~\ref{app:CN}, there exists $\varepsilon_{p}>0$ such that
  OLS with $\yb_{p}=\yb_{p-1}+\varepsilon_{p}\ab_{p}$ as input selects
  the same atoms as with $\yb_{p-1}$ during the first $p-1$
  iterations, \ie $\ab_1,\ldots,\ab_{p-1}$ are successively chosen. At
  iteration $p$, the current active set thus reads
  $\Qc'=\stdacc{1,\dots,p-1}$ and the OLS residual corresponding to
  $\yb_{p}$ takes the form
  \begin{align*}
    \rb_{\Qc'}&=
    \Pb_{\Qc'}^\perp\yb_{p-1}+\varepsilon_{i}\Pb_{\Qc'}^\perp\ab_{p}\,=\,
    \varepsilon_{p}\tilde{\ab}_{p}^{\Qc'}
  \end{align*}
  since $\yb_{p-1}\in\spansub{\Ab_{\Qc'}}$. Thus, $\rb_{\Qc'}$ is
  proportional to $\tilde{\ab}_{p}^{\Qc'}$ and then to
  $\tilde{\bb}_{p}^{\Qc'}$. Finally, the OLS
  criterion~\eqref{eq:ols_rule} is maximum for the atom $\ab_{p}$ and
  the maximum value is equal to
  $\stdbars{\stdscal{\rb_{\Qc'},\tilde{\bb}_{p}^{\Qc'}}}=\|\rb_{\Qc'}\|$
  since $\tilde{\bb}_{p}^{\Qc'}$ is of unit norm.

  Finally, we show that no other atom $\ab_i$ yields this maximum
  value. Apply Lemma~\ref{lem:fullrank} in
  Appendix~\ref{app:reexpress_erc}: the full rankness of
  $\Ab_{\Qc'\cup\stdacc{p,i}}$ (as a family of less than $\spark{\Ab}$
  atoms) implies that
  $\bigcro{\tilde{\bb}_{p}^{\Qc'},\tilde{\bb}_{i}^{\Qc'}}$ is full
  rank, thus $\tilde{\bb}_{p}^{\Qc'}$ and $\tilde{\bb}_{i}^{\Qc'}$
  cannot be collinear.
\end{IEEEproof} 
Using Lemma~\ref{lem:ols_reach}, Proposition~\ref{prop:necess_omp_ols}
simplifies to Theorem~\ref{th:necess_ols} in which the assumption that
\Qc is reachable by OLS is omitted.

\section{Re-expression of the ERC-Oxx formulas}
\label{app:reexpress_erc}
In this appendix, we prove Lemma~\ref{lem:reexpress_erc} by
successively re-expressing
$\tilde{\Ab}_{\Qc^\star\backslash\Qc}^\dag\tilde{\ab}_{j}$
and
$\tilde{\Bb}_{\Qc^\star\backslash\Qc}^\dag\tilde{\bb}_{j}$.
Let us first show that when $\Ab_{\Qc^\star}$ is full rank, the
matrices $\tilde{\Ab}_{\Qc^\star\backslash\Qc}$ and
$\tilde{\Bb}_{\Qc^\star\backslash\Qc}$ are full rank. This result is
stated below as a corollary of Lemma~\ref{lem:fullrank}.
\begin{lemma}
  If $\Qc\cap\Qc'=\emptyset$ and $\Ab_{\Qc\cup\Qc'}$ is full rank,
  then $\tilde{\Ab}_{\Qc'}^\Qc$ and $\tilde{\Bb}_{\Qc'}^\Qc$ are full
  rank.
  \label{lem:fullrank}
\end{lemma}
\begin{IEEEproof} %
  To prove that $\tilde{\Ab}_{\Qc'}^\Qc$ is full rank, we assume that
  $\sum_{i\in\Qc'}\alpha_i\tilde{\ab}_i^\Qc=\zerob$ with
  $\alpha_i\in\Rbb$. By definition of
  $\tilde{\ab}_i^\Qc=\Pb_\Qc^\perp\ab_i=\ab_i-\Pb_\Qc\ab_i$, it
  follows that $\sum_{i\in\Qc'}\alpha_i\ab_i\in\spansub{\Ab_\Qc}$.
  Since $\Ab_{\Qc\cup\Qc'}$ is full rank, we conclude that all
  $\alpha_i$'s are 0.

  The full rankness of $\tilde{\Bb}_{\Qc'}^\Qc$ follows from that of
  $\tilde{\Ab}_{\Qc'}^\Qc$ since for all $i\in\Qc'$,
  $\tilde{\bb}_i^\Qc=\tilde{\ab}_i^\Qc/\stdnorm{\tilde{\ab}_i^\Qc}$ is
  collinear to $\tilde{\ab}_i^\Qc$.
\end{IEEEproof}
The application of Lemma~\ref{lem:fullrank} to
$\Qc'=\Qc^\star\backslash\Qc$ leads to the following corollary.
\begin{corollary}
  Assume that $\Ab_{\Qc^\star}$ is full rank. For
  $\Qc\subsetneq\Qc^\star$, $\tilde{\Ab}_{\Qc^\star\backslash\Qc}$ and
  $\tilde{\Bb}_{\Qc^\star\backslash\Qc}$ are full rank.
  \label{cor:fullrank}
\end{corollary}
\begin{lemma}
  Assume that $\Ab_{\Qc^\star}$ is full rank. For
  $\Qc\subsetneq\Qc^\star$ and $j\notin\Qc^\star$,
  $\tilde{\Ab}_{\Qc^\star\backslash\Qc}^\dag\tilde{\ab}_{j}=
  \bigpth{\Ab_{\Qc^\star}^\dag{\ab}_{j}}_{|(\Qc^\star\backslash\Qc)}$
  where $|$ denotes the restriction of a vector to a subset of its
  coefficients.
  \label{lem:dag}
\end{lemma}
\begin{IEEEproof} 
  The orthogonal decomposition of $\ab_j$ on
  $\spansub{\Ab_{\Qc^\star}}$ takes the form:
  \begin{equation*}
    \ab_j=\Ab_{\Qc^\star}
    \bigpth{\Ab_{\Qc^\star}^\dag{\ab}_{j}}+\Pb_{\Qc^\star}^\perp\ab_j.
  \end{equation*}
  Projecting onto $\spansub{\Ab_\Qc}^{\perp}$, we obtain
  \begin{equation}
    \tilde{\ab}_{j}=\tilde{\Ab}_{\Qc^\star\backslash\Qc}
    \bigpth{\Ab_{\Qc^\star}^\dag{\ab}_{j}}_{|(\Qc^\star\backslash\Qc)}+
    \Pb_{\Qc^\star}^\perp\ab_j
  \label{eq:decompos_abad3}
  \end{equation}
  ($\Pb_{\Qc}^\perp\Pb_{\Qc^\star}^\perp=\Pb_{\Qc^\star}^\perp$
  because
  $\spansub{\Ab_{\Qc^\star}}^{\perp}\subseteq\spansub{\Ab_{\Qc}}^{\perp}$).
  For $i\in\Qc^\star\backslash\Qc$,
  $\tilde{\ab}_i=\ab_i-\Pb_{\Qc}\ab_i\in\spansub{\Ab_{\Qc^\star}}$.
  Thus, we have
  $\spansub{\tilde{\Ab}_{\Qc^\star\backslash\Qc}}\subseteq
  \spansub{\Ab_{\Qc^\star}}$, and
  $\Pb_{\Qc^\star}^\perp\ab_j$ is orthogonal to
  $\spansub{\tilde{\Ab}_{\Qc^\star\backslash\Qc}}$. According to
  Corollary~\ref{cor:fullrank}, $\tilde{\Ab}_{\Qc^\star\backslash\Qc}$
  is full rank. It follows from~\eqref{eq:decompos_abad3} that
  $\tilde{\Ab}_{\Qc^\star\backslash\Qc}^\dag\tilde{\ab}_{j}=
  \bigpth{\Ab_{\Qc^\star}^\dag\ab_j}_{|(\Qc^\star\backslash\Qc)}$.
\end{IEEEproof}
\begin{lemma}
  \label{lem:dag2}
  Assume that $\Ab_{\Qc^\star}$ is full rank. For
  $\Qc\subsetneq\Qc^\star$ and $j\notin\Qc^\star$,
  \begin{align*}
    \|\tilde{\ab}_{j}\|\,
    \tilde{\Bb}_{\Qc^\star\backslash\Qc}^\dag\tilde{\bb}_{j}&=
    \Deltab_{\|\tilde{\ab}_i\|}\,
    \bigpth{\Ab_{\Qc^\star}^\dag{\ab}_{j}}_{|(\Qc^\star\backslash\Qc)}
  \end{align*}
  where $\Deltab_{\|\tilde{\ab}_i\|}$ stands for the diagonal matrix
  whose diagonal elements are
  $\stdacc{\|\tilde{\ab}_i\|,\,i\in\Qc^\star\backslash\Qc}$.
\end{lemma}
\begin{IEEEproof} 
  The result directly follows from $\tilde{\ab}_{j}=
  \stdnorm{\tilde{\ab}_{j}}\, \tilde{\bb}_{j}$,
  $\tilde{\bb}_i=\tilde{\ab}_i/\|\tilde{\ab}_i\|$ for
  $i\in\Qc^\star\backslash\Qc$, and from Lemma~\ref{lem:dag}.
\end{IEEEproof}
\begin{IEEEproof}[Proof of Lemma~\ref{lem:reexpress_erc}]
  The result is obvious when $\tilde{\ab}_{j}=\zerob$. It
  follows from Lemmas~\ref{lem:dag} and~\ref{lem:dag2} when
  $\tilde{\ab}_{j}\ne\zerob$.
\end{IEEEproof}

\section{Technical results needed for the proof of Lemma~\ref{lem:erc_decrease}}
\label{app:etude_fct}
With simplified notations, the expression~\eqref{eq:ols_erc_rec} of
$F_{\Qc^\star,\Qc}^{\mathrm{OLS}}(\ab_j)$ reads
\begin{align}
  \varphi(\eta)&\triangleq\stdbars{\sqrt{1-\eta^2}-C\eta}+D\eta
  \label{eq:varphi}
\end{align}
where $\eta\in(0,1]$ and $C$ and $D$ take the form
\begin{align}
  C&=\sum_{i=1}^N\frac{\beta_i\chi_i}{\eta_i}
  \label{eq:C}\\
  D&=\sum_{i=1}^N\frac{\stdbars{\beta_i}}{\eta_i}
  \label{eq:D}
\end{align}
with $N\geqslant 1$,
$\betab=\stdcro{\beta_1,\ldots,\beta_N}\in\Rbb^N$, and for all $i$,
$\eta_i\in(0,1]$ and $\chi_i\in[-1,1]$ satisfy $\eta_i^2+\chi_i^2=1$.
Note that we can freely assume from~\eqref{eq:ols_erc_rec} that
$\chi_{j}^{\Qc,\Qc'}=
\pm\sqrt{1-\bigpth{\eta_{j}^{\Qc,\Qc'}}^2}\geqslant 0$. When
$\chi_{j}^{\Qc,\Qc'}<0$, one just needs to replace \betab by
$-\betab$ in~\eqref{eq:C} and~\eqref{eq:D}.

The succession of small lemmas hereafter aims at minorizing
$\varphi(\eta)$ for arbitrary values of $\eta,\,\eta_i,\,\chi_i$ and
$\betab$. They lead to the main minoration result of
Lemma~\ref{lem:1Dter}.
\begin{lemma}
  \label{lem:1D}
  Let $\betab\in\Rbb^N$.
  \begin{align}
  \textrm{If}\;C\leqslant0,&\,\forall\eta\in[0,1],\,\varphi(\eta)\geqslant
  1+(\|\betab\|_1-1)\eta.
  \label{eq:minor1Da}\\
  \textrm{If}\;C>0,&\,\min_{\eta\in[0,1]}\varphi(\eta)=\min\Bigpth{1,
  \froc{D}{\sqrt{1+C^2}}}.
  \label{eq:minor1Db}
  \end{align}
\end{lemma}
\begin{IEEEproof} 
  We first study the function
  $f(\eta)\triangleq\sqrt{1-\eta^2}-C\eta$. We have
  $f(0)=1,\,f(1)=-C$, and $f$ is concave on $[0,1]$. To minorize
  $\varphi(\eta)=\bars{f(\eta)}+D\eta$, we distinguish two cases
  depending on the sign of $C$.

  When $C\leqslant 0$, $f(\eta)\geqslant 0$ for all $\eta$. Since
  $|f|=f$ is concave, it can be minorized by the secant line joining
  $f(0)$ and $f(1)$, therefore, $\bars{f(\eta)}\geqslant
  1-(C+1)\eta\geqslant 1-\eta$. \eqref{eq:minor1Da} follows from
  $\varphi(\eta)=\bars{f(\eta)}+D\eta$ and $D\geqslant\|\betab\|_1$
  (because $\eta_i$ are all in $(0,1]$).
  
  When $C> 0$, $f(\eta)\geqslant 0$ for $\eta\in [0,z]$ and $<0$ in
  $(z,1]$, with $z\triangleq 1/\sqrt{1+C^2}$. $D\geqslant 0$ and
  $f(z)=0$ imply that for $\eta>z$,
  $\varphi(\eta)\geqslant\varphi(z)$, thus the minimum of $\varphi$ is
  reached for $\eta\in[0,z]$. On $[0,z]$,
  $\varphi(\eta)=f(\eta)+D\eta$ is concave, therefore the minimum
  value is either $\varphi(0)=1$ or $\varphi(z)=Dz$.
\end{IEEEproof}
The following two lemmas are simple inequalities linking $C$, $D$, and
$\stdnorm{\betab}_1$.
\begin{lemma}
  \label{lem:d2c2}
  $\forall\betab\in\Rbb^N,\,D^2-C^2\geqslant\stdnorm{\betab}_1^2$.
\end{lemma}
\begin{IEEEproof} %
  By developing $C^2$ and $D^2$ from~\eqref{eq:C} and~\eqref{eq:D}, we
  get
  \begin{align*}
    C^2&=\sum_i\frac{\beta_i^2\chi_i^2}{\eta_i^2}+\sum_{i\ne j}
    \frac{\beta_i\beta_j\chi_i\chi_j}{\eta_i\eta_j}\\
    D^2&=\sum_i\frac{\beta_i^2}{\eta_i^2}+\sum_{i\ne j}
    \frac{\stdbars{\beta_i\beta_j}}{\eta_i\eta_j}
  \end{align*}
  Since $\forall i,\,\eta_i^2+\chi_i^2=1$, we have:
  \begin{align}
    D^2-C^2\DRAFT{&}=\sum_i\beta_i^2+\sum_{i\ne j}
    \frac{\stdbars{\beta_i\beta_j}}{\eta_i\eta_j}\,
    (1-\sigma_i\sigma_j\chi_i\chi_j)\nonumber\\
    \DRAFT{&}=\biggcro{\sum_i\stdbars{\beta_i}}^2+\sum_{i\ne j}
    \stdbars{\beta_i\beta_j}\biggcro{
      \frac{1-\sigma_i\sigma_j\chi_i\chi_j}{\eta_i\eta_j}-1}
    \label{eq:d2c2}
  \end{align}
  with $\sigma_i=\sgn(\beta_i)=\pm1$ if $\beta_i\ne 0$, and
  $\sigma_i=1$ otherwise. Because $\eta_i$ and $\chi_i$ satisfy
  $\eta_i^2+\chi_i^2=1$, they reread $\eta_i=\cos\theta_i$ and
  $\chi_i=\sin\theta_i$, so
  $\eta_i\eta_j+\sigma_i\sigma_j\chi_i\chi_j=
  \cos(\theta_{i}\pm\theta_{j})\leqslant 1$ which proves that the last
  bracketed expression in~\eqref{eq:d2c2} is non-negative.
  \eqref{eq:d2c2} yields $D^2-C^2\geqslant\|\betab\|_1^2$.
\end{IEEEproof}
\begin{lemma}
  \label{lem:alpha2}
  $\forall\betab\in\Rbb^N,\,$ $\stdnorm{\betab}_1\leqslant 1$ implies
  that $\stdnorm{\betab}_1\leqslant\froc{D}{\sqrt{1+C^2}}$.
\end{lemma}
\begin{IEEEproof} %
  $(1+C^2)\stdnorm{\betab}_1^2\leqslant
  \stdnorm{\betab}_1^2+C^2\leqslant D^2$ according to
  Lemma~\ref{lem:d2c2}.
\end{IEEEproof}
We can now establish the main lemma that will enable us to conclude
that if $F_{\Qc^\star,\Qc}^{\mathrm{OLS}}(\ab_j)<1$,
$F_{\Qc^\star,\Qc'}^{\mathrm{OLS}}(\ab_j)$ is
monotonically nonincreasing when $\Qc'\supsetneq\Qc$ is growing.
\begin{lemma}
  \label{lem:1Dter}
  $\forall\betab\in\Rbb^N,\,\forall\eta\in[0,1]$, $\varphi(\eta)<1$
  implies that $\|\betab\|_1\leqslant\varphi(\eta)$.
\end{lemma}
\begin{IEEEproof} %
  Apply Lemma~\ref{lem:1D}.

  When $C\leqslant 0$,~\eqref{eq:minor1Da} and $\varphi(\eta)<1$ imply
  that $(\|\betab\|_1-1)<0$. Since $\eta\leqslant 1$, the lower bound
  of~\eqref{eq:minor1Da} is larger than
  $1+(\|\betab\|_1-1)=\|\betab\|_1$.

  When $C>0$,~\eqref{eq:minor1Db} and $\varphi(\eta)<1$ imply that the
  minimum value of $\varphi$ on $[0,1]$ is $D/\sqrt{1+C^2}<1$, then
  $D^2-C^2<1$. Lemmas~\ref{lem:d2c2} and~\ref{lem:alpha2} imply that
  $\stdnorm{\betab}_1\leqslant 1$ and then
  $\stdnorm{\betab}_1\leqslant D/\sqrt{1+C^2}\leqslant\varphi(\eta)$.
\end{IEEEproof}

\section{Behavior of Oxx when the input vector is slightly modified}
\label{app:CN}
\begin{lemma}
  \label{lem:continuity}
  Let $\yb_1$ and $\yb_2\in\Rbb^m$. Assume that the selection
  rule~\eqref{eq:omp_rule}-\eqref{eq:ols_rule} of Oxx with $\yb_1$ as
  input is strict in the first $q>0$ iterations (the maximizer
  is unique). Then, when $\varepsilon>0$ is sufficiently small, Oxx
  selects the same atoms with
  $\yb(\varepsilon)=\yb_1+\varepsilon\yb_2$ as with $\yb_1$ in the
  first $q$ iterations.
\end{lemma} 
\begin{IEEEproof} %
  We show by recursion that there exists $\varepsilon_p>0$ such that
  the first $p$ iterations of Oxx ($p=1,\ldots,q$) with
  $\yb(\varepsilon)$ and $\yb_1$ as inputs yield the same atoms
  whenever $\varepsilon<\varepsilon_p$.

  Let $p\geqslant 1$. We denote by \Qc the subset of
  cardinality $p-1$ delivered by Oxx with $\yb_1$ as input
  after $p-1$ iterations. By assumption, \Qc is also yielded
  with $\yb(\varepsilon)$ when $\varepsilon<\varepsilon_{p-1}$. Since
  $\yb(\varepsilon)=\yb_1+\varepsilon\yb_2$, the Oxx residual takes
  the form $\rb_\Qc=\rb_1+\varepsilon\rb_2$ where $\rb_\Qc,\,\rb_1$
  and $\rb_2$ are obtained by projecting $\yb(\varepsilon)$, $\yb_1$,
  and $\yb_2$, respectively onto $\spansub{\Ab_{\Qc}}^{\perp}$. Hence,
  for $i\notin\Qc$,
  \begin{align}
    \stdscal{\rb_\Qc,\tilde{\cb}_i} &=
    \stdscal{\rb_1,\tilde{\cb}_i}+\varepsilon\stdscal{\rb_2,\tilde{\cb}_i}.
    \label{eq:decomp_resid}
  \end{align}
  Let $\ab_{\ell}$ denote the new atom selected by Oxx in the $p$-th
  iteration with $\yb_1$ as input. By assumption, the atom selection
  is strict, \ie
  \begin{align}
    \stdbars{\stdscal{\rb_1,\tilde{\cb}_{\ell}}}&> \max_{i\ne
      \ell}\stdbars{\stdscal{\rb_1,\tilde{\cb}_i}}.
    \label{eq:pas_exaequo}
  \end{align}
  Taking the limit of~\eqref{eq:decomp_resid} when
  $\varepsilon\rightarrow 0$, we obtain that for any $i$,
  $\stdbars{\stdscal{\rb_\Qc,\tilde{\cb}_{i}}}$ tends toward
  $\stdbars{\stdscal{\rb_1,\tilde{\cb}_{i}}}$.~\eqref{eq:pas_exaequo}
  implies that when $\varepsilon<\varepsilon_{p-1}$ is sufficiently
  small,
  \begin{align*}
    \stdbars{\stdscal{\rb_\Qc,\tilde{\cb}_{\ell}}}&> \max_{i\ne
      \ell}\stdbars{\stdscal{\rb_\Qc,\tilde{\cb}_i}}
  \end{align*}
  by continuity of $\stdbars{\stdscal{\rb_\Qc,\tilde{\cb}_{i}}}$
  ($i\ne\ell$) and $\stdbars{\stdscal{\rb_\Qc,\tilde{\cb}_{\ell}}}$
  with respect to $\varepsilon$. Thus, Oxx with $\yb(\varepsilon)$ as
  input selects $\ab_{\ell}$ in the $p$-th iteration.
\end{IEEEproof}

\section{Bad recovery condition for basis pursuit}
\label{app:brc_l1} 
Contrary to the OMP analysis, the bad recovery analysis of basis
pursuit is closely connected to the exact recovery analysis: in
\S~\ref{sec:overview_necess_reach}, we argued that both analyses
depend on the sign of the nonzero amplitudes, but not on the amplitude
values~\cite{Fuchs04,Plumbley07}. Here, we provide a more formal
characterization of bad recovery for basis pursuit which is based on
the Null Space Property (NSP) given in~\cite[Lemma~1]{Gribonval03}.
The NSP is a sufficient and worst case necessary condition of exact
recovery dedicated to all vectors whose support is equal to
$\Qc^\star$:
\begin{align*}
  \forall\xb\in\Nc(\Ab)\backslash\stdacc{\zerob},\,
  \sum_{i\in\Qc^\star}|x_i|&<\sum_{i\notin\Qc^\star}|x_i|
  \tag*{NSP($\Ab,\Qc^\star$)}
\end{align*}
where $\Nc(\Ab)=\stdacc{\xb:\,\Ab\xb=\zerob}$ is the null space of
\Ab.

Adapting the analysis of~\cite[Lemma~1]{Gribonval03}, we introduce the
following bad recovery condition.
\begin{proposition}
\begin{align*}
  \forall \varepsilonb\in\stdacc{-1,1}^k,\,
  \exists
  \xb\in\Nc(\Ab),\,\sum_{i\in\Qc^\star}\varepsilon_i x_i&>
  \sum_{i\notin\Qc^\star}|x_i|
  \tag*{BRC-BP($\Ab,\Qc^\star$)}
\end{align*}
is a necessary and sufficient condition of bad recovery by basis
pursuit for any $\xb^\star$ supported by $\Qc^\star$.
\label{prop:BRC_BP}
\end{proposition}
This bad recovery condition reads as the intersection of as many
conditions as possibilities for the sign vector
$\varepsilonb\in\stdacc{-1,1}^k$. We will see in the proof below that
$\varepsilonb$ plays the role of the sign of the nonzero amplitudes,
denoted by $\sgn(\xb^\star)\in\stdacc{-1,1}^k$. Therefore, the bad
recovery condition is defined independently on each orthant related to
some sign pattern $\varepsilonb\in\stdacc{-1,1}^k$.
\begin{IEEEproof}
  We first prove that BRC-BP is a sufficient condition for bad
  recovery for any $\xb^\star$ supported by $\Qc^\star$. For such a
  vector $\xb^\star$, let $\yb=\Ab\xb^\star$. Apply the BRC-BP
  condition for $\varepsilonb^\star\triangleq\sgn(\xb^\star)$: there
  exists $\xb\in\Nc(\Ab)$ such that
  $\sum_{i\in\Qc^\star}\varepsilon_i^\star
  x_i>\sum_{i\notin\Qc^\star}|x_i|$. Because this inequality still
  holds when \xb is replaced by $\alpha\xb$ (with $\alpha\ne 0$), we
  can freely re-scale \xb (\ie choose $\alpha$ small enough) so that
  for all $i\in\Qc^\star,\,\sgn(x_i^\star-x_i)=\sgn(x_i^\star)$.
  Then, we have $\stdbars{x_i^\star}=\varepsilon_i^\star
  x_i^\star=\varepsilon_i^\star(x_i^\star-x_i)+\varepsilon_i^\star
  x_i= \stdbars{x_i^\star-x_i}+\varepsilon_i^\star x_i$ and
\begin{align*}
  \|\xb^\star\|_1&=\sum_{i\in\Qc^\star}\stdbars{x_i^\star-x_i}+
  \sum_{i\in\Qc^\star}\varepsilon_i^\star x_i\FINAL{\\
  &}>\sum_{i\in\Qc^\star}\stdbars{x_i^\star-x_i}+
  \sum_{i\notin\Qc^\star}\stdbars{x_i}=\|\xb^\star-\xb\|_1.
\end{align*}
Thus, $\xb^\star$ cannot be a minimum $\ell_1$ norm solution to
$\yb=\Ab\xb$.

Now, let us prove that BRC-BP is also a necessary condition for bad
recovery. Assume that $\xb^\star$ is supported by $\Qc^\star$ and
basis pursuit with input $\yb=\Ab\xb^\star$ yields output $\xb^\star$.
Because basis pursuit yields a minimum $\ell_1$ norm solution to
$\yb=\Ab\xb$, we have for all $\xb\in\Nc(\Ab)$,
$\|\xb^\star-\xb\|_1\geqslant\|\xb^\star\|_1$, \ie
\begin{align}
\forall\xb\in\Nc(\Ab),\,
\sum_{i\notin\Qc^\star}\stdbars{x_i}&\geqslant
\sum_{i\in\Qc^\star}\stdbars{x_i^\star}-
\sum_{i\in\Qc^\star}\stdbars{x_i^\star-x_i}.
\label{eq:CN_bnsp}
\end{align}
Let $\varepsilonb^\star=\sgn(\xb^\star)$ and
$\rho=\min_{i\in\Qc^\star}\stdbars{x_i^\star}$. When
$\stdnorm{\xb}_\infty<\rho$, $x_i^\star-x_i$ and $x_i^\star$ are both
of sign $\varepsilon_i^\star$ when $i\in\Qc^\star$.
Then,~\eqref{eq:CN_bnsp} yields:
\begin{align*}
  \forall\xb\in\Nc(\Ab),\,
  \stdnorm{\xb}_\infty<\rho\,\Rightarrow
  \sum_{i\notin\Qc^\star}\stdbars{x_i}&\geqslant
  \sum_{i\in\Qc^\star}\varepsilon_i^\star x_i.
\end{align*}
This condition also holds when $\stdnorm{\xb}_\infty\geqslant\rho$
because it applies to $\rho\xb/(2\stdnorm{\xb}_\infty)$ whose
$\ell_\infty$ norm is lower than $\rho$. We have shown the
contrapositive of BRC-BP($\Ab,\Qc^\star$), \ie that
BRC-BP($\Ab,\Qc^\star$) does not hold.
\end{IEEEproof}

We performed empirical tests for specific dictionaries of dimension
($m=3, n=5$) where $\Nc(\Ab)$ is of dimension~2 and can be fully
characterized. We checked that the BRC-BP property may indeed be
fulfilled for $\Card{\Qc^\star}=2$.

\bibliographystyle{ieeeji}

\end{document}